\documentclass[%
 reprint,
superscriptaddress,
 amsmath,amssymb,
 aps,
 prx
]{revtex4-2}

\usepackage{graphicx}
\usepackage{dcolumn}
\usepackage{bm}
\usepackage{placeins}

\usepackage{xcolor}
\usepackage{amsmath}
\usepackage{amsfonts}
\usepackage{amssymb}
\usepackage{mathtools}
\usepackage{amsthm}

\begin{document}

\preprint{APS/123-QED}

\newtheorem{theorem}{Theorem}
\newtheorem*{theorem*}{Theorem}

\title{Information bounds the robustness of self-organized systems}

\author{Nicolas Romeo}
\affiliation{%
 Department of Physics, University of Chicago, Chicago, Illinois 60637, USA
}%
\affiliation{%
 Center for Living Systems, University of Chicago, Chicago, Illinois 60637, USA
}%

\author{David G. Martin}
\affiliation{%
 LPTMC, CNRS UMR 7600, Université Pierre et Marie Curie, 75252 Paris, France
}%
\affiliation{%
 Department of Physics, University of Chicago, Chicago, Illinois 60637, USA
}%

\author{Mattia Scandolo}%
\affiliation{%
 Department of Physics, University of Chicago, Chicago, Illinois 60637, USA
}%

\author{Michel Fruchart}
\affiliation{Gulliver, ESPCI Paris, Université PSL, CNRS, 75005 Paris, France}

\author{Edwin M. Munro}
\affiliation{%
 Center for Living Systems, University of Chicago, Chicago, Illinois 60637, USA
}%
\affiliation{
Institute for Biophysical Dynamics,  University of Chicago, Chicago, Illinois 60637, USA
}
\affiliation{%
 Department of Molecular Genetics and Cell Biology, University of Chicago, Chicago, Illinois 60637, USA
}%

\author{Vincenzo Vitelli}
\email{vitelli@uchicago.edu}
\affiliation{%
 Department of Physics, University of Chicago, Chicago, Illinois 60637, USA
}%
\affiliation{%
 Center for Living Systems, University of Chicago, Chicago, Illinois 60637, USA
}%
\affiliation{
Institute for Biophysical Dynamics,  University of Chicago, Chicago, Illinois 60637, USA
}
\affiliation{%
Leinweber Center for Theoretical Physics, University of Chicago, Chicago, Illinois 60637, USA
}%

\date{\today}

\begin{abstract}

Self-organized systems, from synthetic nanostructures to developing organisms, are composed of fluctuating units capable of forming robust functional structures despite noise. Here, we ask: are there fundamental bounds on the robustness of noisy self-organized systems? By viewing self-organization as noisy encoding, we prove that the positional information capacity of short-range classical systems with discrete states obeys a bound reminiscent of area laws for quantum information. We illustrate this principle with lattice models whose dynamics is captured by continuum models derived using exact coarse-graining techniques and validated through Dynamical Renormalization Group calculations. The universal bound is saturated by fine-tuning transport coefficients, which can be rationalized in the continuum limit upon considering the effects of boundaries on domain wall dynamics. 
We illustrate how this limit can be bypassed when long-range correlations are present by investigating a wave-pinning model motivated by biological mechanisms. In this class of models, global constraints reduce the need for fine-tuning by providing effective integral feedback. Our work identifies fundamental limits for the ability of natural and synthetic microsystems to self-assemble into patterns and rationalizes them on purely information-theoretic grounds.

\end{abstract}

\maketitle


\section{Introduction} \label{sec:intro}

\begin{figure*}[t]
    \centering
    \includegraphics[width=\linewidth]{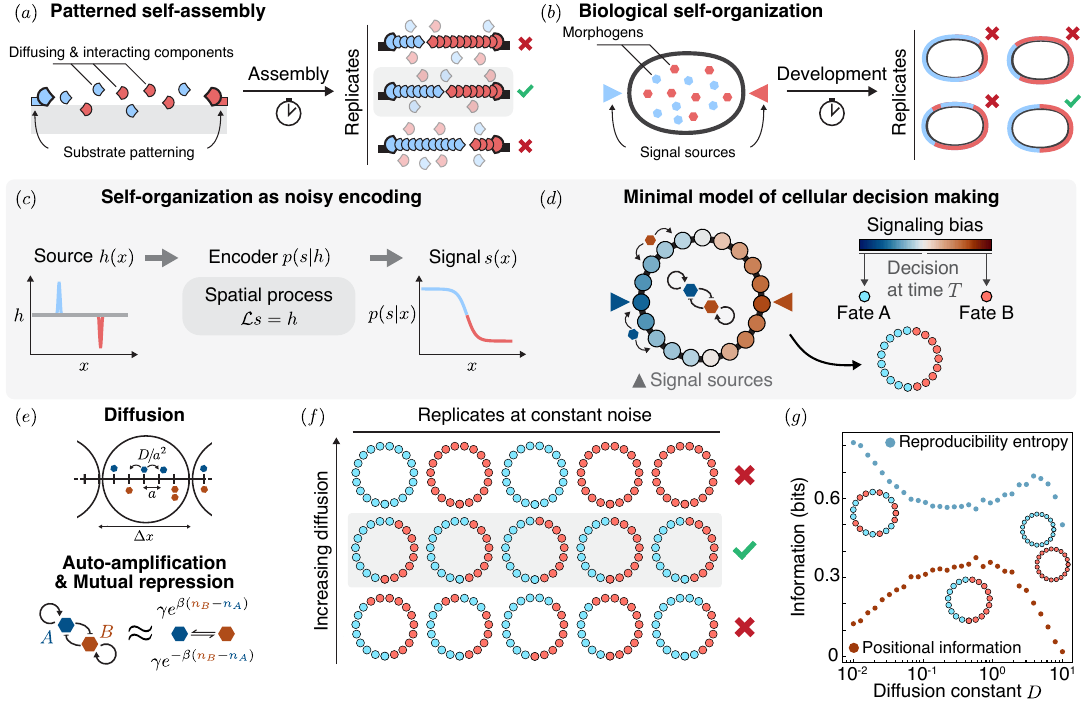}
    \caption{Reproducible self-assembly from the balance between transport and interactions. (a) Synthetic self-assembled systems with prescribed boundaries are subject to noise, which can limit yield and functionality.  (b) Viable biological self-organization requires precise establishment of chemical gradients in response to pre-existing or external symmetry breaking. 
    (c) The self-organizing process can be understood as a noisy encoding process giving rise to a probability distribution $p(s|h)$. Are there limits on the quality of the encoder depending on the nature of the spatial process, represented here by a nonlinear differential operator $\mathcal{L}$? (d) We focus on a minimal model of cellular decision-making, where two interacting species of particles diffuse around and carry opposing signals generated at diametrically opposed signaling sources. At a specified time $T$, the sign of the local difference in signal concentration sets the cell fates. (e) This system is mapped onto a diffusive Ising model, where particles diffuse at a set rate and change type according to local concentrations. (f) At fixed particle numbers, we find that cell fates are most reproducible at intermediate diffusion. (g) This reproducibility is quantified by information-theoretic quantities. While the outcomes are most reproducible at large $D$, the most informative patterns require intermediate diffusion. Simulation in (g) are done with $\Delta x = L/8$, $a=L/48$, $T=500$, $\gamma=1$, $\beta = 2\beta_c$, $h_0=2\beta$, where $\beta_c = \ln(1+\sqrt{2})$; distributions are estimated from $500$ replicates.}
    \label{fig:morpho_sys}
\end{figure*}

The self-organization of synthetic subunits to reliably assemble functional structures~\cite{hagan_equilibrium_2021,mastrangeli_self-assembly_2009, lee_fluidic_2023, miskin_electronically_2020} (Fig.~\ref{fig:morpho_sys}a) or the organization of sub-cellular components to form cells and tissues~\cite{schweisguth_self-organization_2019,collinet_programmed_2021, goehring_polarization_2011} (Fig.~\ref{fig:morpho_sys}b) rely on a complex balance of microscopic interactions \cite{tkacik_information_2025, jacobs_self-assembly_2016, freeman_feedback_2000, veneziano_designer_2016,lin_building_2018,woods_shape-assisted_2022,kim_4-bit_2022,toda_programming_2018, zadorin_synthesis_2017}, material transport \cite{lee_fluidic_2023,li_morphogen_2018,schlissel_diffusion_2024,zhu_reconstitution_2023, gierer_theory_1972} and  signal processing \cite{gregor_probing_2007,green_positional_2015, tripathi_collective_2025}.
A key challenge for robust spatial self-organization of both synthetic and living systems is the presence of noise, either thermal \cite{li_self-assembled_2024} or due to intrinsically stochastic molecular processes or other forms of disorder \cite{bassani_nanocrystal_2024, elowitz_stochastic_2002, battich_control_2015}.

Despite noise, biological signals are known to precisely generate target patterns that encode information about cellular positioning~\cite{wolpert_positional_1969, gregor_probing_2007, merle_precise_2024,moghe_optimality_2025, loo_boundary_2025}, enabling development but also intra- and inter-cellular communication at large~\cite{bryant_physical_2023,hino_erk-mediated_2020, fan_ultrafast_2023}. Take, for example, the determination of the body plan of the fruit fly embryo: Maternal gradients provide an initial signal that activates a set of interacting gap genes whose resulting expression profile is tightly controlled and enables precise localization of body parts~\cite{gregor_probing_2007,little_precise_2013}. 
More generally, biological pattern formation is enabled by both integrating external signals (sometimes called positional information~\cite{wolpert_positional_1969,green_positional_2015}) and various nonlinear self-organizing regulation schemes~\cite{stelling_robustness_2004,corson_self-organized_2017,yang_morphogens_2023, goryachev_dynamics_2008, moghe_optimality_2025, al-mosleh_how_2023, koehler_2026_flowwave}, which feature recurring ingredients such as negative feedback~\cite{freeman_feedback_2000} and the combination of short- and long-range interactions~\cite{gierer_theory_1972}.

Taking inspiration from biological systems, we would like to identify general design principles that enable robust patterning in both natural and engineered systems.
In all of these systems, individual entities (cells, particles, molecules, etc.) have to move or change state in order to make a particular pattern or shape emerge.
To this end, we view self-organization in these complex systems in terms of information processing, with information from external sources encoded into a downstream spatially-patterned signal field (Fig.~\ref{fig:morpho_sys}c)~\cite{schweisguth_self-organization_2019}. 
In this perspective, robust self-organization is equivalent to a faithful encoding of the source signal. 
In this Article, we ask: Are there general bounds on the ability of self-organizing processes to faithfully encode spatial information?

To address this question, we need a way of characterizing how well the eventual state of the entities (their ``fate'') correlates with their position. This can be assessed through a quantity known as the positional information $\mathrm{PI} \equiv I(X:F)$ which is defined as the mutual information between position $X$ and fate $F$.
By a convexity argument, we show the following theorem: in any one-dimensional system exhibiting short-range spatial correlations (exponential or sub-exponential) and with sources at boundaries, the positional information is constrained by the bound
\begin{subequations}
\label{eq:the_bound}
\begin{equation}
    \mathrm{PI} \leq \Pi_{N}^Z
\end{equation}
where 
\begin{equation}
    \Pi_{N}^Z \equiv \log_2 Z + \frac{2}{N}\sum_{i=1}^N \frac{i-1}{N-1}\log_2\left(\frac{i-1}{N-1}\right)
\end{equation}
\end{subequations}
in which $Z$ is the number of possible states (fates) of an entity, while $N$ is the number of spatial sites in the system.
In the limit of continuous space systems ($N \to \infty$), this bound converges to 
\begin{equation}
    \Pi_{\infty}^Z=\log_2 Z - \frac{1}{2\ln2}.
\end{equation}
When it applies, this bound constitutes a fundamental limitation to self-organization. 

In order to bypass the bound \eqref{eq:the_bound}, one needs to avoid the exponential decay of spatial correlations: this can be done, for instance, by putting the system out of equilibrium \cite{Bertini2015, berx_positional_2025}, or at equilibrium by considering long-range interactions~\cite{Campa2009}.
For instance, under constraints on the sources and noise that are set by microscopics which we will discuss later, the bound \eqref{eq:the_bound} applies in a system described in the continuum limit by the local equation
\begin{align}
    \partial_t s = D\nabla^2 s+r s - s^3 +  h(x) + \text{noise}
    \label{eq:phi4_intro}
\end{align}
where the bistable signal field $s$ encodes the $Z=2$ possible fates, while $h(x)$ describes a set of localized external sources.
In contrast, a different bistable system described by
\begin{align}
    \!\!\partial_t s = D\nabla^2 s+r s - s^3     - \epsilon \int \mathrm{d}x' \,s(x')  + h(x) + \text{noise} \!\!
    \label{eq:wp_mag_intro}
\end{align}
can use spatial integral feedback to provide error correction capability and increase positional information. 

From our signal-processing perspective (Fig.~\ref{fig:morpho_sys}c), the sources $h(x)$ are input signals, that are encoded and transmitted through a noisy channel represented by the noisy spatial process $s = \mathcal{L}h + \text{noise}$ [such as Eqs.~\eqref{eq:phi4_intro} or \eqref{eq:wp_mag_intro}], while the resulting field $s(x)$ can be seen as the output signal that is used by the individual entities located at position $x$ to decide what state to adopt. 
The fidelity of the encoding of the input signal depends on the nature of the spatial process $\mathcal{L}$, which strongly affects the robustness of self-organization.

To illustrate our results, we study minimal models of self-organized patterning, which are commonly modeled in the reaction-diffusion framework~\cite{green_positional_2015,kondo_reaction-diffusion_2010}. This framework is relevant for both biological processes and synthetic nanosystems~\cite{epstein_reactiondiffusion_2016}, and can be adapted to incorporate mechanical effects such as coupling to flows or elasticity~\cite{yang_morphogens_2023,mietke_self-organized_2019}. 
After defining a minimal microscopic `Diffusive Ising Model' (DIM) which contains the essential ingredients of the problem, we quantify the robustness of its patterned steady-state in simulation data in terms of information-theoretic quantities measuring the statistical relationship between location and final state~\cite{tkacik_many_2021,dubuis_positional_2013,bruckner_information_2024} (Sec.~\ref{sec:quantifying_robustness}). We find that positional information is maximized in intermediate diffusion regimes, a fact that is rationalized in Sec.~\ref{sec:domain_walls} by studying the interactions between domain walls and sources dynamics in a fluctuating hydrodynamic formulation of the model obtained using exact coarse-graining methods, validated against Dynamical Renormalization Group calculations. Returning to the information-theoretic perspective, we then proceed to show Eq.~\eqref{eq:the_bound} in Sec.~\ref{sec:PI_is_limited}. We there demonstrate this result numerically on both the DIM and the common Ising model, and note that this analytical finding  -- that systems with short-ranged interactions are subject to information theoretic limitations -- is reminiscent of area laws for quantum information~\cite{wolf_area_2008,eisert_colloquium_2010}. Finally, we show that mixing short- and long-range interactions, generically known to lead to pattern formation and a common feature of many self-organized systems, also promotes their robustness. We show this using a `wave-pinning' variant of the DIM model which we study in Sec.~\ref{sec:nonlocal}, which compares to Eq.~\eqref{eq:wp_mag_intro}. Long-range interactions can thus provide a basis for robust self-organization, a finding that could rationalize why scale separation and hierarchical structures -- which induce long-range interactions -- are common motifs in biology, and could underlie a valuable bio-mimetic strategy for engineered nanosystems.


\section{Quantifying robustness of self-organized patterns}
\label{sec:quantifying_robustness}

Before diving into the particulars of our model and analysis, let us introduce a motivating example: our main illustration of self-organization is inspired by the prototypical problem of cell fate patterning on a one-dimensional ring (Fig.~\ref{fig:morpho_sys}d). This situation is relevant to biological development ~\cite{corson_self-organized_2017,goehring_polarization_2011}, in which cells decide on a fate (cell type) based on their readout of self-organized signals. These signals---known in the developmental context as morphogens---are emitted by external or spontaneously generated sources. The morphogens then diffuse and interact, either directly or through state-dependent expression or degradation by the host cell. In our minimal cell fate model, morphogens are subject to mutually repressing and autoamplifying interactions, represented respectively by blunt and pointed arrows in Fig.~\ref{fig:morpho_sys}d. This is a common interaction motif leading to bistability in cell decision. To model cellular decision, at a time $T$ cells pick a fate (color) based on the morphogen that is the most present, sampling the steady state distribution if $T$ is large. Under suitable interpretation of the variables, these elements of diffusion, interactions, and external signal processing qualitatively encapsulate the physics at play in diverse self-organized systems.
 For example, Fig. 1a shows a system of interacting colloidal particles which, given a patterned substrate, assemble into a functional structure: this setup mimics complex self-assembly processes that have been realized in the lab~\cite{yin_formation_2004}. Here, the equivalent of cellular decision is the `freezing-in' of the structure by irreversible chemical reactions.

In the rest of this section, we introduce a minimal lattice model abstracting the situation above, which provides with a tractable system to study the emergence of robust self-organization. To analyze the stochastic outcomes of the resulting dynamics, we then consider information-theoretic measures of robust self-organization; we find that the positional information (PI) is a good measure of robust self-organization, and that it is maximized in intermediate diffusion regimes, which we will examine further in Sec.~\ref{sec:domain_walls}.

 \subsection{Minimal Diffusive Ising model} \label{sec:main_DIM}

We now define our Diffusive Ising lattice model (DIM) which captures the minimal ingredients of sources, transport, and interactions. In this model, sometimes known as the Active Ising Model \cite{solon_flocking_2015,scandolo_active_2023, martin_transition_2025}, the ring of cells, each of size $\Delta x$, is abstracted into a one-dimensional lattice on which morphogens randomly hop between $N_s$ sites of width $a$ with an average rate $D/a^2$, where $a$ stands for the interaction range between particles. The density of morphogens $\rho_0/a$ is taken to be constant such that on average we have $\rho_0 =1$ morphogen per site; there are $\Delta x/a$ sites per cell. Morphogens come in two species $A$ and $B$, which mutually suppress each other by changing their type according to local concentrations $n_A$ and $n_B$: if there are more $A$ than $B$, then morphogens of type $B$ are likely to turn into $A$, and vice versa (Fig.~\ref{fig:morpho_sys}e). These `flipping' dynamics are modified in the presence of sources  which bias the flipping rate.

More precisely, the microscopic evolution rules are the following:
\begin{enumerate}
    \item[\textbf{D1}] Particles hop with equal probability to a neighboring site with rate $D/a^2$,
    \item[\textbf{D2}] $A$ particles turn into $B$ with rate $\gamma e^{\beta(n_B-n_A)+h_i}$ at site $i$, where $n_A$ and $n_B$ are the number of $A$ and $B$ particles on the same site and $h_i$ is the local source bias,
    \item[\textbf{D3}] Similarly, $B$ particles turn into $A$ with rate $\gamma e^{-\beta(n_B-n_A)-h_i}$ at site $i$.
\end{enumerate}
These rules lead to Markovian dynamics expressed in terms of a master equation. We use a tau-leaping scheme implemented in \texttt{julia} to simulate these stochastic dynamics until convergence to steady-state, as described in \cite{martin_transition_2025, solon_flocking_2015}.

We create a symmetric pattern by introducing localized sources at opposing locations, implemented by introducing a biasing signal $h_i = h_0 (\delta_{i,0}-\delta_{i,N_s/2})$ at site $0$ and $N_s/2$ that strongly favors particles of type $A$ at one end and of type $B$ at the other. Simulating this random process at constant noise levels for a duration $T$ and reaching steady-state, we find that the distribution of resulting configurations for different realizations displays variable reproducibility and functionality. At large diffusion, we have reproducible, but non-patterned outcomes; at low diffusion, we obtain patterned but variable outcomes. At intermediate diffusion, we find an optimal regime of patterned and reproducible outcomes, which is characteristic of \emph{robust} patterning. (Fig.~\ref{fig:morpho_sys}f). This observation can be intuitively understood: In systems where $D=0$, there is no neighbor-to-neighbor communication. The signal is then localized at the source points and sites away from the source have little information to work with. If $D\rightarrow\infty$, the system becomes homogeneous and all spatial information is destroyed.

\subsection{Positional information of a pattern}

We now quantitatively characterize this qualitative interpretation by using information-theoretic tools to define reproducibility and pattern robustness. For a system of $N$ cells and $Z$ possible states, we consider the joint distribution $p(s_1,\ldots, s_N) \equiv p(\{s_i\})$ that encodes the probability of finding together cell $1$ at state $s_1$, cell $2$ at state $s_2$ and so on~\cite{tkacik_many_2021,dubuis_positional_2013, bruckner_information_2024}. 
Introducing the marginal probability $p_i(s)$ of finding the $i$-th cell in state $s$, which can be expressed in terms of $p(\{s_i\})$, we consider the positional information (PI) defined as
 \begin{equation}
     \mathrm{PI} = \frac{1}{N}\sum_{i=1}^N\sum_{s=1}^Z p_i(s) \log_2\left(\frac{p_i(s)}{P_s}\right) \leq \log_2Z = 1 \text{ bit} \label{eq:PI_main}
 \end{equation}
where $P_s$ is the probability of finding any cell in the state $s$, reflecting the average fraction of cells of each type. The PI is perhaps best understood as the mutual information between position and state, that is, the information gained on the location of a particle by knowing its fate, and vice versa. Rewriting the PI in terms of the patterning entropy $S_\mathrm{pat} \equiv S(P_s)$ as~\cite{bruckner_information_2024}
\begin{equation}
   \mathrm{PI}= S_\mathrm{pat} - \frac{1}{N}\sum_{i=1}^N S(p_i) \label{eq:PI_methods}
\end{equation}
where $S(p) =-\sum_k p_k \log_2p_k$ is the Shannon entropy of the distribution $p$, we see that $\mathrm{PI}$ reflects both information about the overall system state through $s_\mathrm{pat}$ and local state probabilities through the entropies of the marginal distributions $S(p_i)$. For symmetric systems, the probability of picking a cell in each state, regardless of its position, is equal to $S_\mathrm{pat}=\log_2Z$. Additionally, each cell is equally likely to take any of the $Z$ possible fates, and $\mathrm{PI}=0$ as all marginal entropies $S(p_i)=\log_2Z$. Thus, the maximal value $\mathrm{PI}_{\rm max}=\log_2Z$ of the positional information is achieved for symmetric systems where the marginals $p_i(s)$ are deterministic such that all $S(p_i)=0$.
Reciprocally, the minimum $\mathrm{PI}_{\rm min}=0$ is also reached for symmetric systems where the marginals $p_i(s)$ are maximally random, namely $p_i(s)=1/Z$.

From the biological application side, the PI formalizes early insights by Wolpert, who suggested that morphogen gradients could encode spatial information. 
It has further been shown that PI is experimentally-relevant, defining the optimality of cellular position encoding schemes \cite{wolpert_positional_1969, petkova_optimal_2019, sokolowski_deriving_2025}. Note that our findings remain unchanged with other probabilistic measures to quantify robustness~\cite{peyre_computational_2020,lin_divergence_1991}: we discuss the relative merits of these alternative measures in Appendix~\ref{sec:info_measures}, Fig.~\ref{fig:infomeasures}.
For instance, the reproducibility of outcomes can also be quantified by the reproducibility entropy $S_\mathrm{rep} = - \sum_{\{s_i\}} p(\{s_i\})\log_2 p(\{s_i\})$, but this measure does not account for the presence of a pattern: while $S_\mathrm{rep}$ has a minimum at $D^*$, it also displays a second extremum at large diffusion when the system is uniform~(Fig.~\ref{fig:morpho_sys}g).
At the opposite, we observe that positional information peaks at intermediate diffusion $D^*$, indicating an optimal robustness of the pattern (Fig~\ref{fig:morpho_sys}f).
We thus choose to focus on PI for the remainder of the paper.


\section{Domain wall dynamics set robustness of pattern}
\label{sec:domain_walls}

In our minimal model, microscopic transport and interactions rules conspire, on average, to form an emergent bipartite domain. The robustness of this pattern peaks at intermediate diffusivity $D^*$, as shown in (Fig~\ref{fig:morpho_sys}f).
In this section, we go beyond the intuitive trade-off argument between spatial incoherence VS homogeneity at the origin of this optimum. To this aim, we study the collective dynamics of the DIM through its continuum limit. 
We then validate the ensuing predictions by comparing the measured average response to the theoretical effective response calculated by Dynamical Renormalization Group methods. This hydrodynamic perspective allows us to rationalize the existence of $D^*$ through the influence of external signal sources on domain wall dynamics: while domain walls usually diffuse in noisy environments, the presence of boundaries imposes a confining potential. Importantly, the characteristic width of the domain wall must be of the order of the system size for such confinement to occur.

\subsection{Emergence of collective behavior in finite systems} \label{sec:main_continuumDIM}

To understand the emergent collective behavior of our spatially-extended assemblies of interacting particles, we build a fluctuating hydrodynamics which focuses on the large-scale dynamics of the stochastic system. 
In this description, valid at large spatial scales, the lowest-order contributions from the noise are Gaussian. Starting from diffusive microscopic dynamics described by a master equation, we use exact coarse-graining methods to obtain a fluctuating hydrodynamic description in terms of a Stochastic Partial Differential Equation (SPDE)~\cite{martin_transition_2025,andreanov_field_2006, kourbane-houssene_exact_2018}. 
This procedure is exact in the limit of $a\rightarrow0$~\cite{de_masi_rigorous_1985,de_masi_reaction-diffusion_1986}. To quantify the error introduced by a finite $a$, we compare the observed average response to the effective deterministic response calculated using Dynamical Renormalization Group (DRG) techniques. The overall coarse-graining procedure is summarized in Fig.~\ref{fig:from_lattice_to_continuum}a-b.

\subsubsection{Fluctuating hydrodynamics of the DIM}

We use path-integral techniques to reduce the lattice dynamics described by rules \textbf{D1-3} to a set of nonlinear reaction-diffusion equations for the number density $n_A$ and $n_B$ of particles of type $A$ and $B$. The calculations are reported in Appendix~\ref{sec:appendix_cg}. In a nutshell, this procedure relies on Poisson statistics, which is an exact solution of purely diffusive dynamics. 
By considering interactions as a perturbation to diffusion, we reduce the Markovian dynamics to differential equations on the parameters of the Poisson statistics. The resulting fully nonlinear equations on the density fluctuations $\delta \rho = n_A+n_B - \rho_0 $ and signal $s = n_A-n_B$, valid for all values of $\beta$, are provided in Appendix~\ref{sec:app_DIM}. If $\sinh{\beta} \lesssim 2$, to cubic order in $s$, these equations are given by
\begin{subequations}
    \begin{align}
    \partial_t \delta \rho  = &\, D\partial^2_{xx} \delta\rho + \partial_x \eta_1, \\
    \partial_t s  = & \, D\partial^2_{xx} s - (r+\mu \delta \rho) s - u s^3  \nonumber\\&\, + h(x) + \partial_x \eta_2 +  \eta_3.
\end{align} \label{eq:fluct_hydro_cube}%
\end{subequations}
where $\eta_1,\eta_2,\eta_3$ are Gaussian noise fields, and the source field is given by $h(x) = \bar{h} [\delta(x) - \delta(x-L)]$ for the symmetric source setup with microscopic source intensity $\pm h_0$. Defining $\hat{\gamma} = 2\gamma e^{-\beta +(\cosh{\beta}-1)\rho_0}$, we obtain the coefficients in terms of the microscopic parameters
\begin{subequations}
\begin{align}
    r & = \hat{\gamma} ( 1- \rho_0\sinh\beta)\\
    u & = \hat{\gamma} \sinh \beta /3 \\
    \mu &= (\cosh{\beta}-1)r - 3u.\\
    \bar{h} & = arh_0/\beta
\end{align} %
\end{subequations}
and the noise correlations are then given by $\langle \eta_n(x,t) \eta_{n'}(x',t')\rangle = M_{n,n'} \delta(x-x') \delta(t-t')$
\begin{equation}
    M_{nn'} = \begin{pmatrix}
        2aD\rho_0  &   &  \\
         & 2aD\rho_0  &  \\
         &  & 2a\rho_0\hat{\gamma}
    \end{pmatrix}.
\end{equation}
If $r > 0$, that is, if $\beta < \beta_c =\ln(1+\sqrt{2})$, then the equations only have a stable homogeneous solution at $s=0$. If $\beta > \beta_c$, then the equation has two non-zero homogeneous solutions, which is the regime we focus on in the following. 

In this model, the noise variance is proportional to $a$, or equivalently scales with $1/N_{\Delta x} =  a/\Delta x$ the average number of particles per cell~\cite{weber_master_2017,scandolo_active_2023,martin_transition_2025}. This is essentially a restatement of the central limit theorem: as the number of particles considered grows, the importance of fluctuations diminishes, and the quality of the deterministic continuum description improves.

\subsubsection{Validity range of the fluctuating hydrodynamics}

\begin{figure*}
    \centering
    \includegraphics[scale=0.95]{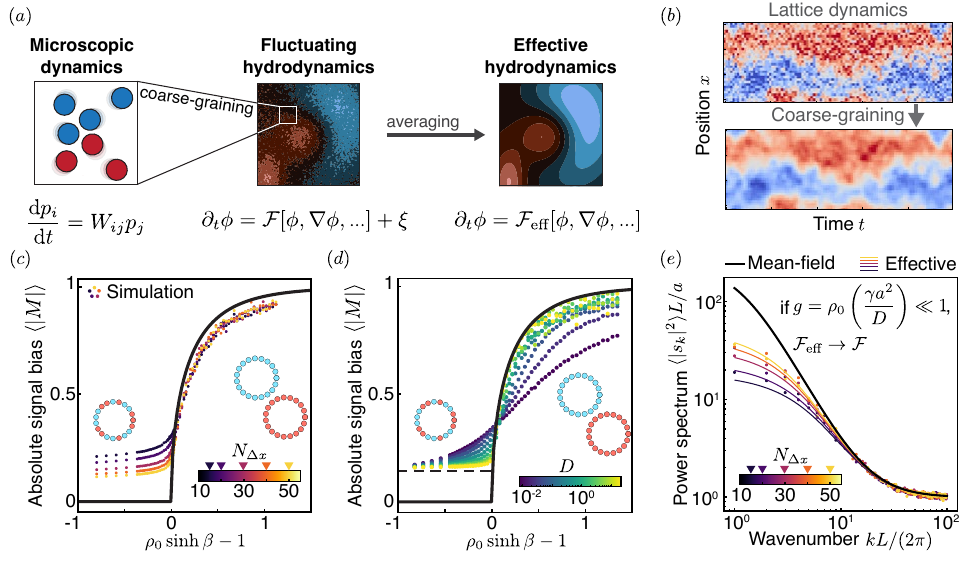}
    \caption{Understanding the emergence of collective behavior in finite systems. (a) Starting from diffusive microscopic dynamics described by a master equation, we use exact coarse-graining methods to obtain a fluctuating hydrodynamics description in terms of a SPDE. Spatio-temporal averaging of fluctuating dynamics gives an effective deterministic hydrodynamic model. (b) This process allows us to understand the emergent collective dynamics and use continuum modeling tools even in small discrete systems. (c-d) We find that the collective response from microscopic simulations rapidly converge to their continuum predictions (black lines) with increasing particle numbers (c) and diffusivity (d) Here, $|M| = \sum_\text{cells}|s_i|/N_\text{cells}$, with $N_\text{cells} = L/\Delta x$. Dashed lines indicates expected absolute signal bias for a finite coarse-grained domain size $\Delta x$. (e) The linear response in Fourier domain from simulations is well-described by the renormalization predictions, validating that the fluctuating hydrodynamic picture is accurate even in small systems with less than tens of particles per coarse-graining domain.}
    \label{fig:from_lattice_to_continuum}
\end{figure*}

The fluctuating hydrodynamics description is valid as $a\rightarrow0$. To get a better sense of the validity range of the SPDE, we perform a self-consistent verification based on estimating the effective deterministic response using Dynamical Renormalization Group (DRG) methods \cite{tauber_critical_2014} . 
In a $d$-dimensional system within the `disordered' regime, where $\beta < \beta_c$, the correlator of the deterministic theory is given, in Fourier space, by
\begin{align}
    \langle s_qs_{q'}\rangle_0 = \delta_{q,q'}a^d\rho_0\frac{Dq^2 +\hat{\gamma}}{Dq^2+r}. \label{eq:corr_mef}
\end{align}
where $q$ is the wavenumber. The correlator in Eq.~\eqref{eq:corr_mef} captures the linear response of the system to fluctuations in the limit $g=\rho_0 (\gamma a^2 /D) \ll 1$ where the amplitude of fluctuations is vanishing. For finite values of $g <1$, the effective linear response is recast in terms of scale-dependent \emph{renormalized} response coefficients $D_m^R(q), \tilde{D}_m^R(q), \tilde{\gamma}(q), r_R(q)$ such that
\begin{equation}
    \langle |s_q|^2\rangle  =  a^d\frac{\tilde{D}_m^R(q) q^2 +\tilde{\gamma}_R(q)}{D_m^R(q) q^2 +r_R(q)}.
\end{equation}
We evaluate the renormalized coefficients to first order in $g$ in Appendix~\ref{sec:RG_calc}.  In practice, we find that the numerical results from the lattice simulations recover the deterministic limit (Fig.~\ref{fig:from_lattice_to_continuum}c-d) as expected when $D\rightarrow \infty$ and $a \rightarrow 0$, and that the effective hydrodynamic equations derived from the fluctuating model are predictive even with systems with around 10 particles per coarse-grained unit length, indicating the validity of this stochastic PDE as a basis for understanding the collective dynamics of these systems. (Fig.~\ref{fig:from_lattice_to_continuum}e).

As an aside, we note that in statistical physics terms the validity criterion  $g=\rho_0 (\gamma a^2 /D) \ll 1$ is a kind of Ginzburg criterion. It delineates the regime in which the mean-field approximation holds by quantifying the crossover regime between Gaussian and non-Gaussian fixed points~\cite{cardy_scaling_1996}. However, unlike the standard Ginzburg criterion, which states that deviation from mean-field behavior depends on the dimension $d$ of the space, the criterion $g\ll 1$ is dimension-independent. Additional discussion of our scaling results is presented in Appendix~\ref{sec:RG_calc}.

\begin{figure*}
    \centering
    \includegraphics[scale=0.95]{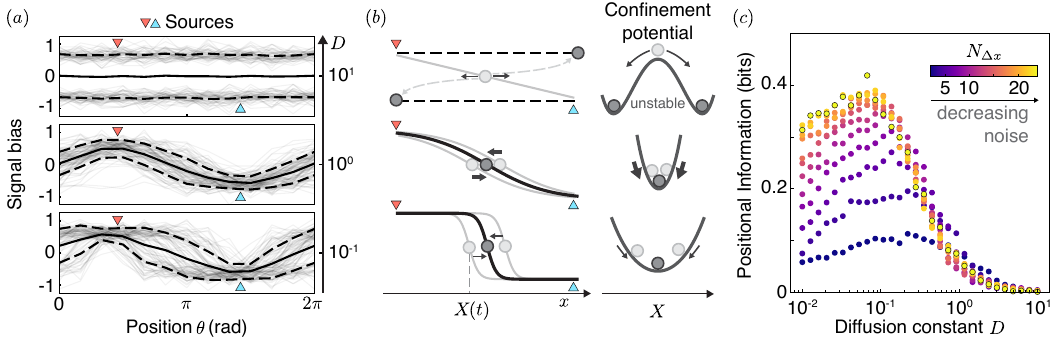}
    \caption{Pattern robustness requires fine-tuned domain wall width in the Diffusive Ising model. (a) Variance in domain position is minimal when domains walls have a width $\ell \approx L$. If $\ell$ is too large, then domain wall solutions are unstable. Signal bias is here normalized by average total density. Full black line indicates mean signal, dashed lines one standard deviation, lighter lines are replicates. (b)  This optimum in variance can be understood through the energetic cost of shifting the domain wall: if $\ell \ll L$, then translating the domain wall changes very little the value of the signal at the boundary $\Delta s\approx 0$. If $\ell > L$ then the front solution is unstable. These effects materialize as a sharply peaked confinement stiffness. (c) Positional information is maximal at intermediate diffusion, and saturate as the particle number density $N_{\Delta x}$ increases. Here, we have $N=16$ cells.
     All simulations have $h_0=3\beta, \beta=3\beta_c/2$, $\gamma=1$ and are averaged over $500$ replicates for information calculations.}
    \label{fig:domain_walls_main}
\end{figure*}

\subsection{Domain wall dynamics and pattern stability}
\label{sec:domainwall}

Now that we have established the continuum limit of the DIM and its region of validity, we rationalize the optimal diffusivity for robust patterning through the lens of continuum domain wall dynamics.
For sufficiently small lattice sizes $a$, the continuum model presented in Eq.~\eqref{eq:fluct_hydro_cube} has the same phenomenology as the following Landau-Ginzburg-type model with dynamics for $s$ given by
\begin{align}
    \partial_t s = D\partial_x^2 s+ r_1 s- u_1 s^3 + h(x) + \sqrt{a\Theta}\zeta_1(t,x),\label{eq:phi4}
\end{align}
with Brownian noise $\zeta_1(x,t)$ and positive coefficients $r_1, u_1,\Theta$ and where the term $h(x)=ah_0\left(\delta(x)-\delta(x-L)\right)$ accounts for the presence of the external signal at $x=0,L$ with strength $\pm h_0$. The coefficients $r_1,u_1$ reflect the timescale of the reactions and the amplitude of the steady state, while $\Theta$ is the reaction noise; as above, the noise variance is scaled by the lattice size $a$. Eq.~\ref{eq:phi4} is an appropriate description of the DIM as long as $a$ is small enough and the coefficients are appropriately scaled by $a$ (Appendix~\ref{sec:app_landau}, Fig.~\ref{fig:phi4}). 

Eq.~\eqref{eq:phi4} allows for homogeneous solutions $s_0=\pm \sqrt{r_1/u_1}$, but also supports the emergence of our desired pattern as a domain wall centered at a position $X(t)$, with morphogens of different types segregating on either side of the wall. Such solutions of the continuum version of this model are given by hyperbolic tangents $s(x) = s_0 \tanh{[(x-X(t))/\sqrt{2}\ell]}$ with a characteristic width $\ell$ set by the diffusion constant as $\ell =\sqrt{D/r_1}$. 

When there are no boundaries or external signals, the presence of noise induces a Brownian diffusion of the wall position with diffusivity $D_f \propto a\sqrt{r_1D}$. Interestingly, the front diffusivity $D_f$ scales as $\sqrt{D}$: this is a consequence of the variance scaling $D_f\sim D/N_\text{int}$, where $N_\text{int}\sim\sqrt{D}$ is the number of particles within interaction range between reactions~\cite{khain_velocity_2013,meerson_velocity_2011,birzu_fluctuations_2018}, or as $D_f \sim a v_F$ where $v_F\sim \sqrt{r_2 D}$ is the characteristic front propagation speed (Appendix~\ref{sec:si_domainwall}). 

In the presence of boundaries, but without noise, the domain wall localizes in the middle by symmetry.
Examining the spatial profiles of the signal in simulations of the DIM with varying diffusion, we observe that maximizing PI configurations require domain wall widths $\ell \approx L$, where $L$ is the distance between sources, so that the entire system has size $2L$. 
If $\ell \ll L$, the domain walls can be anywhere in between the signaling loci. If $\ell$ is too large $(\ell \gg L)$, then the domain wall solutions are unstable (Fig.~\ref{fig:domain_walls_main}a). 

These observations suggest a qualitative picture where the position $X(t)$ in the presence of external signals does not simply diffuse, but has a variance set by the competition between the tendency to conform to the external cue and the strength of the noise. To quantify this competition, we use techniques from nonlinear front propagation theory~\cite{rocco_diffusion_2001,birzu_fluctuations_2018} in Appendix~\ref{sec:si_domainwall} to compute perturbatively the restoring force acting on the domain wall following a displacement from its average middle position $X_0$. This model predicts a confining force $f_\mathrm{c}(X) = - k_\mathrm{c}(X-X_0)$, resulting in an equation~\cite{gardiner_handbook_1994}
\begin{align}
    \frac{\mathrm{d}X}{\mathrm{d}t} & = f_\mathrm{c}(X) +\sqrt{2D_f}\zeta(t),
\end{align}
where $\zeta(t)$ is a standard Brownian noise, and where the stiffness $k_\mathrm{c}$ is given by
\begin{align}
    k_\mathrm{c} &= \frac{3h_0}{\sqrt{2}\ell}\frac{\sinh\left(L/(2\sqrt{2}\ell)\right)}{\cosh\left(L/(2\sqrt{2}\ell)\right)^3}
\end{align}
Hence, the larger $k_\mathrm{c}$, the more confined the domain wall. Our geometric model predicts that the stiffness $k_\mathrm{c}$ is proportional to the curvature of the signal profile at the sources, and has a maximal value $k_c^*  \approx 1.9 h_0/L$  for $\ell \approx 0.35 L$. The variance in the front position $\sigma =\sqrt{D_f/k_\mathrm{c}}$ hence has a sharp minimum at intermediate $\ell/L$ and increases exponentially as $D$ is reduced, but does not vanish as $a\rightarrow0$ (Fig.~\ref{fig:domain_walls_main}b, Appendix \ref{sec:si_domainwall}). If $\sigma$ is of the order of the system size $L$, then the front is unstable. In practice, in direct simulations of the reduced continuum model, we find that there is a weaker scaling of the variance of the front position with $D$, likely due to the front shape deviating from hyperbolic tangent (Appendix~\ref{sec:si_domainwall}, Fig.~\ref{fig:peclet}). 

These results indicate that accurate domain wall positioning requires fine-tuned values of the diffusivity $D$ such that $\ell \sim L$, an undesirable property given that diffusion constants span orders of magnitude depending on the molecular environment~\cite{milo_cell_2015}.

Here, we remark that in Eq.~\eqref{eq:phi4} with microscopically derived parameters, since both $h_0$ and $D_f$ scale linearly with the noise amplitude $a$, the variance $\sigma$ is independent of $a$ at first order: there is a non-vanishing optimal confinement length as the noise amplitude goes to 0, in accordance with our numerical observations. 
The existence of this optimum depends on the coupling between source strength and lattice size: in non-equilibrium models where this microscopic constraint is not present, 
it is possible to control $D$ and noise amplitude separately. Such unconstrained systems can have $|s|>s_+$, which allows the variance $\sigma$ to vanish for any fixed value of $D$ as the noise amplitude goes to $0$, leading to arbitrarily strong domain wall confinement and large positional information (Appendix~\ref{sec:app_landau}, Fig.~\ref{fig:phi4}).

\section{Positional information is limited in short-range systems}
\label{sec:PI_is_limited}

Since, to first order, the variance of the front position does not vanish as $a\rightarrow 0$, reducing the noise intensity by increasing the particle number does not nail the desired pattern: diffusion always need to be fine-tuned in the DIM. 
In fact, we find numerically that the maximum value of PI saturates as the noise amplitude decreases ($N_{\Delta x} \rightarrow \infty$) at a value smaller than its theoretically maximum value of $\log_2 Z = 1$ bit for our two-state system (Fig.~\ref{fig:domain_walls_main}c).
In this section, we show that this saturation of positional information is a non-perturbative feature of systems with short-range correlations. 
First, we show that it is present in other systems as well, namely an Ising magnet with external fields, a prototypical model with short-range interaction for which marginal probabilities can be exactly computed. It also appears in generalizations of the Ising model with $Z>2$ (Appendix~\ref{sec:potts_si}, Fig.~\ref{fig:potts}). We then show that it is a consequence of the correlation structure of the models: the existence of a finite correlation length leads to bounded positional information, which is reminiscent of area laws in quantum information theory~\cite{eisert_colloquium_2010,wolf_area_2008}. Finally, we show that different correlation structures lead to different bounds on PI, with `more concave' correlations leading to higher PI.

\subsection{Alternative model of short-range patterning}
\label{sec:ising}

\begin{figure*}
    \centering
    \includegraphics[scale=0.95]{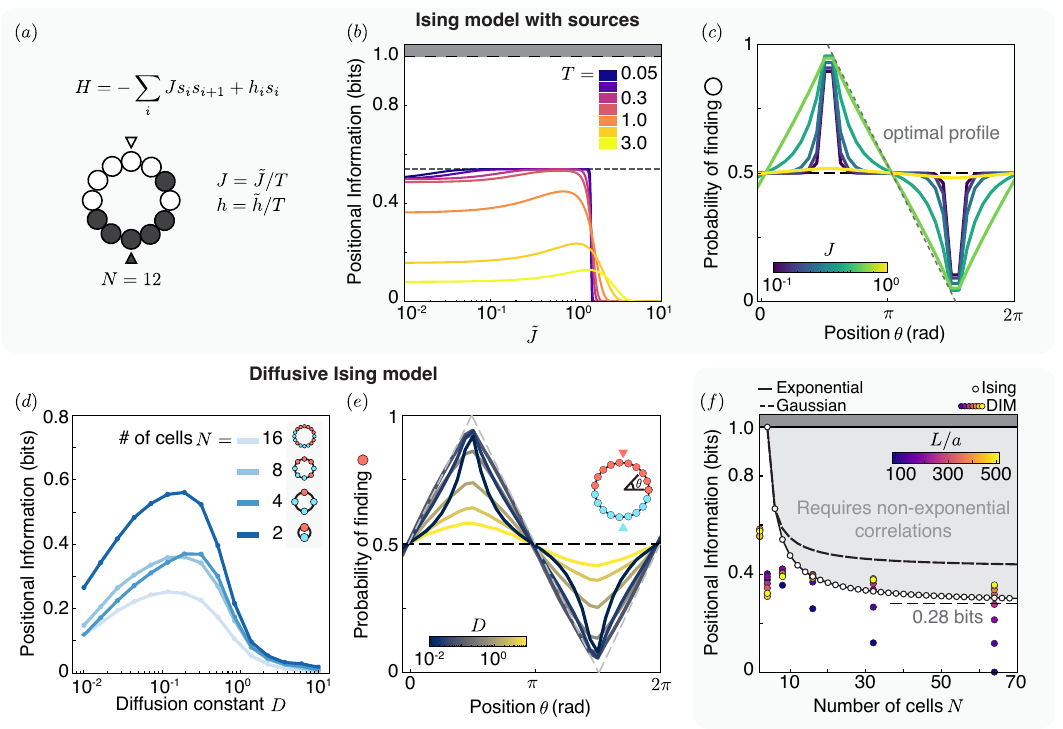}
    \caption{A universal bound on positional information in short-ranged correlated systems. (a) We consider a 1D Ising chain with broken translational symmetry by having non-zero opposing magnetic field $\pm h$ at two diametrically-opposed points.  $J$ is related to the correlation length in this model by $\ell \sim 1/\ln J$. Here states are represented as white and black dots. (b) As we vary the coupling strength for a given magnetic field strength $
    \smash{\tilde{h}}=3$ for variable temperature $T$, we find a similar optimal coupling $\tilde{J}^*$ as in the DIM. Interestingly, as we reduce temperature (noise amplitude), the optimal range of $\tilde{J}$ expands but the PI stays bounded. Here, $N=8$. Dashed line indicates theoretical prediction for the bound. (c)  This saturation can be explained by considering the relationship between positional information and the marginal probability $p(\theta)$ of finding a red region at $\theta$: in short-ranged correlated systems, PI is maximal when $p(\theta)$ is piecewise linear; in diffusive systems, a different optimum exists. Small $J$ lead to exponentially-peaked probabilities around source points, while large $J$ homogenize the system. Optimal values of $J$ lead to a sawtooth profile.
    (d) In the DIM, the maximal value of PI at constant particle density $a = L/96$ depends on the number of cells $N=L/\Delta x$. This is also true for the Ising model (Fig.~\ref{fig:Ising_varN}).
    (e) Marginal probability profiles are similar to the ones in the Ising, albeit with some slight non-convexity that stems from the diffusive aspects of the model. (f) In short-ranged correlated systems, PI is maximal when $p(\theta)$ is piecewise linear; in diffusive systems, a different optimum exists. This constraint leads to a system-size dependent bound on PI for systems with short-ranged correlations. Simulations in (d-e) have $h_0=3\beta, \beta=3\beta_c/2$, $\gamma=1$ and are averaged over $500$ replicates.}
    \label{fig:morpho_analysis}
\end{figure*}

To show that the saturation of $\mathrm{PI}$ with decreasing noise is a feature of short-range correlated models, we consider the Ising model with an external but localized magnetic field. In a periodic lattice with $N=2M$ sites, the Ising Hamiltonian $H$ is given by~\cite{kardar_statistical_2007}
\begin{equation}
    H(\{s_i\}) = -\sum_{i=1}^N Js_i s_{i+1} + h_is_i
\end{equation}
with the convention that $s_{N+1} \equiv s_1$, and where the source is patterned according to $h_i = h ( \delta_{i, 1} - \delta_{i, M})$, where $\delta_{i,j}$ is the Kronecker symbol (Fig.~\ref{fig:morpho_analysis}a).
The steady-state probability distribution $p(\{s_i\})=e^{-H(\{s_i\})}/\mathcal{Z}$, where $\mathcal{Z}=\sum_{\{s_i\}}e^{-H(\{s_i\})}$, is computable using the transfer matrix formalism, which allows us to efficiently compute all probabilities and information-theoretic quantities of interest (Fig.~\ref{fig:morpho_analysis}b-c; see Appendix~\ref{sec:app_infoIsing} for computational details and Fig.~\ref{fig:Ising_varN} for results with variable $N$).  We also note that other lattice systems, such as symmetric exclusion processes, have been found to be bounded in their positional information~\cite{singh_limits_2025}.

\subsection{Classical area laws for positional information}

The saturation of positional information thus seems to be a feature of multiple models. How can we understand such saturation?

Consider the PI written in terms of the patterning entropy $S_\mathrm{pat} \equiv S(P_s)$ as
\begin{equation}
   \mathrm{PI}= S_\mathrm{pat} - \frac{1}{N}\sum_{i=1}^N S(p_i). 
\end{equation}
If the system is stochastic, we examine the influence of correlations on the maximum value of $\mathrm{PI}$ by considering the mutual information $I(F_i : F_j)$ or connected correlation $C(F_i, F_j)$ between $F_i$ and $F_j$ the random variables encoding the fates (states) of the cells at positions $i$ and $j$.
We find that the presence of pairwise correlations bounds the PI below its maximum value $\log_2Z$ as 
\begin{subequations}
\begin{align}
    \mathrm{PI} &\leq \log_2Z -  \frac{1}{N^2}\sum_{i,j} I(F_i:F_j) \\
    &\leq \log_2Z  -\frac{1}{2N^2Z^2}\sum_{i,j} C(F_i,F_j)^2.
\end{align}\label{eq:ineqs_PI} %
\end{subequations}
which stems from the inequality $0\leq I(F_i:F_j) \leq S(p_i)$ for any $j\in \{1,\ldots,N\}$~\cite{cover_elements_2005}. A more detailed derivation is presented in Appendix~\ref{sec:app_PIspatial}.

The presence of spatial correlations in a stochastic system can thus counter-intuitively decrease positional information: Informally, to have high PI, one needs signals that are as distinct as possible in every place, which is difficult to achieve in strongly spatially-correlated systems: uniform systems have maximal spatial correlations, but have $\mathrm{PI}=0$.

It is difficult to say more without further constraints on the structure of spatial interactions: increasing spatial correlations without an overall reduction in uncertainty in the marginals $p(F_i)$ reduces PI. In the Ising model, by examining the marginal probability $p_i = p_i(z=+)$ of finding the cell at position $i$ in state +, we find that $\mathrm{PI}$ is maximized in this system when it takes on a sawtooth (piecewise linear) profile, which does not saturate the $1$-bit bound (Fig~\ref{fig:morpho_analysis}c). The saturating value for the optimal positional information depends on the number of coarse-grained states $N$, which set the resolution of the spatial patterns. A larger $N$ implies a larger possible state space, and the patterns must be more precise to obtain a similar $\mathrm{PI}$, as seen in Fig.~\ref{fig:morpho_analysis}d for the DIM (Fig.~\ref{fig:Ising_varN} shows the equivalent results for the Ising model).

\subsection{Linear bound in locally-interacting discrete-state systems} \label{sec:sawtooth_bound}

We now show that this sawtooth  profile is optimal for short-range correlated systems: More precisely, we prove that for short-range correlated systems where each cell can take on $Z$ distinct states and there exists a length $\xi$ for which the correlation $\langle s_i s_{i+k}\rangle \leq e^{-|k|/\xi}$, where $s_i$ is the signal at cell $i$, then the $\mathrm{PI}$ for a system of size $N$ is bounded by the positional information $\Pi_N^Z$ of the sawtooth profile which converges to a finite value
\begin{align}
    \Pi_N^Z  \xrightarrow[N\rightarrow\infty]{}  \Pi_\infty^Z &= \log_2 Z - \frac{1}{2\ln 2} \\& \approx 0.28 \text{ bits for } Z=2. \nonumber
\end{align}
We prove below the inequality $\mathrm{PI} \leq \Pi_N^Z$ for the case $Z=2$, and postpone its extension and numerical results for $Z>2$ to Appendix~\ref{sec:potts_si}. We also note that this bound is consistent with the results of Ref.~\cite{singh_limits_2025, berx_positional_2025} for classes of lattice processes around equilibrium.

\subsubsection{Proof of the sawtooth bound}

Consider a periodic system of $N=2M$ cells. To maximize $S_\mathrm{pat}$, we consider a situation similar to an Ising model where $h_i=0$ except at two diametrically opposite points where $h_1 = - h_M = h$, such that the system is fully symmetric $P_+ = P_- = 1/2$. In this case, we write (in bits)
\begin{equation}
    \mathrm{PI} = 1  - \frac{1}{N}\sum_i S_i  \label{eq:PI_ineq1}
\end{equation}
where $S_i = -p_i \log_2 \left(p_i\right) - (1-p_i) \log_2\left(1-p_i\right) \equiv S(p_i) \geq 0 $ is the binary entropy of the $i$-th marginal, with equality if $p_i =0 $ or $1$. Since $S_i \geq 0$, we have $ \mathrm{PI} \leq 1 \text{ bit}$ for all situations where $Z=2$.
We now consider systems with local couplings such that correlations are short ranged $\langle s_i s_{j}\rangle \sim e^{-d(i,j)/\xi}$ where $d(i,j)$ is the distance between cells $i$ and $j$, and  without loss of generality, we consider here that the system takes values $s_i\in\{0,1\}$.  We note that if $p_i$ is maximal at $i=1$ and minimal at $i=M$, maximizing PI requires $p_1 = 1$ and $p_{M}=0$. By symmetry, $p_{M/2} = p_{3M/2} = 1/2$, and we restrict our attention to the sector between $1$ and $M/2$. Since $p_1 = 1$, we have $\langle s_1 s_i \rangle = p_i$, and since the correlations are bounded by an exponentially decaying function, $p_i$ is bounded by a convex and monotonically decreasing function between $p_1$ and $p_{M/2}$. For $p_i \in [1/2,1]$, $S_i$ decreases monotonically with $p_i$, such that maximizing $\mathrm{PI}$ is now equivalent to minimizing the entropy for $(p_i)_{1\leq i \leq M/2}$.
By the definition of convexity, we have the bound $p_i \leq Ai + B$, where $A$ and $B$ are set by the two constraints $p_1 = 1, p_{M/2} =1/2$.

The maximum $\mathrm{PI}$ is achieved thus when the marginal probability is linear in the position $p_i = 1-(i-1)/(M-1)$ between $0$ and $M$. This implies that the PI is bounded by
\begin{align}
    \mathrm{PI} \leq &\,1 + 2\times\frac{1}{2M}\sum_{i=1}^M\left[ \frac{i-1}{M-1}\log_2 \frac{i-1}{M-1} \right.\nonumber\\ & \left.+ \left(1- \frac{i-1}{M-1}\right) \log_2\left(1- \frac{i-1}{M-1} \right) \right]\nonumber \\
    =& 1+ \frac{2}{M}\sum_{i=1}^{M} \frac{i-1}{M-1}\log_2\left(\frac{i-1}{M-1}\right) \equiv \Pi_N \label{eq:Pi_N}
\end{align}
where we use the symmetry of the terms in the last line. When $N\rightarrow\infty$, this sum is a Riemann sum of negative terms and the PI is thus bounded by
\begin{align}
    \mathrm{PI}(N\rightarrow\infty) \leq\mathrm{\Pi}_\infty & = 1 +2\int_{0}^1\mathrm{d}x\,x\log_2(x)  \nonumber\\ &=1-\frac{1}{2\ln2}\approx 0.28 \text{ bit}
\end{align}
For finite $N$, the corrections to the integral from the finite sum lead to a less constraining bound on the PI, which is then bounded by a larger value $\Pi_N  \geq \Pi_\infty$. The computed value is consistent with the numerically obtained value for the Ising model and close to the saturating value of the DIM (Figs.~\ref{fig:morpho_analysis}f, \ref{fig:Ising_varN}).

\subsubsection{Distinct ``universality classes" depending on dynamics}
\label{sec:conservednoise}

The linear bound of Eq.~\eqref{eq:Pi_N} is tight for the Ising model, as expected from its exponential correlation structure, while the DIM --- which lightly violates the exponential correlation structure assumption due to conserved noise (Fig.~\ref{fig:morpho_analysis}e, Eq.~\ref{eq:corr_mef}) --- is slightly above the exponential bound (Fig.~\ref{fig:morpho_analysis}f).

\begin{figure*}
    \centering
    \includegraphics[scale=0.95]{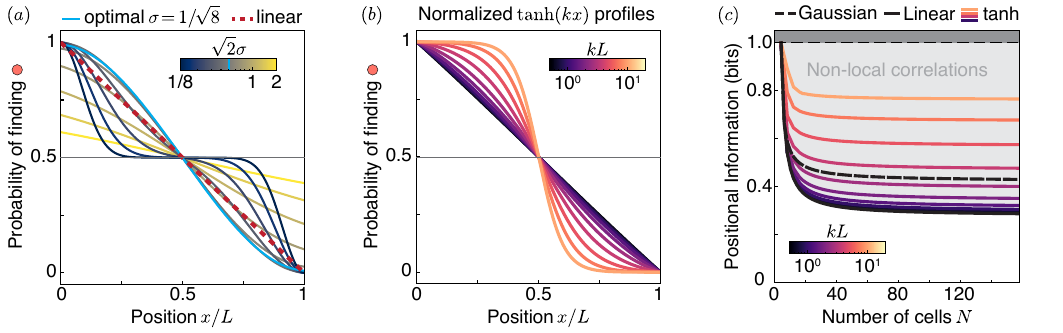}
    \caption{Positional information and convexity of the marginals. (a) Marginal probabilities for a Gaussian profile described by Eq.~\eqref{eq:gaussian_profile}. There is an optimal profile for $\sigma = 1/\sqrt{8}= 1/(2\sqrt{2})$. (b) Hyperbolic tangent profiles $\tanh(kx)$ linearly scaled such that $p(0)=1$ and $p(1)=0$. The more step-like the profile, the higher the positional information. (c) Corresponding bound on the positional information for the optimal Gaussian profile (rescaled as done for hyperbolic profiles), the sawtooth linear profile for exponentially-correlated systems and for the hyperbolic profile with varying $k$.}
    \label{fig:bound_convexity}
\end{figure*}

The marginal probabilities $p_i$ measured in the DIM do not exactly satisfy the convexity requirements for the bound to hold. Unfortunately, the correlation functions in the presence of conserved noise are singular, limiting our analytical reach; nevertheless, we can still obtain a bound on PI in this model by turning to the limiting case of diffusive dynamics for which after a finite time $\sim 1/\gamma$, the particles released at source points lose memory of their past: the PI for the DIM sits below the bound for this `Gaussian' diffusive model. In this limiting regime, the marginals have the form 
\begin{equation}
    p_\sigma(x) = \frac{1}{2} \left( 1+ e^{-\frac{x^2}{2\sigma^2} }  - e^{-\frac{(1-x)^2}{2\sigma^2} }\right) \label{eq:gaussian_profile}
\end{equation}
where $0 \leq p_\sigma(x) \leq 1$ but there is no normalization requirement on $p_\sigma$ (Fig.~\ref{fig:bound_convexity}a). To optimize positional information, we seek $\sigma$ such that $\partial_x p_\sigma(x=1/2)$ is largest in magnitude by solving $\partial_\sigma \partial_xp_\sigma = 0$: We find an optimum for $\sigma=1/(2\sqrt{2})$.  In general, PI is maximized when the marginal profile is the most `step-like' (Fig.~\ref{fig:bound_convexity}b). PI is thus bounded by a different asymptotic value in the limit of large system size depending on the convexity profile of the marginals, reflecting the nature of the dynamics (Fig.~\ref{fig:bound_convexity}c).

In summary, locally interacting systems with a finite correlation range bound the PI below its maximal value. This is reminiscent of area laws in quantum systems, which generically appear in locally-interacting systems with finite correlation lengths~\cite{eisert_colloquium_2010,wolf_area_2008}.  However, if spatial interactions have the additional effect of reducing overall entropy by, for instance, averaging signals in space or imposing feedback control, then the PI can reach its maximal value. This latter scenario is similar in flavor to results in field theory which state that systems with many neighbors or long-range interactions suppress fluctuations and thus behave closer to mean-field~\cite{kardar_statistical_2007, baker_ising_1963}.


\section{Non-locality enables robust patterning} \label{sec:nonlocal}

\begin{figure*}
    \centering
    \includegraphics[scale=0.95]{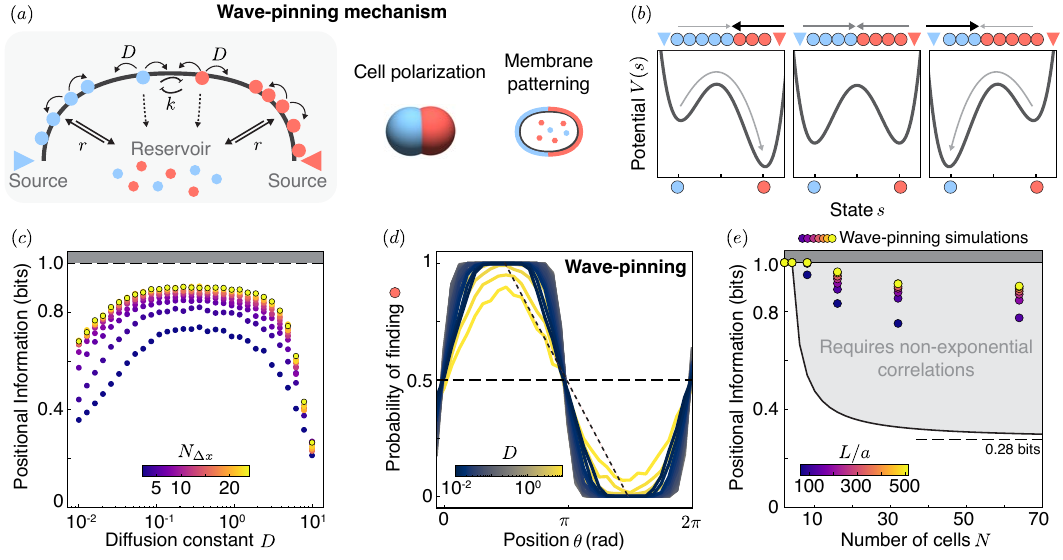}
    \caption{Long-ranged interactions can stabilize patterns, allowing for increased robustness. (a) A conserved number of molecules of two types bind to a membrane with a rate proportional to their concentration in the reservoir. Once on the membrane, the molecules diffuse with constant $D$ and unbind at a rate $r$ in the absence of molecules of the other type in the vicinity. When the other molecule type is present, the unbinding rate increase nonlinearly, leading to bistable surface concentration dynamics. (b) The dependence of the binding rate on the reservoir particle number leads to global coupling which stabilizes the front position: if there are more $A$ (blue) than $B$ (red) bound, then binding of $B$ is favored, and vice versa. (c) The positional information as a function of $D$ for variable noise is now much higher, and stays at its maximal value for a wider range of $D$. (d) The marginal probability of finding $B$ at position $\theta$ can now be non-convex. Here we keep $a=L/96$. (e) The maximal PI can now reach the theoretical maximum, beating the bound for short-ranged systems, with increasing particle numbers $L/a$ (lower noise) leading to better PI. In all simulations $r=1, k=4/3, N_0^{A,B} = 6(L/a), T=5\cdot 10^3/D, L=4, \alpha=30$, and results are averaged over $1000$ replicates.}
    \label{fig:wave-pinning}
\end{figure*}

Our results in the previous section show that short-range correlated models have a maximal positional information capacity, which is reached for fine-tuned parameter values corresponding to optimal confinement of the domain walls. This is undesirable in a biological scenario: Cells in developing embryos have positional information that corresponds to errors in position of the order of a single cell size~\cite{gregor_probing_2007,merle_precise_2024} while biological transport coefficients vary strongly depending on the chemistry of the local environment~\cite{milo_cell_2015, schlissel_diffusion_2024}. Are there simple ways to make the system maximally robust, which, by definition, here means $\mathrm{PI} = 1$ bit? In this section, we answer by the positive by considering a class of biologically-inspired models for which non-local interactions allow robust patterning, which is mechanistically understood as a consequence of integral feedback on the domain wall position.

In our previous model, PI is limited by the short-range correlation structure of the diffusive Ising model. In contrast, most self-organizing models of pattern formation develop long-ranged correlations by a combination of long-range diffusion~\cite{gierer_theory_1972,schweisguth_self-organization_2019} or the presence of higher spatial derivatives~\cite{swift_hydrodynamic_1977}. For our purposes, the short-range correlations limit positional information by providing only weak confining forces on the domain wall position. A well-studied way to impose tighter control on the domain wall position is wave-pinning \cite{mori_wave-pinning_2008, mori_asymptotic_2011}, which has been identified as a biologically relevant mechanism in the anterior-posterior patterning of the \emph{Caenorhabditis elegans} embryo \cite{goehring_polarization_2011, hubatsch_cell-size_2019}, and other instances of cell polarization \cite{miller_forced_2022,zhu_developmental_2020, goryachev_dynamics_2008,otsuji_mass_2007} (see Appendix~\ref{sec:appendix_wp} for a plausible adaptation of the lattice model to \emph{C. elegans} embryogenesis). 

In wave-pinning systems, signaling molecules of a given kind bind to a membrane with a rate proportional to the number of molecules available in a homogeneous reservoir and unbind with a rate that depends on the local bound concentration of signals, giving two possible locally stable states on the membrane, with zero or a finite bound concentration state. The assumption that the reservoir is homogeneous captures the empirical fact that bulk diffusion is often much faster than membrane-bound diffusion.
Importantly, conservation of the molecular number implies that the binding rate depends on the state of the entire system. 

\subsection{Wave-pinning dynamics}
\label{sec:wpdynamics}

We here write down a minimal microscopic model of wave-pinning, which we dub the wave-pinning Ising model as it shares similarities with the Diffusive Ising model discussed above (Fig.~\ref{fig:wave-pinning}a). Particles move according to the following rules:
\begin{itemize}
    \item[\textbf{W1}] Particles on the surface hop with equal probability to the left or right neighboring site, at rate $D/a^2$
    \item[\textbf{W2}] Particles of type $A$ (resp. $B$) unbind from the surface into the reservoir with rate $R_A = r + kn_B(n_B - 1)$ (resp. $R_B = r + kn_A(n_A - 1)$).
    \item[\textbf{W3}] Particles in the reservoir bind to the membrane at a rate $(M+\alpha)r$, with $M$ the number of sites and $\alpha$ the strength of the bias to bind to the source point. Particles bind to a non-source site with probability $1/(M+\alpha)$, and bind to the source site with probability $(1+\alpha)/(M+\alpha)$.
\end{itemize}
We again use a tau-leaping scheme to simulate these stochastic dynamics until steady-state. This system, like the DIM, is amenable to path-integral-based coarse-graining as $a\rightarrow0$, and we find the integro-differential equations (Derived in Appendix~\ref{sec:wp_cg_derivations})
\begin{subequations}
\begin{align}
    \partial_t n_A &= D\partial_x^2n_A - rn_A - kn_B^2n_A + rN_A^c +h_A(x)\\
    \partial_t n_B &= D\partial_x^2n_B - rn_B - kn_A^2n_B + rN_B^c +h_B(x)
\end{align} \label{eq:wp_1} %
\end{subequations}
with the cytoplasmic concentrations $N^c_X$ given by the conservation laws
\begin{subequations}
    \begin{align}
        N_X^c = N^0_X - \frac{1}{a}\int_0^L \mathrm{d}x'\, n_X(x')
    \end{align}
\end{subequations}
with $X=A$ or $B$, and where $a$ is the lattice spacing, standing-in for the interaction range. $N_A^0$ and $N_B^0$ are the total numbers of particles of type $A$ and $B$, respectively.
To cast these dynamics into a more familiar form, we can rewrite Eq.~\eqref{eq:wp_1} in terms of the total density $\rho = n_+ + n_-$ and signal $s= n_+ - n_-$ as
\begin{subequations}
    \begin{align}
        \partial_t s & = D\partial_x^2s + 
        \left(\frac{k}{4} \rho^2 -r\right) s - \frac{k}{4} s^3 + rs^c \\
        \partial_t \rho & = D\partial_x^2\rho - r\rho - \frac{k}{4}\rho(\rho^2 - s^2) +   r \rho^c \\
        s^c & = s^0 - \frac{1}{a}\int \mathrm{d}x\,s \\
        \rho^c & = \rho^0 - \frac{1}{a}\int \mathrm{d}x\,\rho
    \end{align}
\end{subequations}
When $D$ is sufficiently small so that domain walls are very narrow and density fluctuations around its steady-state value $\bar{\rho}$ can be safely ignored, the signal then obey the reduced Landau-Ginzburg-like equation
\begin{subequations}
\begin{align}
    \partial_t s & = D\partial_x^2s + r_2 s - u_2 s^3 + h \\
    h & = rs^0 - \frac{r}{a}\int \mathrm{d}x\,s
\end{align}\label{eq:wp_mag}%
\end{subequations}
with $r_2 = k(N^0_A+N_B^0)^2/(4+2/a)^2-r$ and $u_2 = k/4$. We now see that the flux from the reservoir can be understood in this Ising-like picture as a uniform, self-organized magnetic field. Similarly to Eq.~\eqref{eq:phi4} this equation exhibits bistable dynamics and therefore admits homogeneous and domain wall solutions between values $s_\pm = \pm \sqrt{r_2/u_2}$. Unlike Eq.~\eqref{eq:phi4}, Eq.~\eqref{eq:wp_mag} is a nonlocal integro-differential equation: The dynamics of $s(x)$ depend not only on the value of $s(x)$ and its derivatives, but also on the global state of the system via the integral term. Nonlocality commonly appears in systems where fast spatial degrees of freedom are integrated out~\cite{couder_single-particle_2006,chen_evolving_2025}, where topological constraints lead to global couplings~\cite{moroz_entropic_1998}, or in the presence of quantum entanglement, which allows `volume law' scaling~\cite{eisert_colloquium_2010,wolf_area_2008}. Here, nonlocality stems from the conservation of particle number and the separation of timescale between bulk and surface diffusion. 

 The effect of this nonlocal term is understood as follows: if the average magnetization $\langle s \rangle_x = \frac{1}{L}\int \mathrm{d}x\,s $ is greater than $s^0$, then $h<0$, biasing the effective reaction free energy towards the negative state. If the system contains domain walls, this asymmetry then leads the domain walls to drift until $\langle s \rangle_x = s^0$, imposing a set ratio of positive and negative domains. If $s^0 = 0$, this dynamical balance is achieved when $\langle s \rangle_x = 0$, which happens when the positive and negative domains are the same size (Fig.~\ref{fig:wave-pinning}b).

\subsection{Spatial feedback stabilizes domain wall position}

 The coupling to the reservoir thus provides a form of spatial integral feedback, whose effects can be seen in the wave-propagation picture~\cite{mori_wave-pinning_2008,mori_asymptotic_2011}:
 In this picture, the position of the front at position $X(t)$ propagating at velocity $c$ evolves under the dynamics
 \begin{equation}
     \frac{\mathrm{d}X}{\mathrm{d}t} = c,
 \end{equation}
 and we insert the front-propagation ansatz $s(x,t)=s(x-X(t))$ into Eq.~\eqref{eq:wp_mag} to calculate the front velocity $c$ as a function of the profile $s(x,t)$. Assuming that the domain size $L$ is large enough compared to the width of the domain wall $\ell = \sqrt{D/r_2}$, we assume the profile to be an hyperbolic tangent $s(x,t)=s_+ \tanh[(x-X(t))/(\sqrt{2}\ell)]$. Neglecting for now the effect of sources and noise, we find that the front velocity depends on its position as
 \begin{align}
     c &= - \frac{\int_{s_-}^{s_+} \mathrm{d}s\, \left( r_2 s-u_2s^3+h\right)}{\int_{-\infty}^\infty\mathrm{d}x \,(\partial_x s)^2}\\
     & = - \frac{3\sqrt{r_2/u_2}}{\sqrt{2}\ell} h \\
     & = -3r\sqrt{\frac{r_2 D}{2u_2^2}}\left(\frac{s^0}{s_+} + \frac{1}{a}(2X(t)-L)\right)
 \end{align}
 using the definition of $h$ from Eq.~\eqref{eq:wp_mag} and with $\int \mathrm{d}x \,s(x,t) \approx s_+ X(t) + s_-(L-X(t))$ under the assumption that the domain wall is very thin and therefore the solution is approximately constant on either side of the domain wall. When $s^0 = 0$, we finally find a restoring force
 \begin{align}
     \frac{\mathrm{d}X}{\mathrm{d}t} = f_\mathrm{wp}(X) =&  -\frac{3}{\sqrt{2}}\frac{r}{u_2}\sqrt{r_2 D}\left(X(t)-\frac{L}{2}\right) \nonumber\\  \equiv &-k_\mathrm{wp}(X(t)-X_0)
 \end{align}
 with wave-pinning-induced confinement stiffness 
\begin{align}
    k_\mathrm{wp} &=\frac{6\sqrt{2}\,r}{k}\sqrt{r_2 D}.
\end{align}

Assuming that the confining force $f_\mathrm{c}$ due to the external sources is additive to this wave-pinning confinement, the domain wall position $X(t)$ now follows ~\cite{mori_wave-pinning_2008,mori_asymptotic_2011}
\begin{align}
    \frac{\mathrm{d}X}{\mathrm{d}t} = f_\mathrm{c}(X) + f_\mathrm{wp}(X) + \sqrt{2D_f}\zeta(t)
\end{align}
where $f_\mathrm{wp}(X)= -k_\mathrm{wp}(X-X_0)$ is the additional confinement due to the wave-pinning feedback. 
Notably, while $k_\mathrm{c}$ decays exponentially when $D\rightarrow0$,  this wave-pinning-induced stiffness $k_\mathrm{wp}$ only decays as $\sqrt{D}$.  The presence of an additional global conservation constraint thus realizes a form of spatial integral feedback, allowing for precise localization of the domain wall with much weaker dependence on the diffusion constant: effectively, the energetic cost of being in the minority state is lowered, and the domain wall moves to expand the minority region (Fig.~\ref{fig:wave-pinning}b). We note that wave-pinning does not remedy the instability of front solutions for large $D$, but nonetheless acts to reduce the variance of domain wall position, especially at small $D$~\cite{hubatsch_cell-size_2019}.

Including this mechanism, microscopic simulations show much higher PI, with a flat maximum region achieved for values of $D$ that span an order of magnitude (Fig.~\ref{fig:wave-pinning}c). Indeed, this system has correlations that do not vanish at infinity and thus do not satisfy the assumptions underlying the bounds derived above: if we consider a small perturbation $\phi(x)$ around the steady-state $s_0(x)$ approximated as a step function, we find in Appendix~\ref{sec:appendix_wp} that there are constants $A$ and $B$ such that $\langle \phi(x) \phi(y)\rangle = A + Be^{-|x-y|/\ell}$, which does not vanish as $x$ and $y$ are taken far apart, violating the conditions for the bound to apply. As expected, the nonlocal coupling through the reservoir makes it possible for the marginal probabilities $p_i$ to now have saturating profiles, allowing the PI to beat the bound imposed by exponential or Gaussian correlations (Fig.~\ref{fig:wave-pinning}d-e). This is again reminiscent of area laws for classical mutual information: in a similar way that quantum entanglement can `break the area law', classical long-range correlations can lead to volume-law scaling. Similarly, for positional information, long-range correlations enable maximal PI.


\section{Implications for realistic systems} \label{sec:implications}

\begin{figure}
    \centering
    \includegraphics[scale=1]{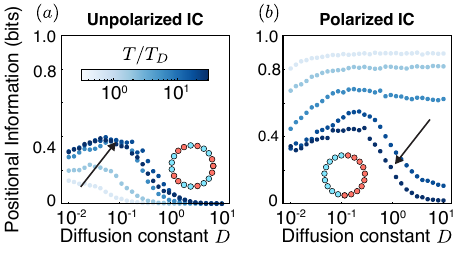}
    \caption{Initial condition dependence of outcomes with finite decision times in the DIM. (a) The positional information starting from unpolarized (random) initial conditions (IC) increases with decision time $T$ (a), while it decreases when the system is initialized in a polarized state for which $\mathrm{PI}=1\text{ bit}$ (b). $T$ is normalized by the time scale $T_D = L^2/D$ of diffusive dynamics. Arrows indicate increasing time. Here $a= L/384, \gamma=1, \beta=3\beta_c/2, h_0=3\beta, L=4, N=32$, and results are averaged over $500$ replicates.}
    \label{fig:varT}
\end{figure}

Small systems that integrate information from their surroundings in time are fundamentally limited in their error rate by the Berg-Purcell limit as their sensing range diminishes~\cite{berg_physics_1977,vergassola_infotaxis_2007, tkacik_information_2025}. In spatially-extended systems that have no memory, our information-theoretic quantification of robustness shows a similar limitation to the information capacity of spatial patterns depending on the nature of their spatial correlations at steady state.

We note that outside of stationary regimes, whether this limitation is important depends on context: If the noise amplitude is small and the observation timescale short, a `good enough' initial condition might still lead to an acceptable final state for practical purposes. We find that the positional information starting from unpolarized (random) initial conditions (IC) increases with the decision time $T$ as the system reaches steady state. In contrast, when the system is initialized in a polarized state for which $\mathrm{PI}=1\text{ bit}$, PI decreases to its steady-state value as the decision time $T$ increases. (Fig.~\ref{fig:varT}).

Long-range interactions are known to help induce and stabilize pattern formation~\cite{gierer_theory_1972,al-mosleh_how_2023, coropceanu_self-assembly_2022, angerpointner_delay-facilitated_2025}. Although the patterns we considered here were linearly stable even in the absence of nonlocal correlations, we show that long-range phenomena greatly extend the stability regions of these patterns. In light of this, incorporating long-range interactions in synthetic self-organized systems might be a productive strategy for increasing yield and functionality.

\begin{figure}
    \centering
    \includegraphics[width=0.97\linewidth]{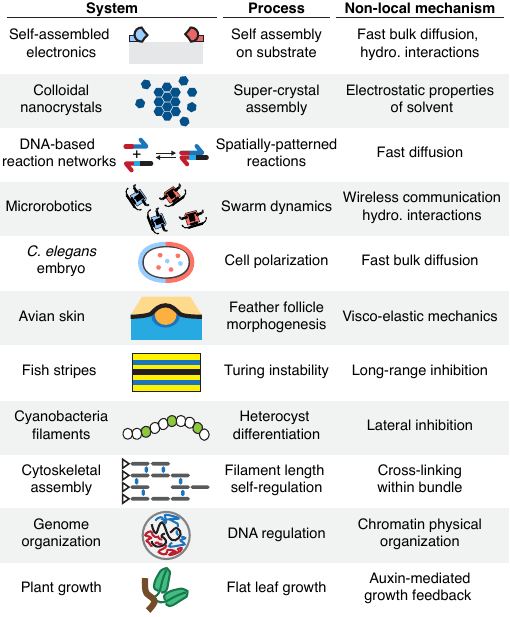}
    \caption{Illustrations of self-organized systems where long-range interactions can enable robust patterning. Self-assembly of electronic components could benefit from asymmetric diffusion properties \cite{yin_formation_2004} or hydrodynamic interactions \cite{lee_fluidic_2023}. Colloidal nanocrystals self-assembly can be controlled by tuning long-range interactions through the dielectric properties of the solvent~\cite{coropceanu_self-assembly_2022}. DNA toolkit reactions can benefit from engineering molecular mobility to establish reliable spatial patterns \cite{zadorin_synthesis_2017}, while microrobotic systems~\cite{miskin_electronically_2020} can benefit from long-range wireless communication. In biology, \emph{C. elegans} embryo patterning is stabilized by wave-pinning~\cite{goehring_polarization_2011}, avian skin morphogenesis benefits from long-range mechanical effects~\cite{yang_morphogens_2023}, and Turing patterning (or lateral inhibition) is a common motif to describe the emergence of fish stripes and other self-organized systems such as cyanobacteria heterocyst differentiation~\cite{turing_chemical_1952, corson_self-organized_2017, schweisguth_self-organization_2019}. Nonlocal coupling can also help regulate filament length in cytoskeletal assembly~\cite{mcinally_length_2024}. Physical organization of the genome allows distant regions in the genome to be in physical proximity, effectively creating long-range interactions, a feature that is evolutionarily conserved~\cite{dekker_exploring_2013, kim_chromatin_2025}. Growing flat surfaces such as leaves requires long-range feedback to be stable \cite{al-mosleh_how_2023}.}
    \label{fig:table_systems}
\end{figure}

The necessity of building long-range correlations for accurate spatial information processing might explain the evolutionary advantage of hierarchical organizations and scale separations in biological systems. For example, fast information transport has been found to be essential for wound healing \cite{fan_ultrafast_2023}, and averaging signals over large numbers of cells distributed in space can allow for more precise sensory readouts~\cite{graf_bifurcation_2024}. Since mechanical responses can be long-ranged, mechano-chemical interactions can enable long-range communication and help complement molecular signaling mechanisms in contexts from development to inflammation~\cite{gross_guiding_2019, boocock_theory_2021, yang_morphogens_2023, moghe_optimality_2025}. In Fig.~\ref{fig:table_systems} we list some examples of engineered and living systems where long-range interactions could or are known to help generate spatial organization.

Other ways to have long-range communication are possible in non-equilibrium physical systems. Systems with conserved degrees of freedom can display non-exponential correlations, as we saw in the DIM~\cite{garrido_long-range_1990}, and a hallmark of self-organized criticality is the persistence of long-ranged correlations in complex systems, although criticality is also associated with larger sensitivity to noise~\cite{bak_self-organized_1988, mora_are_2011}. It is also known that in active matter, boundary effects can be long-ranged \cite{baek_generic_2018,ben_dor_disordered_2022}. 

Finally, while designing long-range couplings can provide an alternative path to internal information processing and memory to enable robust spatial patterning, other solutions exist: coupling to the surrounding environment in a state-dependent manner can also enable greater positional information at the cost of increased dissipation~\cite{berx_positional_2025}.

\section{Conclusion}

In summary, in this article we studied the robustness of pattern formation in minimal lattice and continuum models of self-organization using information-theoretic methods. We find that in general, self-organized systems whose dynamics induce local (exponential or sub-exponential) correlations are limited in their ability to robustly pattern as their positional information is bounded below its absolute maximum. Intermediate spatial coupling strengths optimize robust patterning, a consequence of the effect of localized sources on domain wall dynamics. To circumvent the limitation on positional information, we investigate a biologically-inspired model that includes long-range interactions. By introducing effective integral feedback, these interactions both lift the theoretical limit imposed by local correlation structures and extend the range of spatial coupling strengths that allow for optimal robustness. This fact could provide a strategy to engineer more reliable self-organized synthetic systems, from self-assembled colloidal nanocrystals~\cite{coropceanu_self-assembly_2022} to microrobotic systems~\cite{miskin_electronically_2020}; It could also rationalize the ubiquity of effectively nonlocal interactions in biological settings, from regulation of cytoskeletal filament lengths~\cite{mcinally_length_2024} and skin patterning~\cite{corson_self-organized_2017} to genome organization~\cite{dekker_exploring_2013} and plant growth~\cite{al-mosleh_how_2023} (Fig.~\ref{fig:table_systems}).

\begin{acknowledgments}
We acknowledge helpful discussions with Noah Mitchell, Peter Satterthwaite, Lara Koehler, Daniel Seara, and Yael Avni. NR acknowledges support from the University of Chicago Biological Physics and Center for Living Systems Fellowships. This research was supported by the Physics Frontier Center for Living Systems funded by the National Science Foundation (PHY-2317138). Computing resources were provided by the University of Chicago Research Computing Center. Code is available at \url{https://github.com/NicoRomeo/RobustPatterns}.
\end{acknowledgments}




\appendix

\section{Diffusive Ising model}
\label{sec:app_DIM}

To model the transport and interaction of particles interacting at a finite range $a$ and with a finite average density $\rho_0/a$, in the main text we considered a microscopic lattice model in which particles hop between sites at rate $D/a^2$ and change type $s=\pm$ (or equivalently $A$ and $B$) to align with the locally more prevalent species with a rate $r$ dependent on the number of particles of each type in the vicinity. \cite{solon_flocking_2015, scandolo_active_2023,martin_transition_2025}.

We consider reaction rates $r = \gamma \exp\left(-s (\beta m_i +h_i) \right)$, where $m_i = n_i^+ - n_i^-$ is the population difference in $+$ and~$-$ particles at site $i$, and $h_i$ is the local external source field biasing towards a specific species. In the main text $n_i^+ - n_i^-$ is called $s$ for \emph{signal}; in the Appendices,  we will sometimes denote $s$ by $m$ as it is similar to the \emph{magnetization} familiar in statistical physics, while $\rho = n_+ + n_-$ is the (unitless) number density, or simply density.

\subsection{Fluctuating hydrodynamics equations}

We use a coarse-graining method developed in \cite{andreanov_field_2006,lefevre_dynamics_2007,kourbane-houssene_exact_2018} and applied in \cite{martin_transition_2025} for the more complex case of two non-reciprocally interacting species to coarse-grain the lattice model described by rules \textbf{D1-3} of Sec.~\ref{sec:main_DIM}.

This method is exact in the limit of short-interaction range $a\rightarrow0$, where the dynamics are dominated by diffusion and the master equation solved by a Poissonian distributions. With finite $a$, we can use this Poissonian ansatz to  to compute higher-order correlations and derive a closed fluctuating hydrodynamic equation. To understand the limits of validity of this procedure, we use renormalization group theory (RG) to derive the effective finite-$a$ response and compare it to our numerical solutions. The RG analysis also provides an analytical criterion of validity for the fluctuating hydrodynamics: the perturbative corrections are valid as long as the coupling constant $g = \rho_0 \left(\hat{\gamma} a^2/D\right)$ of the RG is small, meaning that $g \ll 1$ (Details and derivation in Appendix~\ref{sec:RG_calc}).

To this order, the full dynamics are given by
\begin{subequations}
    \begin{align}
    \partial_t \delta \rho   = &\, D\partial^2_{xx} \delta\rho + \sqrt{a}\partial_x \eta_1 \\
    \partial_t s  = &\, D\partial^2_{xx} s - g(\delta\rho, s+h(x)) \nonumber\\ &\, +\sqrt{a} \partial_x \eta_2 + \sqrt{a \bar{g}(\delta\rho, s+h(x))} \eta_3
\end{align} \label{eq:fluct_hydro_full}%
\end{subequations}
with the functions
\begin{subequations}
    \begin{align}
        g(\delta \rho , s) = & \, 2\gamma e^{-\beta +(\cosh{\beta}-1)(\rho_0 + \delta\rho)} \nonumber\\ &\times \left[\cosh{(\sinh(\beta)s)}s \right.\nonumber\\ & \left.\quad- \sinh{(\sinh(\beta)s)}(\rho_0 + \delta \rho) \right] \\ 
        \bar{g}(\delta \rho , s)  = & \, 2\gamma e^{-\beta +(\cosh{\beta}-1)(\rho_0 + \delta\rho)}
    \end{align}
\end{subequations}
The noises $\eta_n$ are correlated by the matrix $\langle \eta_n(x,t) \eta_{n'}(x',t')\rangle = M_{n,n'} \delta(x-x') \delta(t-t')$, with 
\begin{equation}
    M_{nn'} = \begin{pmatrix}
        2D(\rho_0 + \delta \rho) &  2D s & 0 \\
        2D s & 2D(\rho_0 + \delta\rho) & 0 \\
        0 & 0 & 2S_f
    \end{pmatrix}
\end{equation}
with $S_f = \left[ \rho \cosh{(m \sinh{\beta})} - m \sinh{m\sinh{\beta}} \right]$.

In the vicinity of the phase transition for small enough $a$, such that $\rho_0 \gg |\delta\rho|$, we can reduce this model to the cubic order model with additive noise presented in Sec.~\ref{sec:main_continuumDIM}. We use the cubic reduction for the dynamical RG analysis in Appendix~\ref{sec:RG_calc}.

To show that the qualitative conclusions of this model are independent of the precise form of the local interactions defined by the lattice rules, in Appendix \ref{sec:appendix_cg} we also consider the situation where particles interact with others at neighboring sites $a$, which leads to a transition rate $r = \gamma \exp\left(-\beta \sum_{\langle i,j \rangle} m_{i,j} s\right)$, where $\sum_{\langle i,j\rangle}$ denotes the sum over nearest-neighbor sites (not including the site $i$). In the limit $\rho_0 \gamma a^2/D \ll 1$, we find that the resulting macroscopic (hydrodynamic) model is unchanged from the standard same-site DIM, up to some small modifications in the relationship between microscopic parameters and hydrodynamic coefficients, allowing us to focus on the case of same-site interactions in the main text.

\subsection{Comparison between Diffusive Ising and Landau-Ginzburg phenomenology} \label{sec:app_landau}

In Sec.~\ref{sec:domain_walls} we rationalize the existence of an optimal diffusion constant in terms of domain wall dynamics in a Landau-Ginzburg ($\phi^4$) theory. Here we discuss the differences between the hyrodynamics of the DIM and Landau-Ginzburg  phenomenology.

A first discrepancy between the Landau-Ginzburg dynamics [Eq.~\eqref{eq:phi4}] and the DIM comes from neglecting density dynamics: the density equation for $\delta \rho$ mostly contributes to additional noise since the average $\langle \delta \rho \rangle$ vanishes. . The main effect of the density dynamics is to introduce conserved noise since the system is diffusive at short time scales: this induces some non-exponential correlation behavior, and as such the correlation function in Fourier space is not a Lorentzian and reads
\begin{align}
    \langle s_qs_{q'}\rangle = \delta_{q,q'}a\rho_0\frac{Dq^2 +\hat{\gamma}}{Dq^2+r}
\end{align}
in the limit $\rho_0(\gamma a^2/D) \ll 1$ of validity of the coarse-graining scheme. The diffusive effects effectively lead the DIM to be weakly non-local, and it does not have perfectly exponentially-decaying correlation functions, and as such is not strictly subject to the bound presented in Sec.~\ref{sec:sawtooth_bound}. It is instead subject to a different bound for purely diffusive systems with Gaussian spatial correlations, which we derive in Sec~\ref{sec:conservednoise}. Conserved noise also affects the quantitative prediction of domain wall dynamics, although it has no qualitative effect, as it simply changes the numerical value of the diffusion constant but not its scaling with parameters (Appendix~\ref{sec:si_domainwall}).

A second discrepancy between the Landau-Ginzburg and Diffusive Ising models is that the field in Landau-Ginzburg dynamics is not constrained to be bounded, a fact that already appears at the cubic-order approximation of Eq.~\eqref{eq:fluct_hydro_cube}. 
To understand the effects of this difference, we directly simulate the Landau-Ginzburg equation [Eq.~\eqref{eq:phi4}]. The numerical implementation uses a Euler-Mayurama scheme with standard second-order finite-difference discretization of the Laplacian. 

\begin{figure*}
    \centering
    \includegraphics[scale=0.95]{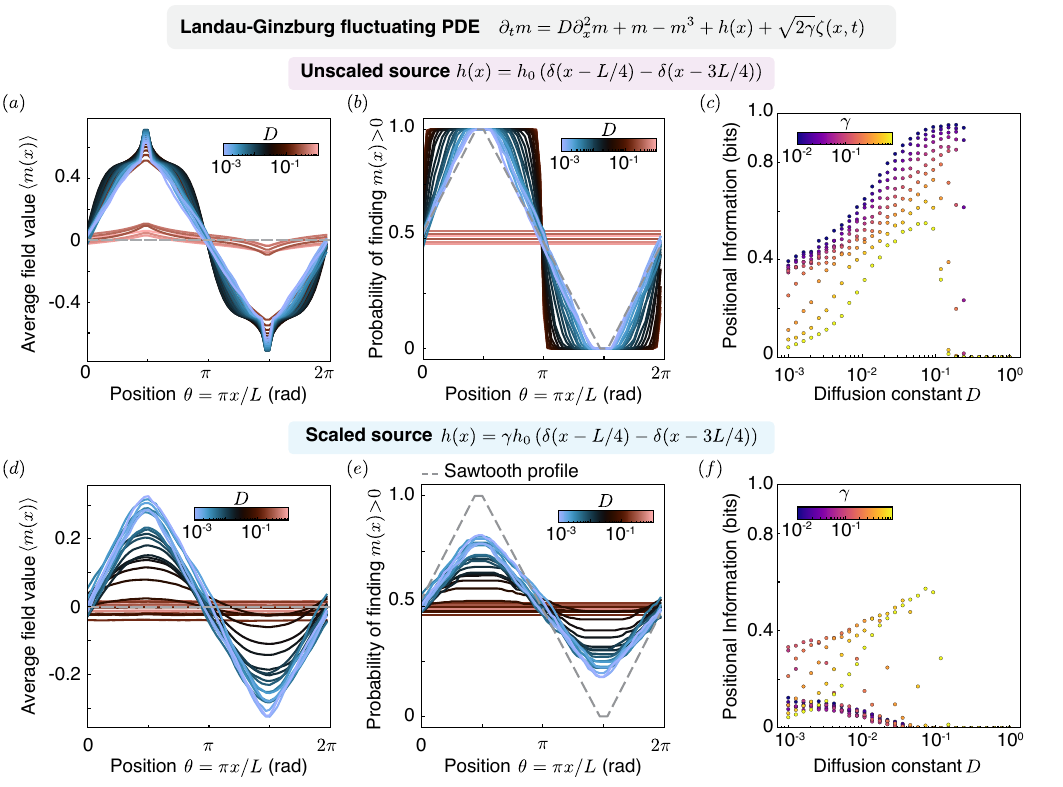}
    \caption{Landau-Ginzburg theory recapitulates results from lattice simulations when constrained by microscopic rules. Without any microscopic constraint, (a-b) The average field value $\langle m(x)\rangle$ peaks at sources, and the marginal probability can now become step-like (c) The positional information now increases with decreasing noise, still showing a sharp optimum for intermediate $D$. (d-e) With the relationship between noise and source strength imposed by microscopic dynamics, positional information is limited  Average field value $\langle m(x)\rangle$ shows a sawtooth-like profile, which is reflected in the marginal probability. (f) Positional information saturates to a low value as noise is decreased.  In (a,b,d,e), $\gamma=10^{-3}$. All simulations have $h_0=10$, $L = 4$, are integrated over a duration $T=100/D$ in dimensionless units and are discretized in space in time using finite differences $\Delta x = L/128$, $\Delta t = 0.03 (\Delta x)^2/D$.}
    \label{fig:phi4}
\end{figure*}

We find that the Landau-Ginzburg dynamics, unconstrained to be discrete-valued, are generally not subject to the information-theoretic bounds we derived in Sec.~\ref{sec:sawtooth_bound} (Fig.~\ref{fig:phi4}a-c). However, when the model coefficients are constrained by the microscopic dynamics as in Eq.~\eqref{eq:fluct_hydro_cube} (ignoring density dynamics), the model reproduces the information saturation phenomenon even at low noise, as long as the source strength is small enough to avoid saturation (Fig.~\ref{fig:phi4}d-f): Since the variance of the noise in the DIM is proportional to the lattice length scale $a$ and the effective macroscopic source intensity is also proportional to $a$, the noise and the source strengths are intrinsically linked in the microscopic model.

\section{Information theory for self-organization}
\label{sec:info_theory}

In this Appendix, we define the information-theoretic measures used to quantify the robustness of self-organization, discuss the relative merits of these measures, and prove the inequalities of Eq.~\eqref{eq:ineqs_PI}. We consider a similar formalism as in Ref.~\cite{bruckner_information_2024}: We consider the state of a self-organized system of $N$ cells, or more generally components, to be described by the state vector $\mathbf{z} = (z_1, ..., z_N)$ where $z_i \in \{1,\ldots, Z\}$ is the state of cell $i$ chosen among $Z$ possible states. We consider that the cells are fixed in space: thus the index $i$ reflects spatial position. Given a stochastic patterning process, we obtain or estimate from experiments the probability of observing a given pattern of cell states $P(\mathbf{z})$. 

\subsection{Information-theoretic measures of robust self-organization}
\label{sec:info_measures}
As a first way to assess the reproducibility of the patterns, we consider the entropy of this distribution
\begin{equation}
    S_\mathrm{rep}= \frac{1}{N}S[P(\mathbf{z})] = - \frac{1}{N}\sum_{\mathbf{z}} P(\mathbf{z})\log_2P(\mathbf{z}) 
\end{equation}
which following Ref.~\cite{bruckner_information_2024} we term \emph{reproducibility entropy}. $S_\mathrm{rep}$ ranges between $0$, achieved for a perfectly reproducible system where every replicate is identical, and $\log_2Z$ for a system where every one of the $Z^N$ possible states is equiprobable.

$S_\mathrm{rep}$ alone cannot be a perfect measure of robust patterning: for instance, a perfectly uniform system has no spatially-varying pattern but is perfectly reproducible.  

To account for these spatial correlations, the \emph{positional information} (PI) is defined as the mutual information between position and state, which can be understood as the information one gains on the cell state by knowing the position of the cell \cite{dubuis_positional_2013}. To compute PI, we need to assign a probability of picking the cell at position $i$, which we take as uniform $P(i) = 1/N$. 
\begin{align}
    \mathrm{PI} & = \sum_{i=1}^N \sum_{z=1}^Z P(z,i) \log_2\left(\frac{P(z,i)}{P_z(z)P(i)}\right) \nonumber\\ &= \frac{1}{N} \sum_{i=1}^N\sum_{z=1}^Z P_i(z_i) \log_2\left(\frac{P_i(z_i)}{P_z(z)}\right) \label{eq:PI}
\end{align}
with $P(z,i) = P_i(z_i) P(i)$ the joint distribution of states and positions and 
\begin{equation}
    P_z(z) = \frac{1}{N}\sum_{i=1}^N \sum_{z_i=1}^Z P(\mathbf{z})\delta_{z_i,z}
\end{equation}
is the pooled distribution of cell states, giving the probability of finding a cell in state $z$ across all positions.

To quantify the diversity of the cell population, Ref.~\cite{bruckner_information_2024} defines the patterning entropy
\begin{equation}
    S_\mathrm{pat} = - \sum_{z=1}^Z P_z(z)\log_2P_z(z).
\end{equation}
If all cells have identical states, then $S_\mathrm{pat} = 0$ while if all states are identical $S_\mathrm{pat} = \log_2 Z$. For all distributions, $0\leq S_\mathrm{pat} \leq \log_2Z$. We can then rewrite the PI as 
\begin{equation}
    \mathrm{PI} = S_\mathrm{pat} - \frac{1}{N}\sum_{i=1}^N S[p_i(z_i)] \equiv S_\mathrm{pat} - S_\mathrm{cf}
\end{equation}
where we have defined the correlation-free entropy $S_\mathrm{cf} = \frac{1}{N}\sum_{i=1}^N S[p_i(z_i)]$. $S_\mathrm{cf}$ can be understood as the (normalized) entropy of the product of the marginals, ignoring the correlations between states at different sites.

In the main text, we consider symmetric systems for which both states are equally likely across replicates, leading this patterning entropy to always be $S_\mathrm{pat}=1$ bit. This is without loss of generality as in general, $S_\mathrm{pat}$ is maximized in a symmetric system where, averaged over all positions, particles are equally likely to be in every state. In such symmetric systems, we have
\begin{equation}
    \mathrm{PI} = \log_2Z - \frac{1}{N}
    \sum_i S[p_i(z_i)] \label{eq:PI_sym}
\end{equation}

\begin{figure*}
    \centering
    \includegraphics[scale=0.95]{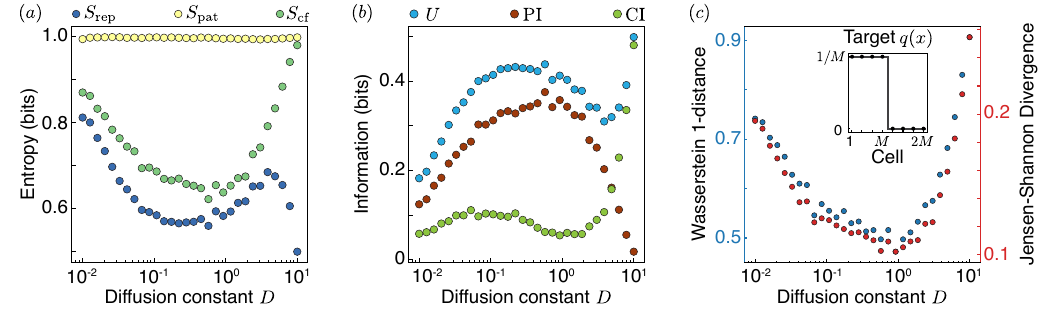}
    \caption{Information-theoretic measures of self-organization. \emph{(a)} Reproducibility, Patterning and Correlation-free entropies for the DIM system. \emph{(b)} Utility, Positional Information and Correlation Information for the DIM system. \emph{(c)} Wasserstein 1-distance and Jensen-Shannon Divergence for the observed probability distributions, with target distribution $q(x)$ plotted in inset. Simulation parameters: $N=2M=8$, $L=4$, $a=L/48$, $\beta = 2\beta_c$, $h=3\beta$.}
    \label{fig:infomeasures}
\end{figure*}

Additionally, Ref.~\cite{bruckner_information_2024} defines the correlational information $\mathrm{CI} = S_\mathrm{cf}-S_\mathrm{rep}$ to measure the reduction in entropy due to the information contained in the spatial correlations, and propose the utility $U=S_\mathrm{pat}-S_\mathrm{rep} = \mathrm{CI}+\mathrm{PI}$ as a measure of robust self-organization. We note that the utility $U$ cannot distinguish between self-organized homogeneous states and patterned states, which limits its use in our case.  We show these quantities computed on the DIM in Fig.~\ref{fig:infomeasures}a-b: To compute the PI in the diffusive Ising model, we partition the spatial domain into $N$ bins of equal width (``cells'') and binarize the signal field as $b_i = \mathrm{sgn}\left(s_i\right)$. We estimate the marginal probability $P_i^+$ of finding $b_i = 1$ by its observed frequency, along with the pooled probability $P_+$ to find $b=1$, irrespective of position. This allows us to compute the PI following Eq.~\eqref{eq:PI_main}
\begin{align}
    \mathrm{PI} = \frac{1}{N} \sum_{i=1}^N &\left[P_i^+ \log_2\left(\frac{P_i^+}{P_+}\right)+P_i^- \log_2\left(\frac{P_i^-}{P_-}\right)\right]
\end{align}
with $P_- = 1-P_+, P_i^- = 1-P_i^+$.

An alternative way to measure the robustness of the patterns comes from considering the defined deterministic pattern of the system.  We can then define the robustness of the pattern distribution by defining probabilistic distances between the observed distributions and the target distribution $Q(\mathbf{z})$ corresponding to the deterministic pattern. Here we explore two such measures, the Wasserstein and Jensen-Shannon distances.

The Wasserstein-1 metric between the distributions $p(x)$ and $q(y)$ is defined in terms of a transport plan $\gamma(x,y)$ associated with the cost function $c(x,y) = |d(x-y)|$, where the difference $d(x-y)$ accounts for the periodicity of the domain. The metric is defined as the solution to a constrained optimization problem \cite{peyre_computational_2020}
\begin{equation}
    W_1(p,q) = \min_{\gamma} \int \gamma(x,y)c(x,y)\,\mathrm{d}x\mathrm{d}y
\end{equation}
with $\int \gamma(x,y)\mathrm{d}x = p(x)$ and $\int \gamma(x,y)\mathrm{d}y = q(x)$. Computing this distance requires solving a linear program to obtain the optimal transport plan $\gamma^*(x,y)$.

The (squared) Jensen-Shannon distance between distributions $p$ and $q$ is defined in terms of the symmetrized Kullback-Leibler divergence between $p$, $q$ and their mixture distribution $\mu = (p+q)/2$~\cite{lin_divergence_1991}
\begin{equation}
    \mathrm{JSD}(p||q) = \frac{1}{2}D(p || \mu) + \frac{1}{2}D(q||\mu).
\end{equation}

For both of these approaches, we use as $p(x)$ the normalized marginal probability $p_i/(\sum_i p_i)$ and define the target $q$ as the step pattern associated with the deterministic bipartite solution (Fig.~\ref{fig:infomeasures}c, inset).

The positional information has the advantage over these two measures that it is agnostic to a target pattern. Since we find qualitative agreement between $\mathrm{PI}$ and both the $\mathrm{JSD}$ and $W_1$ distances (Fig.~\ref{fig:infomeasures}b-c), in the main text we report only the positional information.

\subsection{Positional information and spatial correlation}\label{sec:app_PIspatial}

In Sec.~\ref{sec:PI_is_limited}, we highlight that spatial correlations can reduce positional information. In this section, we prove the inequalities Eq.~\eqref{eq:ineqs_PI}. 
We start by the inequality $0\leq I(F_i:F_j) \leq S(F_i)$ for any $j\in \{1,\ldots,N\}$~\cite{cover_elements_2005}. Inserting this inequality in Eq.~\eqref{eq:PI_methods}, we find
\begin{align}
       \mathrm{PI} \leq S_\mathrm{pat} - \frac{1}{N}\sum_i I(F_i:F_j) \label{eq:ineq_PI}
\end{align}
which can be itself bounded using the connected correlation function $C(A,B) = \langle AB\rangle -\langle A\rangle \langle B\rangle$ 
\begin{align}
    \mathrm{PI} \leq S_\mathrm{pat} -\frac{1}{N}\sum_i \frac{1}{2Z^2}C(F_i,F_j)^2 \label{eq:ineq_C}
\end{align}
The inequality~\eqref{eq:ineq_C} follows from Pinsker's inequality~\cite{cover_elements_2005} which captures that the mutual information is larger than $C(A,B)$ since it accounts for all correlations between two random variables, including those that are invisible to two-point correlation functions~\cite{wolf_area_2008}
\begin{equation}
    I(A,B) \geq \frac{1}{2}\frac{(\langle fg\rangle-\langle f\rangle \langle g\rangle )^2}{||f||_\infty ||g||_\infty}
\end{equation}
In particular, for $f=g=\mathrm{id}$ and $F_i$ taking values in $\{1,\ldots,Z\}$, we have $I(F_i:F_j) \geq C(F_i,F_j)^2/(2Z^2)$.

Summing the inequalities (\ref{eq:ineq_PI}-\ref{eq:ineq_C}) for all values of $j$ and dividing by $N$, we find a symmetrized version of these inequalities showing that the PI is reduced from its maximal value by at least the average pairwise information
\begin{align}
    \mathrm{PI} &\leq S_\mathrm{pat} -  \frac{1}{N^2}\sum_{i,j} I(F_i:F_j) \\
    &\leq S_\mathrm{pat} -\frac{1}{2N^2Z^2}\sum_{i,j} C(F_i,F_j)^2.
\end{align}
which, since $S_\mathrm{pat} \leq \log_2 Z$, leads to Eq.~\eqref{eq:ineqs_PI}.

\section{Positional information in the 1D Ising model} \label{sec:app_infoIsing}

Here we detail the computation of the positional information in the 1D Ising model considered in Sec.~\ref{sec:PI_is_limited}. We consider an Ising model with a site varying magnetic field $h_i$, such that the Hamiltonian is given by
\begin{equation}
    H(\{s_i\}_{i=1,\ldots,N}) = - \sum_{i=1}^N Js_i s_{i+1} + h_i s_i  \label{eq:Ising1d_H}
\end{equation}
with periodic boundary conditions such that $s_{N+1} = s_N$. The 1D Ising model can be solved exactly using transfer matrices \cite{kardar_statistical_2007}, with the transfer matrix at the $i$-th bond given by
\begin{equation}
    T_i = \begin{pmatrix}
        e^{J+h_i} & e^{-J} \\ e^{-J} & e^{J-h_i}
    \end{pmatrix}. \label{eq:transfer_mat}
\end{equation}
The partition function is given by 
\begin{equation}
    \mathcal{Z} = \mathrm{tr}\prod_{i=1}^N T_i
\end{equation}
from which the probability of a particular particle configuration $\{s_i\}$ is given by
\begin{equation}
    p(\{s_i\})  = \frac{e^{-H(\{s_i\})}}{\mathcal{Z}}.
\end{equation}

\begin{figure*}
    \centering
    \includegraphics[scale=0.95]{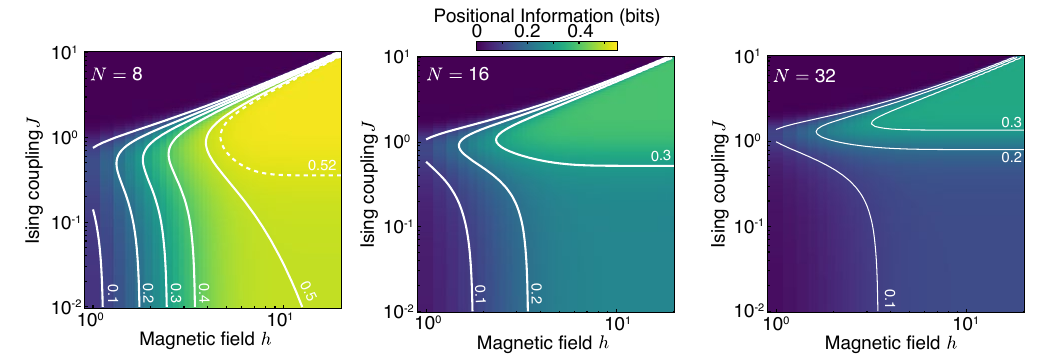}
    \caption{The maximal positional information in the Ising model decreases with increasing $N$. Parameter search of $J$ and $h$ for variable $N$ show that the optimal region of $J,h$ and optimal value of PI both become smaller with increasing $N$.}
    \label{fig:Ising_varN}
\end{figure*}

From the probability distribution,  give the marginal probability of finding a positive particle at site $i$ and its mean value can be obtained by some transfer matrix manipulations as 
\begin{subequations}
\begin{align}
    p_i \equiv p_i(s_i=+) = \frac{1}{\mathcal{Z}} \mathrm{tr}\left[\left(\prod_{j=1}^{i-1} T_j\right) P  \left(\prod_{j=i}^{N} T_j\right) \right], \\  \langle s_i \rangle = \frac{1}{\mathcal{Z}} \mathrm{tr}\left[\left(\prod_{j=1}^{i-1} T_j\right) \sigma_Z  \left(\prod_{j=i}^{N} T_j\right) \right]
\end{align}
\end{subequations}
with the matrices 
\begin{align}
 P = \begin{pmatrix}
     1 & 0 \\ 0 & 0
 \end{pmatrix}, \quad \quad \sigma_z = \begin{pmatrix}
     1 & 0 \\ 0 & -1
 \end{pmatrix}
\end{align}
These marginal probabilities can be numerically evaluated, allowing us to compute the positional information in the Ising model for a given set of $h_i$ and $J$. To do so, we define the pooled probability of finding a particle in the $+$ state
\begin{equation}
    P_+ = \frac{1}{N}\sum_i p_+(i)
\end{equation}
which allows us to compute the positional information exactly as
\begin{equation}
    \mathrm{PI} = \frac{1}{N}\sum_i  p_i \log \left(\frac{p_i}{P_+}\right) + (1-p_i) \log\left(\frac{1-p_i}{1-P_+}\right).
\end{equation}
We show results of the numerical evaluation of the marginals in Fig.~\ref{fig:morpho_analysis}, and show $\mathrm{PI}$ against $J,h$ for variable $N$ in Fig.~\ref{fig:Ising_varN}. We find that as the effective temperature $T$ defined by $J=\tilde{J}/T$, $h=\tilde{h}/T$ is reduced, the PI saturates to a value $\Pi_N$ dependent on the number of sites $N$ [Eq.~\eqref{eq:Pi_N}], in agreement with the numerical results to high accuracy.

\section{Extending to multiple species: Potts model and bound extension}
\label{sec:potts_si}

\begin{figure*}
    \centering
    \includegraphics[scale=0.95]{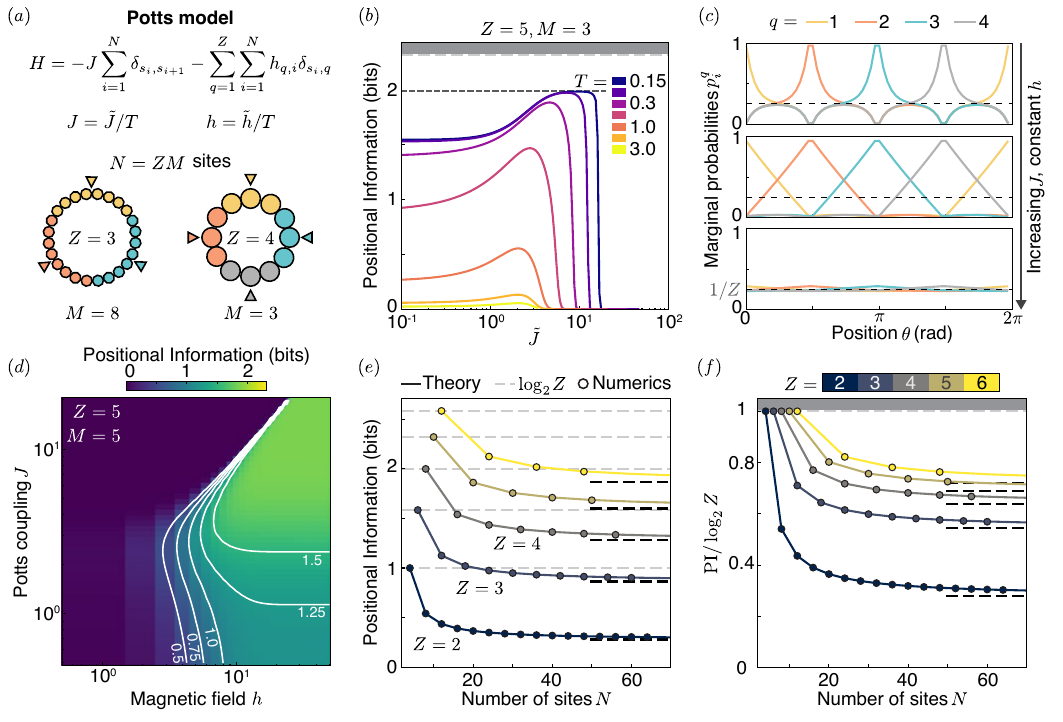}
    \caption{Positional information saturation in the Potts model. (a) The Potts model with symmetric sources (triangles) as described in the main text. (b) Similarly as the Ising model, the Potts model shows a saturation of the PI with decreasing temperature at constant temperature-scaled magnetic field $\tilde{h}$. Greyed region indicates maximal PI for $Z=5$, while dashed line indicates the theoretical bound of Eq.~\eqref{eq:bound_PI_potts} for local coupling. (c) Marginal probabilities again display a sawtooth profile at optimal $J$. Here, $Z=4, M=16, h=6, J\in\{2,5,8\}$ with $J$ increasing from top to bottom rows. (d) The $\mathrm{PI}$ as a function of $J, h$ saturates around $1.6$ bits for $Z=5, M=5$. (e-f) The maximal PI from numerics is in excellent agreement with the theoretical bound. Light dash lines indicate absolute maximum $\log_2Z$, while black dash lines indicate asymptotic value $\Pi_\infty^Z = \log_2Z-1/(2\ln2)$. (f) The maximum PI as $Z$ increases becomes closer to $\log_2Z$, as the limit imposed by the sawtooth profile becomes relatively less important as $\sim1/
\ln Z$.}
    \label{fig:potts}
\end{figure*}

We consider the Potts model with $Z$ states on a periodic 1D lattice, which has a Hamiltonian
\begin{equation}
    H(\{ s_i\}_{i=1,
    \ldots,N}) = -J\sum_{i=1}^N \delta_{s_i, s_{i+1}} - \sum_{q=1}^Z \sum_{i=1}^N h_{q,i} \delta_{s_i,q}.
\end{equation}
where again we use the convention $s_{N+1}=s_1$. The $Z\times Z$ transfer matrix at position $i$ is then given by its components
\begin{equation}
    T^{k,q}_i = e^{J\delta_{kq}}e^{h_{k,i}/2+h_{q,i}/2}
\end{equation}
where $k,q\in \{1,\ldots,Z\}$. The partition function is again $\mathcal{Z} = \mathrm{tr}\left(\prod_{i=1}^N T_i\right)$, allowing us to obtain the steady-state probability distribution. Defining the projection matrix on the $q$-th species $[\mathcal{P}_q]_{ij} = \delta_{iq}\delta_{jq}$,
we can again obtain the marginal probability for the $q$-th species at site $i$ by
\begin{align}
    p_{q,i} \equiv p_i(s_i = q) = \frac{1}{\mathcal{Z}} \mathrm{tr}\left[\left(\prod_{j=1}^{i-1} T_j\right) \mathcal{P}_q \left(\prod_{j=i}^{N} T_j\right) \right].
\end{align}
We can then compute $\mathrm{PI}$ from the probabilities $p_{q,i}$ as 
\begin{align}
    \mathrm{PI} = \frac{1}{N} \sum_{i=1}^N \sum_{q=1}^Z p_i^q \log_2\left(\frac{p_i^q}{P_q}\right)
\end{align}
where $P_q$ is the expected fraction of cells in state $q$.
We compute these probabilities for symmetric systems with $h_{q,i} = h\delta_{i,qM}$, with $h$ a prescribed intensity. In Fig~\ref{fig:potts}b-d we show the steady-state probabilities and the PI as a function of $J$ and $h$. In particular, we again find that for a given system size $N=ZM$, the PI is bounded below its maximal value of $\log_2Z$.

Examining the marginals $p_i^q$, we find again an optimal sawtooth profile (Fig.~\ref{fig:potts}c). We now prove the optimality of the sawtooth profile for the systems with $Z>2$ and compute the corresponding optimal value of the PI by an extension of the previous argument.

If there are $Z > 2$ possible fates (outcomes) for each cell, then the maximum PI is achieved for a symmetric system with $N = ZM$ sites, with sources of different species at $i=qM$, $q=1,\ldots, Z$. By symmetry, we focus on the source at $i=1$ and the interval $i\in\{1,\ldots,M\}$. Then to be as deterministic as possible and satisfy the exponential correlation decay, each source leads to an exponentially-decaying probability profile $p_i^1$ of finding species $1$, while $p_i^2$ correspondingly rises as $p_i^2 = 1-p_i^1$ and all other $p_i^q=0$ for $2<q\leq Z$. We are thus reduced to the $Z=2$ case on this interval $i\in\{1,\ldots,M\}$, which leads to a contribution to the PI bounded by the sawtooth profile 
\begin{align}
    \frac{1}{N} \sum_i^M \left[p_i^1\log_2 p_i^1 + (1-p_i^1)\log_2(1-p_i^1) \right] \nonumber \\\leq \frac{2}{N}\sum_{i=1}^{M}\frac{i-1}{M-1}\log_2\left(\frac{i-1}{M-1}\right)
\end{align}
By symmetry, all intervals $i\in\{(q-1)M+1,\ldots,qM\}$ with $q\in\{1,\ldots,Z\}$ add an identical contribution, which sum up to
\begin{align}
    \mathrm{PI} &\leq \log_2 Z + \frac{2Z}{N}\sum_{i=1}^M \frac{i-1}{M-1}\log_2\left(\frac{i-1}{M-1}\right) \\
    &=\log_2 Z + \frac{2}{M}\sum_{i=1}^M \frac{i-1}{M-1}\log_2\left(\frac{i-1}{M-1}\right) = \Pi^Z_N    \label{eq:bound_PI_potts}
\end{align}
Similar to the $Z=2$ case, we recognize a Riemann sum and hence find the asymptotic bound
\begin{equation}
    \Pi^Z_N \rightarrow \Pi_\infty^Z = \log_2 Z - \frac{1}{2\ln2}.
\end{equation}
with $\Pi^Z_N \geq \Pi_\infty^Z$. The predicted curves $\Pi^Z_N$ match accurately our numerically determined values, and the bound becomes relatively less constraining as $Z$ increases (Fig.~\ref{fig:potts}ef).

\section{Pattern robustness and stability of domain wall dynamics} \label{sec:si_domainwall}

We show in Sec.~\ref{sec:PI_is_limited}, Fig.~\ref{fig:morpho_analysis} that we can justify the existence of an optimal diffusion constant as a consequence of Ising physics. Intuitively, information from the boundaries can only penetrate a distance of the order of the correlation length, and thus robust patterning requires larger correlation lengths, but systems with too large correlation lengths become homogeneous as the pattern becomes unstable.
Here, we present a mechanistic view of this result as a consequence of domain wall dynamics discussed in Sec.~\ref{sec:domain_walls}, which also helps us to generalize our results to other front-sustaining systems (Sec.~\ref{sec:nonlocal}; \cite{di_talia_waves_2022}). In particular, while we consider in the main text $\mathbb{Z}_2$-symmetric systems, in experimentally relevant settings there might be imperfections that break the symmetry in rates between $A$ (+) and $B$ (-) states. 
After presenting general results on the dynamics of asymmetric fronts, we compute the front diffusion constant in asymmetric systems and derive the confinement force acting on the front in the presence of boundaries.  Finally, we derive a quantitative criterion under which we can safely ignore asymmetries in rates between $A$ and $B$ states by comparing the velocity due to the relative stability of states to the noise amplitude.

\subsection{Domain wall dynamics in bistable systems}

We will consider simplified cubic order dynamics and neglect density fluctuations $\rho=\rho_0$, but extend to the general cubic 1D reaction-diffusion system~\cite{van_saarloos_front_2003,petrovskii_exactly_2005,birzu_fluctuations_2018} considered on the unbounded real line $x\in \mathbb{R}$
\begin{equation}
    \partial_t m = D \partial_x^2 m + f(m) =  D \partial_x^2m + r(m-m^*)(1-m^2)
\end{equation}
where $f(m)$ denotes the reaction term. We note that by using the transformation $m\leftarrow2\rho-1$, $m^*\leftarrow2\rho^*-1$, $r \leftarrow 8r$ we recover the model studied in Ref.~\cite{birzu_fluctuations_2018, petrovskii_exactly_2005}. To derive the effects of noise and external sources on front dynamics, we here present and extend relevant results of the theory of wave propagation in metastable states in infinite domains~\cite{meerson_velocity_2011, khain_velocity_2013,petrovskii_exactly_2005, mori_asymptotic_2011}. If $m^* = 0$, we recover the symmetric case, which admits stable domain-wall solutions connecting stable domains at $m=\pm 1$ with characteristic width $\ell = \sqrt{D/r}$.  If $m^* \neq 0$, the two stable solutions at $m_- = -1$ to $m_+ = 1$ are no longer equivalent, and the domain walls of width $\ell_f$ can have a non-zero front velocity $c$.
To find $c$, we consider solutions of the form $m(\xi) = m(x-ct)$ which must satisfy
\begin{equation}
   D m'' + cm' + f(m) = 0. \label{eq:front_eq}
\end{equation}
For fronts connecting the stable states $m_- = -1$ to $m_+ = 1$, we can multiply Eq.~\eqref{eq:front_eq} by $m'(\xi)$ and integrate from $\xi = -\infty$ to $+\infty$. 
Multiplying by $m'$ and integrating over the domain, we thus find the relationship
\begin{equation}
    c = - \frac{\int_{m_-}^{m_+} f(m)\mathrm{d}m}{\int_{-\infty}^{+\infty} (m')^2 \mathrm{d}\xi} \label{eq:c_front}
\end{equation}
where we assume $m'\rightarrow 0$ at infinities. The front profile is given by 
\begin{equation}
    m(\xi) = -\tanh\left(\frac{\xi}{\sqrt{2}\ell}\right) 
\end{equation}
where $\ell = \sqrt{D/r}$. By changing variables and using that $m' = (1-m^2)/(\sqrt{2}\ell)$, the denominator above can be computed as
\begin{equation}
    \int_{-\infty}^{+\infty} (m')^2 \mathrm{d}\xi = \int_{-1}^1 \frac{1}{\sqrt{2}\ell}(1-m^2)\mathrm{d}m = \frac{2\sqrt{2}}{3\ell}.
\end{equation}
For the cubic reaction term considered here, the numerator is equal to 
\begin{equation}
    -\int_{m_-}^{m_+} f(m)\mathrm{d}m = \frac{4r}{3} m^*
\end{equation}
indicating that $c$ is proportional to $m^*$ as
\begin{equation}
    c = \sqrt{2rD}m^*.
\end{equation}
If $m^* > 0$, the wave travels forward ($c >0$), while $m^* <0$ leads to backward propagating waves ($c < 0$). The direction of travel can be understood by interpreting the numerator as a difference in potential $\int_{m_-}^{m_+} f(m)\mathrm{d}m = V(m_+) - V(m_-)$, with $f(m) = - \mathrm{d}V/\mathrm{d}m$: the more stable steady state (deeper potential) invades the less stable state (shallower potential) \cite{di_talia_waves_2022, petrovskii_exactly_2005}.

The motility of the domain walls can destroy bistable patterns, by sending the fronts all the way to the edges of the domain. However, as we we detail below, in the presence of noise the front position fluctuates about its average position $x = c t$ with a diffusion constant $D_f$ scaling with $\sqrt{D}$
~\cite{khain_velocity_2013, birzu_fluctuations_2018}.

\subsection{Front diffusivity}

\begin{figure*}
    \centering
    \includegraphics[scale=0.95]{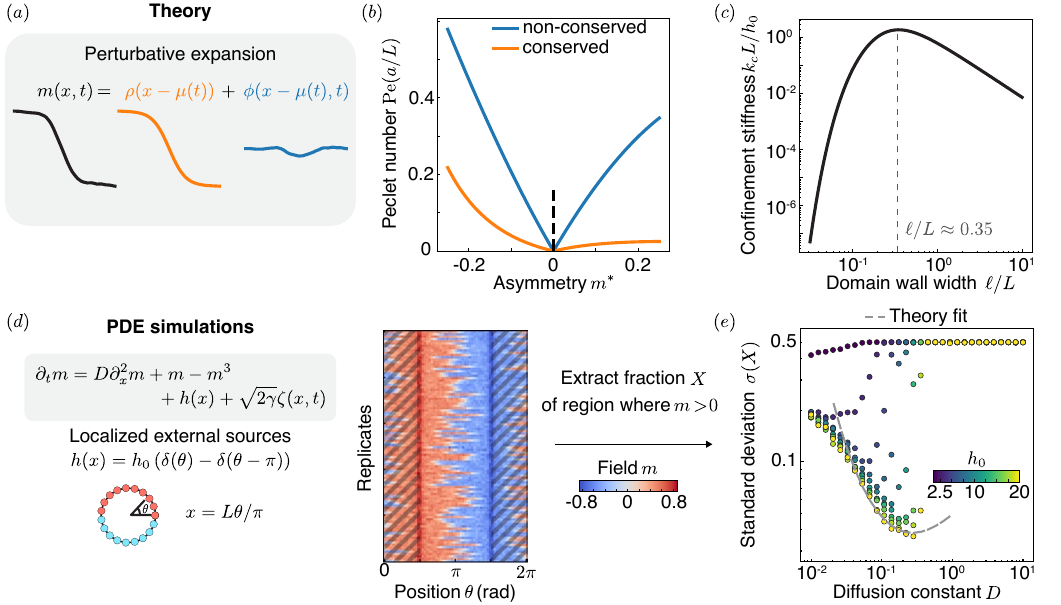}
    \caption{Results from nonlinear front propagation theory, and comparison to simulations. (a) Perturbative expansion used to compute the front diffusivity $D_f$ and the confinement stiffness $k_\mathrm{c}$. (b) P\'eclet number (rescaled by $a/L$) as a function of the asymmetry parameter $m^*$, for non-conserved and conserved noise considered separately. Here $\rho_0=\chi=1$. (c) Calculated confinement stiffness at first order, showing a rapid exponential decay as $\ell/L\rightarrow 0$. (d) We simulate the cubic polynomial reaction-diffusion system, and measure the fraction of positive region for $\theta\in[\pi/2, 3\pi/2]$ (excluding the hashed region). (e) We find that while the confinement stiffness computed perturbatively gives the correct trend, other effects effectively stiffen the confinement at low $D$. Results computed using $500$ replicates, $\gamma=0.1$.}
    \label{fig:peclet}
\end{figure*}

We now present the classical calculation of the front diffusivity $D_f$ by perturbative methods, which will lay out the strategy we use later for the derivation of the confinement force, following Refs.~\cite{rocco_diffusion_2001, meerson_velocity_2011, khain_velocity_2013, birzu_fluctuations_2018}. Counterintuitively, the front diffusivity $D_f$ scales as $\sqrt{D}$. This can be understood as a reduction in variance where the emergent diffusion averages over the particles present in the vicinity of the front as $D_f\sim D/N_\text{int}$, where $N_\text{int}\sim\sqrt{D}$ is the number of particles within interaction range between reactions. In the presence of non-conserved noise, the dynamics of the system are given by
\begin{equation}
    \partial_t m = D\partial_x^2 m + f(m) + \eta(x,t)
\end{equation}
where $\langle \eta(t_1,x_1) \eta(t_2,x_2)\rangle = \Gamma(\phi(t_1,x_1))^2 \delta(t_1-t_2)\delta(x_1-x_2)$.  For conciseness, we refer to Ref.~\cite{birzu_fluctuations_2018} for the derivation in the presence of conserved noise which follows similar steps as the one for non-conserved noise. In the co-moving frame $\xi=x-ct$, the profile now satisfies
\begin{equation}
    Dm''+cm'+f(m)+\eta(t,\xi) = 0.\label{eq:fluct_front_dyn}
\end{equation}
Note that in the co-moving frame the noise is still explicitly time-dependent and satisfies $\langle \eta(t_1,\xi_1) \eta(t_2,\xi_2)\rangle = \Gamma(\phi(\xi_1))^2 \delta(t_1-t_2)\delta(\xi_1-\xi_2)$.
The effect of noise is twofold: noise shifts the position of the front $\xi=x-ct$ by a time-dependent term $\mu(t)$, and changes the shape of the front away from its unperturbed shape $\rho$. To solve Eq.~\eqref{eq:fluct_front_dyn} perturbatively, we thus write the solution in terms of the shifted unperturbed solution $\rho$ and the first-order perturbation of the front profile $\phi$ (Fig.~\ref{fig:peclet}a)
\begin{equation}
    m(t,\xi) = \rho(\xi-\mu(t)) + \phi(t,\xi-\mu(t)). \label{eq:decomp_m}
\end{equation}
Note, again, that $\phi(t,\xi)$ is explicitly time-dependent. Introducing the decomposition of Eq.~\eqref{eq:decomp_m} into Eq.~\eqref{eq:fluct_front_dyn}, we have
\begin{equation}
    \partial_t \phi - \mathcal{L}\phi =  \rho'\dot{\mu}+\eta(t,\xi) \label{eq:1st_order_phi}
\end{equation}
where $\dot{\mu} = \mathrm{d}\mu/
\mathrm{d}t$ and we introduce the differential operator $\mathcal{L} = \frac{\mathrm{d}^2}{\mathrm{d}\xi^2 } + c\frac{\mathrm{d}}{\mathrm{d}\xi} +f'(m)$. Reflecting the translational invariance of the dynamics, we have $
\mathcal{L}\rho' = 0$. The adjoint $\mathcal{L}^\dagger = \frac{\mathrm{d}^2}{\mathrm{d}\xi^2 } - c\frac{\mathrm{d}}{\mathrm{d}\xi} +f'(m)$ thus has a zero eigenvector $\mathcal{L}^\dagger \psi = 0$, with $\psi = e^{c\xi}\rho'(\xi)$. To exclude translations of the profiles from being included in $\phi$, as they should only be incorporated through $\mu(t)$, we require the solvability condition (which implicitly defines $\mu(t)$)
\begin{equation}
    \int_{-\infty}^{+\infty} \mathrm{d}\xi \,\psi(\xi) \phi(\xi) =0.
\end{equation}
Multiplying Eq.~\eqref{eq:1st_order_phi} by $\psi$ and integrating over $\xi$, we find by using the solvability condition
\begin{align}
    \dot{\mu}(t) = - \frac{\int_{-\infty}^{+\infty}\mathrm{d}\xi\, \psi(\xi) \eta(t,\xi)}{\int_{-\infty}^{+\infty}\mathrm{d}\xi\, \psi(\xi) \rho'(\xi)}.
\end{align}
Since $\langle \eta(t,\xi)\rangle = 0$, we remark that $\langle \dot{\mu}(t)\rangle = 0$. Integrating $\dot{\mu}(t)$ in time, we now have an expression for the shift in position $\mu(t)$ as a function of the noise dynamics
\begin{align}
    \mu(t) = X(t)-ct = - \int_0^t  \frac{\int_{-\infty}^{+\infty}\mathrm{d}\xi \,\psi(\xi) \eta(t',\xi)}{\int_{-\infty}^{+\infty}\mathrm{d}\xi \,\psi(\xi) \rho'(\xi)}\mathrm{d}t'. \label{eq:h_t_front}
\end{align}
The diffusivity can now be obtained by computing the mean squared displacement at a time $t$
\begin{equation}
    D_f^\mathrm{nc} = \frac{\langle X_f^2(t)\rangle-\langle X_f\rangle^2}{2t} = \frac{\langle \mu^2(t) \rangle}{2t}.
\end{equation}
\begin{widetext}
Inserting the expression of $\mu(t)$ in Eq.~\eqref{eq:h_t_front} to compute the variance, we have
\begin{align}
    D_f^\mathrm{nc} = & \frac{1}{2t} \left\langle  \int_0^t  \frac{\int_{-\infty}^{+\infty}\mathrm{d}\xi\, \psi(\xi) \eta(t_1,\xi)}{\int_{-\infty}^{+\infty}\mathrm{d}\xi \,\psi(\xi) \rho'(\xi)} \mathrm{d}t_1 \int_0^t \frac{\int_{-\infty}^{+\infty}\mathrm{d}\xi\, \psi(\xi) \eta(t_2,\xi)}{\int_{-\infty}^{+\infty}\mathrm{d}\xi\, \psi(\xi) \rho'(\xi)}\mathrm{d}t_2\right\rangle \\
    =& \frac{1}{2t}\frac{\int_0^t \mathrm{d}t_1 \int_0^t \mathrm{d}t_2 \int_{-\infty}^{+\infty}\mathrm{d}\xi_1 \int_{-\infty}^{+\infty}\mathrm{d}\xi_2 \, \psi(\xi_1)\psi(\xi_2)\langle \eta(t_1, \xi_1)\eta(t_2, \xi_2)\rangle}{\left(\int_{-\infty}^{+\infty}\mathrm{d}\xi\, \psi(\xi)\rho'(\xi) \right)^2} \\
    = & \frac{1}{2} \frac{\int_{-\infty}^{+\infty}\mathrm{d}\xi\, \psi(\xi)^2 \Gamma(\phi(\xi))^2}{\left(\int_{-\infty}^{+\infty}\mathrm{d}\xi\, \psi(\xi)\rho'(\xi) \right)^2}
\end{align}
For additive noise $\Gamma^2  = a\rho_0 \hat{\gamma}$ as we find in the DIM, we then have
\begin{equation}
    D_f^\mathrm{nc} = \frac{3a\rho_0\hat{\gamma}}{4\pi} \sqrt{\frac{D}{r}} \frac{(4(m^*)^2-1)}{m^*(1+m^*)(m^*-1)^2}\tan{\pi m^*} = \frac{3a\rho_0}{4\chi}\sqrt{r D} + O(m^*)
\end{equation}
where we define $\chi = r/\hat{\gamma}$.

In the presence of conserved noise $\partial_x(\Gamma(\phi)\eta)$, one finds that $D_f$ is given by \cite{birzu_fluctuations_2018}
\begin{equation}
    D_f^\mathrm{c} = \frac{a\rho_0 }{5\pi}\sqrt{rD} \frac{(3+m^*)(4(m^*)^2-1)(1+2m^*)(9+7m^*)}{m^*(1+m)(1-m)^2}\tan\pi m^* = \frac{27a\rho_0}{5}\sqrt{rD}+O(m^*)
\end{equation}
when $\Gamma^2= a\rho_0D$ as in the (simplified) Diffusive Ising model. If both conserved and non-conserved noises are present and uncorrelated, then those contributions are additive 
\begin{equation}
    D_f = D_f^\mathrm{nc} + D_f^\mathrm{c}.
\end{equation}
and we see that $D_f$ overall scales with $a\rho_0\sqrt{rD}$.
\end{widetext}

To compute the relative importance of the state asymmetry compared to the diffusion of the front, we can compute the P\'eclet number $\mathrm{Pe} = cL/D_f$ where $L$ is the size of the domain. In both conserved and non-conserved cases, $\mathrm{Pe}$ scales as $L/(\rho_0 a)$ with a $m^*$  dependent scaling factor, with an additional scaling $\mathrm{Pe}\propto \chi$ in the non-conserved case: as the system gets closer to criticality ($\chi=0$), the effect of the noise is amplified (Fig.~\ref{fig:peclet}b).

\subsection{Front confinement by external sources}

Now we consider an external symmetry-breaking field $h(x) = h_0 \left[ \delta(x-L/2)-\delta(x+L/2)  \right]$, and we restrict ourselves for simplicity to the case $c=0$. While we still consider here that the domain is unbounded, translational symmetry is now broken and we stay in $(x,t)$ coordinates. The stationary front solution now satisfies
\begin{equation}
    D\rho''(x)+f(\rho(x)) = -h(x) \label{eq:affine_stab}
\end{equation}
where we emphasize the functional dependence on the position. We now look for the restoring force on the front when the front is perturbed and shifted away from its equilibrium position. Decomposing again $m(x,t) = \rho(x-\mu(t)) + \phi(x-\mu(t), t)$, at first order
\begin{widetext}
\begin{align}
    \partial_t \phi(x-\mu(t)) - \mathcal{L}\phi(x-\mu(t)) = D\rho''(x-\mu(t)) +f(\rho(x-\mu(t))+h(x) + \rho'(x-\mu(t))\dot{\mu(t)}. \label{eq:perturb_h}
\end{align}
\end{widetext}
By introducing $h(x-\mu(t))$ in Eq.~\eqref{eq:perturb_h}, we can use Eq.~\eqref{eq:affine_stab} and the operator $\mathcal{L}$ defined previously to write
\begin{equation}
    \partial_t \phi - \mathcal{L}\phi = \rho'\dot{\mu} + (h(x)-h(x-\mu)) \approx \rho'\dot{\mu} +h'(x)\mu \label{eq:perturb_h_2}
\end{equation}
where we now suppress the position dependence except on the source field and use that $\mu(t)$ is assumed small. Since $c=0$, the operator $\mathcal{L}=\mathcal{L}^\dagger$ is now self-adjoint and has a zero-mode given by $\rho_0'$, where $\rho_0(x) = -\tanh(x/\sqrt{2}\ell)$ is the steady profile in the absence of sources. After projecting onto this zero mode by multiplying Eq.~\eqref{eq:perturb_h_2} by $\rho_0'(x)$ and integrating over $x$ and using the solvability condition $\int \mathrm{d}x \rho_0(x)\phi(x-\mu(t),t) = 0$, we find the first-order differential equation for $\mu(t)$
\begin{equation}
    \dot{\mu}(t) = -\frac{\int_{-\infty}^{+\infty} \mathrm{d}x \,\rho'_0 h'}{\int_{-\infty}^{+\infty} \mathrm{d}x \, \rho_0'\rho'} \mu(t) \equiv -k_c \mu(t).
\end{equation}
The solvability condition effectively defines $\mu(t)$ so that $\phi(x-\mu(t))$ has no overlap with $\rho_0$.
If we assume that the source field strength $h_0$ is small enough so that the solution of Eq.~\eqref{eq:affine_stab} is close to the solution of the homogeneous problem, we can approximate $\rho\approx \rho_0$.  We then find a confinement stiffness
\begin{equation}
    k_c = \frac{\int_{-\infty}^{+\infty} \mathrm{d}x\, \rho'_0 h'}{\int_{-\infty}^{+\infty} \mathrm{d}x \, \rho_0'\rho_0'} = \frac{2\rho_0''(L/2) h_0}{\int \mathrm{d}x \, (\rho_0')^2}.
\end{equation}
where the last equality follows from the oddness of the unperturbed profile $\rho(-L/2)=-\rho(L/2)$. With $\rho_0(x) = - \tanh(x/\sqrt{2}\ell)$, we thus find
\begin{equation}
    k_c = \frac{3h_0}{\sqrt{2}\ell}\frac{\sinh\left(L/(2\sqrt{2}\ell)\right)}{\cosh\left(L/(2\sqrt{2}\ell)\right)^3} 
\end{equation}
which has a maximum at $\ell \approx 0.35 L$, tends to zero for both $\ell\rightarrow0$ and $\ell \rightarrow +\infty$ and asymptotically scales as 
\begin{equation}
    k_c L \sim \left\{ \begin{matrix}
       6\sqrt{2}h_0\left(\frac{L}{\ell}\right)e^{-L/(\sqrt{2}\ell)} &\text{ if } \ell \ll L \\
        \frac{3h_0}{4\sqrt{2}}\left( \frac{L}{\ell}\right)^2 &\text{ if } \ell \gg L
    \end{matrix}\right.
\end{equation}
This exponential decay as $\ell/L \ll 1$ is very strong, and we expect other effects such as front deformation at the edges to dominate the confinement effects. In direct simulations of the noisy reaction-diffusion system with non-conserved noise, we fit the position and amplitude of the confinement force to the observed variance, which relates to the confinement stiffness as $\sigma = \sqrt{D_f/k_\mathrm{c}}$ (Fig.~\ref{fig:peclet}d-e). More precisely, to allow some margin for deviations of the front profile from the hyperbolic tangent shape, we fit the constants $\alpha$ and $\beta$ where $\ell = \alpha \sqrt{D}$ and $\sigma = \beta /\sqrt{k_\mathrm{c}[\ell]}$ such that the minimum coincides with the observed one. We find that while there is a clear minimum variance (maximum confinement stiffness) with varying $D \propto \ell^2$, the variance does not grow as fast as predicted as $D\rightarrow0$. We also remark that while the source strength $h_0$ sets the width of the region where the front solutions exist, the front variance only weakly varies with  $h_0$ in this existence region. In conclusion, while the confinement force derived above seems to capture some of the observed phenomenology, it is not fully quantitative, perhaps because it fails to account for the stability of the homogeneous solutions.

\section{Wave-pinning as integral feedback} \label{sec:appendix_wp}

In this appendix, we specify how we choose parameters for the wave-pinning model described in Sec.~\ref{sec:nonlocal}, which is given from the microscopic rules \textbf{W1-3}. The derivation of the continuum model is postponed to Appendix~\ref{sec:wp_cg_derivations}. 
We then compute the two-point correlation function for the system linearized about its steady-state, and then show how the model can be adapted to describe \emph{C. elegans} antero-posterior patterning.

\subsection{Parameter selection}

 To compute the coefficients $r_2, u_2$ and choose simulation parameters, we examine steady homogeneous solutions for which $\rho=\bar{\rho}$, $s=\bar{s}$. For a system in dynamical balance, in the homogeneous domains $s^c = 0$ and $\rho^c = \rho^0 - \bar{\rho}/a$. The steady state values thus satisfy
 \begin{subequations}
     \begin{align}
         \bar{s}\left( \frac{k}{4r}\bar{\rho}^2 - 1  - \frac{k}{4r}\bar{s}^2\right) & = 0\\
         \bar{\rho} \left( 2 + \frac{k}{4r}(\bar{\rho}^2-\bar{s}^2)\right) &= \rho^0
     \end{align}
 \end{subequations}
To impose $\bar{s} \neq 0$, this imposes 
\begin{align}
    \bar{\rho} = \frac{\rho^0}{2+1/a}, \quad \bar{s}  = \pm \sqrt{\bar{\rho}^2-\frac{4r}{k}}.
\end{align}

The steady state density in the WPI model is $\bar{\rho} = \rho^0/(2+1/a)$. For simulations, we want the bistable solutions $\bar{s} \neq 0$ to exist. For $\rho^0$ large enough, $k = (4/3) r$ satisfies the requirement.  

 \subsection{Two-point correlations in the Wave-Pinning Ising model}
\label{sec:2point_corr_wp}
 
Here we compute the two-point correlation function for the Wave-pinning model, confirming that this model has non-vanishing long-range correlations.
To proceed, we consider the linearization of the dynamics around a steady-state front solution $s_0(x) = \mathrm{sgn}(x-L/2)s_+$ with $s_+ = \sqrt{u_2/r_2}$ on a periodic domain of length $L$. We take $s^0=0$ We write the magnetization $s(x,t) = s_0(x) + \phi(x,t)$, which leads to the dynamics 
\begin{equation}
     \partial_t \phi = D \partial_{xx}^2 \phi - 2 r_2 \phi - \frac{r}{a}\int_0^L \mathrm{d}x \,\phi + f(x,t)
\end{equation}
To compute the two-point correlation function, we can write the Fourier transform of $\phi$ in time and space
\begin{subequations}
     \begin{align}
         \phi(x,t) & = \sum_q \int_{-\infty}^{+\infty} \frac{\mathrm{d}\omega}{2\pi} \phi_{q,\omega} e^{-i(qx-\omega t)} \\
         \phi_{q,\omega} & = \frac{1}{L}\int_0^L\mathrm{d}x \int_{-\infty}^{+\infty}\mathrm{d}t\, \phi(x,t) e^{i(qx-\omega t)} 
     \end{align}
\end{subequations}%
where we note that the sum over the wavenumber $q=2\pi n/L, n\in\mathbb{Z}$ is discrete due to the periodicity of the domain.
 \begin{widetext}
In Fourier space, the equations for $q=0$ and $q\neq 0$ modes read
 \begin{subequations}
     \begin{align}
         i\omega \phi_{0,\omega}&= -\left(2r_2+\frac{Lr}{a}\right)\phi_{0,\omega} +f_{0,\omega}, \\
         i\omega \phi_{q,\omega} & = -(Dq^2 +2r_2) \phi_{q,\omega} + f_{q,\omega},
     \end{align}
 \end{subequations}

 For Gaussian driving noise $\langle f_{q,\omega} f_{q',\omega'}\rangle = \theta \delta(q+q')\delta(\omega+\omega')$, we thus find that the correlation function in Fourier space is given by
 \begin{subequations}
     \begin{align}
         \langle \phi_{q,\omega}\phi_{q', \omega'}\rangle =  &\,\delta(q+q') \delta(\omega+\omega')  \left( \frac{\delta_{q,0}}{\omega^2+(2r_2+Lr/a)^2}+\frac{1-\delta_{q,0}}{\omega^2+(Dq^2+2r_2)^2} \right)
     \end{align}
 \end{subequations}
 As $t\rightarrow\infty$, we can find the steady-state correlation function
 \begin{equation}
     \langle \phi(x)\phi(x')\rangle = \sum_q \langle \phi_{q,0}\phi_{-q,0}\rangle \approx \frac{\theta}{2r_2+Lr/a} + \theta\int_{-\infty}^{+\infty}\frac{\mathrm{d}q}{2\pi} \frac{e^{-iq(x-x')}}{D q^2 + 2r_2} = \frac{\theta}{2r_2+Lr/a} + \frac{\theta}{2\sqrt{2Dr_2}}e^{-|x-x'|/\ell}
 \end{equation}
 with $\ell = \sqrt{D/(2r_2)}$ and the approximation comes from approximating the finite sum by an integral. The coupling to the spatially uniform $(q=0)$ mode thus leads to long-range coupling by preventing the vanishing of correlations in fluctuations as $|x-x'|\rightarrow+\infty$. 
 \end{widetext}

\subsection{Experimental example: \emph{C. elegans} antero-posterior patterning} \label{sec:celegans_numbers}

Wave-pinning models have found important use in explaining \emph{C. elegans} antero-posterior patterning. In this section, we develop a microscopically-plausible lattice model that coarse-grains to the continuum model introduced by Goehrings et al \cite{goehring_polarization_2011}. In this model, the embryo is modeled as a prolate spheroid with radii $27 \times 15\times 15$ $\mu$m and by axisymmetry the system is reduced to one-dimensional dynamics. We here use the notation from that reference.
\begin{widetext}

In the lattice version of the model, the binding and unbinding reactions for the anterior PARs $A$ (aPARs: PAR-3, PAR-6, and atypical protein kinase C) $A$ and posterior PARs $P$ (pPARs: PAR-1, PAR-2, and LGL)
 \begin{subequations}
     \begin{align}
        A \leftrightharpoons & A_\text{cyto} \\ 
        P \leftrightharpoons & P_\text{cyto}
     \end{align}
 \end{subequations}
have reaction rates in lattice units given by $k_\text{off}^A + \tilde{k}_{AP}P$, $k_\text{off}^A + \tilde{k}_{PA}A(A-1)$. The goal here is to relate the lattice units to the measured experimental values.

From these reactions, the coarse-graining procedure performed on the wave-pinning Ising (Appendix.~\ref{sec:wp_cg_derivations}) can be readily adapted, with the Poissonian ansatz giving 
\begin{subequations}
\begin{align}
    \partial_t N_A = & D_A \partial_x^2 N_A - k_\text{off}^A N_A -\tilde{k}_{AP} N_P N_A + \tilde{k}_\text{on}^A N_\text{A,cyto} \nonumber\\
    &+ \partial_x \left[\sqrt{2D_A N_A \Delta x}\zeta_1 \right] + \sqrt{(k_\text{off}^A N_A +\tilde{k}_{AP} N_P N_A + \tilde{k}_\text{on}^A N_\text{A,cyto})\Delta x}\zeta_2 \\
    \partial_t N_P = & D_P \partial_x^2 N_P - k_\text{off}^P N_P -\tilde{k}_{PA} N_A^2 N_P + \tilde{k}_\text{on}^P N_\text{P,cyto} \nonumber\\
    &+ \partial_x \left[\sqrt{2D_P N_P \Delta x}\zeta_3 \right] + \sqrt{(k_\text{off}^P N_P +\tilde{k}_{PA} N_A^2 N_P + \tilde{k}_\text{on}^P N_\text{P,cyto})\Delta x}\zeta_4 
\end{align}
\end{subequations}
Where $N_A$ and and $N_P$ the number of a- and p-PARs on a given lattice site, and $N_{X,cyto}$ their number in the cytoplasm of the embryo. To relate the model parameters to the experimental concentration units, we seek the conversion factor $\Omega$ (which has units of surface) such that $A = N_A/\Omega$, $P = N_P /\Omega$. We use the number conservation laws written in lattice and continuum units to find $\Omega$. 
 
Number conservation in the continuum model reads
\begin{equation}
    A_\text{cyto} V = \rho^A_\text{tot} V - S \langle A \rangle
\end{equation}
or equivalently
\begin{equation}
    A_\text{cyto} = \rho^A_\text{tot} - \psi \langle A \rangle
\end{equation}
where $\psi = S/V = 0.174$ $\mu\text{m}^{-1}$ for the prolate spheroid representing the embryo, and $\langle A \rangle = L^{-1} \int_0^L \mathrm{d}x A(x)$ is the average membrane-bound density, with $L = 135$ $\mu$m. Written as is, this conservation law neglects the curvature of the embryo, which according to Ref.~\cite{goehring_polarization_2011} does not significantly affect results.
In terms of molecule numbers, we have
\begin{equation}
    N^A_\text{cyto}= N^A_\text{tot} - S \langle A \rangle \label{eq:num_cons_cont}
\end{equation}
In the microscopic lattice model, we have
\begin{equation}
    N_\text{cyto}^A = N_\text{tot}^A - \left(\frac{L}{\Delta x}\right) \langle N_A \rangle \label{eq:num_cons_lattice}
\end{equation}
For number conservation to be consistent between Eqs.~\eqref{eq:num_cons_cont}~and~\eqref{eq:num_cons_lattice}, we thus find that
\begin{equation}
    \Omega = \left(\frac{\Delta x}{L}\right)S.
\end{equation}
We thus get the fluctuating continuum equations in the experimental units of \cite{goehring_polarization_2011}, 
\begin{subequations}
    \begin{align}
        \partial_t A = & D_A \nabla^2 A - k_\text{off}^A A - k_{AP} P A + k_\text{on}^A A_\text{cyto} \nonumber\\ 
        & + \partial_x \left[ \sqrt{2D_A A \left(\frac{\Delta x}{\Omega}\right)}\zeta_1\right] + \sqrt{\frac{\Delta x}{\Omega} (k_\text{off}^A A + k_{AP}AP + k_\text{on}^A A_\text{cyto}) }\zeta_2\\
        \partial_t P = & D_P \nabla^2 P - k_\text{off}^P P - k_{PA} A^2 P + k_\text{on}^P P_\text{cyto} \nonumber\\ 
        & + \partial_x \left[ \sqrt{2D_P P \left(\frac{\Delta x}{\Omega}\right)}\zeta_3\right] + \sqrt{\frac{\Delta x}{\Omega} (k_\text{off}^P P + k_{PA}A^2P + k_\text{on}^P P_\text{cyto}) }\zeta_4
    \end{align} \label{eq:SPDE_AP}
\end{subequations}
with $\tilde{k}_{AP}= (k_{AP}/S)(L/a)$, $\tilde{k}_{PA}=  (k_{PA}/S^2)(L/a)^2$, $\tilde{k}_\text{on}^A = k_\text{on}^A (a/L) \psi$,  $\tilde{k}_\text{on}^P = k_\text{on}^P (a/L) \psi$.
\end{widetext}

If we wanted to simulate these dynamics at full scale, how many particles do we need to include in our microscopic dynamics?

For our simulations to be faithful, we need our simulations to be in a regime where there exists homogeneous steady states with positive concentrations. Let $\phi^A$ and $\phi^P$ be the average number of particles per site of type $A$ or $P$. Then in steady state, in homogeneous domains, we have 
\begin{align}
\phi^A = & \frac{k_\text{on}^A}{k_\text{off}^A}\left(\frac{a}{L}\right) \psi  N^A_\text{cyto}, \\
\phi^P = &\frac{k_\text{on}^P}{k_\text{off}^P} \left(\frac{a}{L}\right) \psi  N^P_\text{cyto}.
\end{align}
Total number conservation implies 
\begin{equation}
    \frac{N_\text{cyto}^A}{N_\text{tot}^A} = 1- \frac{k_\text{on}^A}{k_\text{off}^A}\psi
\end{equation}
and likewise for $P$
\begin{equation}
    \frac{N_\text{cyto}^P}{N_\text{tot}^P} = 1- \frac{k_\text{on}^P}{k_\text{off}^P}\psi.
\end{equation}
For the values of the rate constants tabulated in Ref.~\cite{goehring_polarization_2011}, $k^A_\text{off}=5.4\cdot10^{-3}$ $s^{-1}$, $k^P_\text{off}=7.3\cdot10^{-3}$ $s^{-1}$, $k^A_\text{on}=8.58\cdot10^{-3}$ $\mu \text{m}/s$, $k^P_\text{on}=4.74\cdot10^{-2}$ $\mu \text{m}/s$, $k_{AP} = 0.19$ $\mu\text{m}^2/s$, $k_{PA} = 2.0$ $\mu\text{m}^4/s$  and geometric factor $\psi$ and ratio $N^A_\text{tot}/N^P_\text{tot} = 1.56 \approx 3/2$, we have $\psi k_\text{on}^A/k_\text{off}^A\approx 1/3$, $\psi k_\text{on}^P / k_\text{off}^P\approx 1/10$, and thus that to obtain $\phi^P \approx 1$, we need $\phi^A \approx 5 \phi^P = 5$, which requires 
\begin{equation}
    N^A_\text{tot} =  15 \left(\frac{L}{a}\right) \phi^P, \quad
    N^P_\text{tot} = 10 \left(\frac{L}{a}\right) \phi^P.
\end{equation}
where $L/a$ is the number of lattice sites. To get the experimental number of particles $N^A_\text{tot}= 4.0\cdot10^4$, $N^P_\text{tot}= 2.5\cdot 10^4$, we thus need $a = 4\cdot10^-4 L$, or $a = 5\cdot10^{-2}$ $\mu$m. Such a simulation would require $\approx 2.7\cdot10^3$ lattice sites, which while computationally expensive is within reach of tau-leaping schemes used in this article.

\section{Coarse-graining diffusive lattice models}
\label{sec:appendix_cg}

We follow the method outlined in \cite{martin_transition_2025} to obtain fluctuating hydrodynamics descriptions of our microscopic lattice dynamical models. Briefly, the method consists of deriving a field-theoretic form of the chemical master equation, and then leverage a Poissonian ansatz (when diffusion dominates as $a\rightarrow0$) to compute higher-order correlations and derive a closed hydrodynamic equation. 

In what follows, we consider particles of type $+$ and $-$ which can hop to a neighboring site with probability rate $D/a^2$ and change type at site $i$ with rate $g_i^\pm(\{n_+, n_-\})$ which can depend on populations at the same site or neighboring sites. For the DIM, the case considered in \cite{martin_transition_2025}, the flip rate from $+$ to flip to $-$ is $g_i^+ = \gamma e^{-\beta(n_+-n_-)}$ while the flip rate from $-$ to $+$ is $g_i^- = \gamma e^{+\beta(n_+-n_-)}$. 

We also consider the case of nearest-neighbor interactions in the DIM, where the flip rate for particles at site $i$ is given by $g_\pm(\{n_+, n_-\}) = \gamma \prod_{\langle i,k\rangle }e^{\mp \beta(n_+^k-n_-^k)}$. Here the product is taken over the sites $k$ which are nearest-neighbors of $i$ (not including $i$ itself). We find that this case recovers, up to a rescaling of the reaction timescale, the coarse-graining of the same-site DIM, and thus does not greatly affect the phenomenology.

Finally, we coarse-grain the lattice version of the wave-pinning system described in Sec.~\ref{sec:nonlocal} in which particles diffuse and bind or unbind from the lattice into a reservoir, with rates varying depending on the local environment. The phenomenology of this model is detailed in Appendix~\ref{sec:appendix_wp}.

\subsection{From master equation to mean-field hydrodynamics}

To construct the master equation, we consider the random variables $n_i^\pm(t_j)$ counting the number of particles of type $\pm$ at site $i$ at time $t_j$ for $(i,j)\in \{1, \ldots, L\}\times \{1, \ldots, M\}$. The probability of observing any configuration $\{n\} \equiv \left\{n_i^\pm(t_j), (i,j)\in \{1, \ldots, L\}\times \{1, \ldots, M\} \right\}$ is given by
\begin{equation}
    P[\{n\}] = \left\langle \prod_{i=i}^L \prod_{j=1}^M \prod_{\sigma=\pm} \delta\left( n_i^\sigma(t_{j+1}) - n_i^\sigma(t_j)-J_i^\sigma(t_j)\right) \right\rangle_{\{J\}}
\end{equation}
where $J_i^\pm(t_j) \in \{-1, 0, 1 \}$ is the change in particle number of type $\pm$ at site $i$ in the infinitesimal time $t_{j+1}-t_j = \mathrm{d}t$, and the average is taken over all possible such changes in configuration. For our set of reactions, there are four cases:
\begin{enumerate}
    \item[(i)] A particle hops from site $i$ to site $i+1$: $J_i^\pm(t_j) = -1$, $J_{i+1}^\pm(t_j) = 1$. 
    \item[(ii)] A particle hops from site $i$ to site $i-1$: $J_i^\pm(t_j) = -1$, $J_{i-1}^\pm(t_j) = 1$. 
    \item[(iii)] A $+$ particle flips at site $i$: $J_i^+(t_j)  = +1 = - J_i^-(t_j)$.
    \item[(iv)] A $-$ particle flips at site $i$: $J_i^-(t_j)  = +1 = - J_i^+(t_j)$.
\end{enumerate}
For each of these events, all others $J_{k\neq i}^\pm(t_j) =0 $.

\begin{widetext}
To proceed, we use the integral expression of the Dirac function $\delta(s) = \int (2\pi)^{-1} \mathrm{d}\hat{s}\, e^{i s\hat{s}}$, which allows us to rewrite the probability as 
\begin{align}
    P[\{n^\pm\}] =  \int\prod_{j=1}^M  \prod_{i=i}^L   \mathrm{d}\hat{n}_i^+ (t_j) \mathrm{d}\hat{n}_i^- (t_j) e^ {\hat{n}_i^+(t_j) \left[ n_i^+(t_{j+1}) - n_i(t_j)^+\right]}e^ {\hat{n}_i^-(t_j) \left[ n_i^-(t_{j+1}) - n_i(t_j)^-\right]} \left\langle e^{-\hat{n}_i^+ (t_j)J_i^+(t_j)-\hat{n}_i^- (t_j)J_i^-(t_j)} \right\rangle_{\{J^j\}}
\end{align}
where the imaginary unit is absorbed into the (now imaginary-valued) fields $\hat{n}^\pm_i(t_j)$, and the average is taken over the possible changes in particle states $\{J^j\}\equiv\left\{J_i^\pm(t_j), i\in \{1,\ldots,L\}\right\}$.

To evaluate this average, we denote $f(C) = \prod_{i=1}^L e^{-\hat{n}_i^+(t_j)J_i^+(t_j)-\hat{n}_i^-(t_j)J_i^-(t_j)}$ for a given particle configuration $C$, and write out
\begin{align}
    \langle f \rangle_{\{J^j\}} = \sum_{C\in\{J^j\} } f(C) P(C|\{n^j\}).
\end{align}
To proceed, we decompose this sum into the terms corresponding to different types of moves:
\begin{enumerate}
    \item $C_0$: all $J_i(t_j)=0$, there are no moves during $\mathrm{d}t$. $f(C_0) = 1$.
    \item $\mathcal{N}_d^j$ Diffusive move: for $i\rightarrow i\pm1$, $J_i^\sigma(t_j)=-1, J_{i\pm1}^\sigma(t_j)=+1$, $P(C|\{n^j\}) = \frac{D}{a^2} n_i^\sigma(t_j)\mathrm{d}t$
    \item $\mathcal{N}_{f,+}^j$ Particle at site $i$ flips from $+$ to $-$: $J_i^+(t_j) = -1$,$J_i^-(t_j) = +1$, and 
    \begin{equation}  
        P(C|\{n^j\}) =  n_i^+(t_j) g_i^+(\{n_+(t_j), n_-(t_j)\})\mathrm{d}t
    \end{equation}
    \item $\mathcal{N}_{f,-}^j$ Particle at site $i$ flips from $-$ to $+$: $J_i^+(t_j) = +1$,$J_i^-(t_j) = -1$, and 
    \begin{equation}
        P(C|\{n^j\}) = n_i^-(t_j)  g_i^-(\{n_+(t_j), n_-(t_j)\})\mathrm{d}t
    \end{equation}
\end{enumerate}
\begin{align}
    \langle f \rangle_{\{J^j\}} = f(C_0) P(C_0|\{n^j\}) & +  \sum_{C\in\mathcal{N}_d^j } f(C) P(C|\{n^j\}) \nonumber\\
    & +\sum_{C\in\mathcal{N}_c^j } f(C) P(C|\{n^j\}) \nonumber\\
    & +\sum_{C\in\mathcal{N}_a^j } f(C) P(C|\{n^j\}) 
\end{align}
Since $P(C_0|\{n^j\}) = 1 - \sum_{C\in\mathcal{N}_d^j } P(C|\{n^j\}) - \sum_{C\in\mathcal{N}_c^j } P(C|\{n^j\}- \sum_{C\in\mathcal{N}_a^j } P(C|\{n^j\}$, we have
\begin{equation}
    \langle f \rangle_{\{J^j\}} = 1+ T_d + T_{f,+} + T_{f,-}
\end{equation}
with $T_x = \sum_{C\in\mathcal{N}_x^j } (f(C)-1)P(C|\{n^j\}) $, $x\in\{d,f+,f-\}$. Since $T_x \propto \mathrm{d}t$ since the probabilities are themselves proportional to $\mathrm{d}t$, we can re-exponentiate to find
\begin{equation}
    \langle f \rangle_{\{J^j\}} =  \exp(T_d + T_{f,+} + T_{f,-}) + O(\mathrm{d}t^2)
\end{equation}
We can now write down each term in the sum:
\begin{subequations}
    \begin{align}
        T_d  = \sum_i & \frac{D}{a^2}\mathrm{d}t \left( \left[ e^{\hat{n}^+_i(t_j) - \hat{n}^+_{i+1}(t_j)}-1\right]n^+_{i}(t_j) + \left[ e^{\hat{n}^+_{i+1}(t_j) - \hat{n}^+_{i}(t_j)}-1\right]n^+_{i+1}(t_j) \right) \nonumber\\
        & + \frac{D}{a^2}\mathrm{d}t \left( \left[ e^{\hat{n}^-_i(t_j) - \hat{n}^-_{i+1}(t_j)}-1\right]n_{i}^-(t_j) + \left[ e^{\hat{n}^-_{i+1}(t_j) - \hat{n}_{i}^-(t_j)}-1\right] n^-_{i+1}(t_j) \right)\\
        T_{f,+} + T_{f,-} = \sum_i &  \mathrm{d}t \left( e^{\hat{n}_i^+(t_j)-\hat{n}_i^-(t_j)}-1 \right) g_i^+(\{n_+, n_-\})n_i^+(t_j) \nonumber\\
        & + \mathrm{d}t \left( e^{-\hat{n}_i^+(t_j)+\hat{n}_i^-(t_j)}-1 \right) g_i^-(\{n_+, n_-\})n_i^-(t_j)
    \end{align}
\end{subequations}
Putting everything together, we now have
\begin{align}
    P[\{n\}] =  \int\prod_{j=1}^M  \prod_{i=i}^L \mathrm{d}\hat{n}_i(t_j)e^S
\end{align}
With the exact microscopic action $S$ given by
\begin{align}
    S = \sum_{i,j} & \hat{n}_i^+(t_j) \left( n_i^+(t_{j+1})-n_i^+(t_j)\right) + \hat{n}_i^-(t_j) \left( n_i^-(t_{j+1})-n_i^-(t_j)\right) \nonumber\\
    &+ \frac{D\mathrm{d}t}{a^2}\left[n_i^+(t_j) \left( e^{\hat{n}^+_i(t_j) - \hat{n}^+_{i+1}(t_j)}-1\right) + n^+_{i+1}(t_j) \left( e^{-\hat{n}^+_i(t_j) + \hat{n}^+_{i+1}(t_j)} -1\right)  \right] \nonumber\\
    & + \frac{D\mathrm{d}t}{a^2}\left[n_i^-(t_j) \left( e^{\hat{n}^-_i(t_j) - \hat{n}_{i+1}^-(t_j)}-1\right) + n_{i+1}^-(t_j) \left( e^{-\hat{n}^-_i(t_j) + \hat{n}^-_{i+1}(t_j)} -1\right)  \right] \nonumber\\
    & + \mathrm{d}t \left( e^{\hat{n}_i^+(t_j)-\hat{n}_i^-(t_j)}-1 \right) g_i^+(\{n_+,n_-\})n_i^+(t_j) \nonumber\\
    & + \mathrm{d}t \left( e^{-\hat{n}_i^+(t_j)+\hat{n}_i^-(t_j)}-1 \right) g_i^-(\{n_+,n_-\}) n_i^-(t_j)
\end{align}

To obtain a continuum (in space) model, we have to switch from a discrete number representation to a continuum parametrization. To this end, we first replace the integer-valued $n_i^\pm$ by real-valued averages. We remark that as $a\rightarrow0$ and the system is dominated by diffusion, the random variables $n_i(t_j)^\pm$ are Poisson-distributed: we denote the parameters of these distribution $\rho_i^\pm(t_j)$, such that, for instance,
\begin{subequations}
    \begin{align}
        \langle n_i(t_j)^\pm \rangle &= \rho_i^\pm(t_j) \\
        \langle n_i^+(t_j)(n_i^+(t_j)-1)\rangle & = \sum_{l=0}^\infty \frac{[\rho_i(t_j)]^l}{l!}l(l-1)e^{-\rho_i(t_j)} = \rho_i(t_j)^2
    \end{align}
\end{subequations}
and we denote the averaged reaction term by 
\begin{equation}
    \langle g_\pm(\{n_+,n_-\})n_i(t_j)^\pm \rangle = f_\pm(\{\rho_+,\rho_-\})
\end{equation}
This implies that the average action, taken over this factorized Poisson distribution is given by 
\begin{align}
    \langle S \rangle = &\sum_{i,j} \hat{n}^+_i(t_j) (\rho^+_i(t_{j+1}) -\rho^+_i(t_j)) + \hat{n}^-_i(t_j) (\rho^-_i(t_{j+1}) -\rho^-_i(t_j)) \nonumber \\
    &+ \frac{D\mathrm{d}t}{a^2}\left[\rho_i^+(t_j) \left( e^{\hat{n}^+_i(t_j) - \hat{n}^+_{i+1}(t_j)}-1\right) + \rho_{i+1}^+(t_j) \left( e^{-\hat{n}^+_i(t_j) + \hat{n}^+_{i+1}(t_j)} -1\right)  \right] \nonumber\\
    &+ \frac{D\mathrm{d}t}{a^2}\left[\rho_i^-(t_j) \left( e^{\hat{n}^-_i(t_j) - \hat{n}^-_{i+1}(t_j)}-1\right) + \rho_{i+1}^-(t_j) \left( e^{-\hat{n}^-_i(t_j) + \hat{n}^-_{i+1}(t_j)} -1\right)  \right] \nonumber\\
    & + \mathrm{d}t \left( e^{\hat{n}_i^+(t_j)-\hat{n}_i^-(t_j)}-1 \right) f_+(\{\rho_+,\rho_-\}) \nonumber\\
    & +  \mathrm{d}t \left( e^{-\hat{n}_i^+(t_j)+\hat{n}_i^-(t_j)}-1 \right) f_-(\{\rho_+,\rho_-\})
\end{align}
Since the $\rho_i(t_j)$ are real-valued, we can now Taylor-expand in time $\rho_i(t_{j+1}) = \rho_i(t_{j})+\dot{\rho}_i\mathrm{d}t + O(\mathrm{d}t^2)$ to find
\begin{align}
    \langle S \rangle = \int \mathrm{d}t \sum_{i} \Bigg[ &\hat{n}_i^+(t_j) \dot{\rho}^+_i +\hat{n}_i^-(t_j) \dot{\rho}^-_i+ \frac{D}{a^2}\left[\rho_i^+(t_j) \left( e^{\hat{n}^+_i(t_j) - \hat{n}^+_{i+1}(t_j)}-1\right) + \rho_{i+1}^+(t_j) \left( e^{-\hat{n}^+_i(t_j) + \hat{n}^+_{i+1}(t_j)} -1\right)  \right] \nonumber\\
    & + \frac{D}{a^2}\left[\rho_i^-(t_j) \left( e^{\hat{n}^-_i(t_j) - \hat{n}^-_{i+1}(t_j)}-1\right) + \rho_{i+1}^-(t_j) \left( e^{-\hat{n}^-_i(t_j) + \hat{n}^-_{i+1}(t_j)} -1\right)  \right] \nonumber\\
    & + \left( e^{\hat{n}_i^+(t_j)-\hat{n}_i^-(t_j)}-1 \right) f_+(\{\rho_+,\rho_-\}) \nonumber  + \left( e^{-\hat{n}_i^+(t_j)+\hat{n}_i^-(t_j)}-1 \right) f_-(\{\rho_+,\rho_-\})\Bigg]
\end{align}
To simplify the action, we can change variables to the density $\rho$ and magnetization $m$ and their corresponding response fields $\hat{\rho}, \hat{m}$
\begin{equation}
    \rho_i = \rho_i^++\rho_i^-, \quad m_i = \rho_i^+ - \rho_i^-, \quad \hat{\rho}_i = \frac{\hat{n}_i^+ + \hat{n}_i^-}{2}, \quad \hat{m}_i = \frac{\hat{n}_i^+ - \hat{n}_i^-}{2}.
\end{equation}
With those new variables, we can Taylor-expand the particle and response fields as 
\begin{subequations}
    \begin{align}
        \rho_{i+1} &= \rho_i + a\partial_x \rho_i + \frac{a^2}{2}\partial_x^2 \rho_i + o(a^2) \\
        \hat{\rho}_{i+1} &= \hat{\rho}_i + a\partial_x \hat{\rho}_i + \frac{a^2}{2}\partial_x^2 \hat{\rho}_i+ o(a^2)
    \end{align}
\end{subequations}
and likewise for $m,\hat{m}$. Neglecting terms of order $\nabla^3$ and above, we have an effective continuum action
\begin{equation}
    \langle S \rangle = \int \mathrm{d}t \frac{\mathrm{d}x}{a} S[\rho, m, \hat{\rho}, \hat{m}] + o(a)
\end{equation}
with
\begin{align}
    S =  &\hat{\rho} \partial_t\rho + \hat{m} \partial_t m  + D(\partial_x \hat{\rho})(\partial_x \rho)  + D(\partial_x \hat{m})(\partial_x m) \nonumber \\ & + \frac{D}{2}[(\partial_x \hat{\rho} + \partial_x\hat{m})^2 + (\partial_x \hat{\rho} - \partial_x\hat{m})^2 ]\rho + \frac{D}{2}[(\partial_x \hat{\rho} + \partial_x\hat{m})^2 - (\partial_x \hat{\rho} - \partial_x\hat{m})^2 ] m \nonumber \\
    & + \left(e^{2\hat{m}}-1 \right)f_+\left(\frac{\rho+m}{2}, \frac{\rho-m}{2}\right) \nonumber + \left(e^{-2\hat{m}}-1 \right)f_-\left(\frac{\rho+m}{2}, \frac{\rho-m}{2}\right)
\end{align}
For higher-dimensional settings, the same construction leads to the same action with all derivatives replaced by gradients $(\partial_x \hat{\rho})(\partial_x \rho) \rightarrow (\nabla \hat{\rho})^T (\nabla \rho)$, $\partial_{xx} \rightarrow \nabla^2$.

In the presence of external sources localized at single sites, the contributions from a single term $\langle S \rangle = \sum_i \delta_{ij}$ in the average action become $S_\text{source} = a\delta(x)$ in the continuum action, leading to a source strength scaled by the lattice width $a$. In the microscopic rules used in main text where the reaction rate term at site $i$ is $g_i^\pm = \gamma e^{\mp \beta(n_+-n_-) + h_0 \delta_{ij}}$, we thus have contributions to the action $f_\pm\left(\frac{\rho+m+h(x)}{2}, \frac{\rho-m-h(x)}{2}\right)$ with $h(x) = a(h_0/\beta) \delta(x)$.

To find the equations of motion, we look for saddle-point solutions of 
\begin{equation}
    \frac{\delta \langle S\rangle }{\delta \rho } = \frac{\delta \langle S\rangle }{\delta m } =\frac{\delta \langle S\rangle }{\delta \hat{\rho} } =\frac{\delta \langle S\rangle }{\delta \hat{m} } =0
\end{equation}
The conditions a$\frac{\delta \langle S\rangle }{\delta \rho } = \frac{\delta \langle S\rangle }{\delta m }=0$ are satisfied for $\hat{\rho}=\hat{m}=0$, and the remaining conditions give $\partial_t \rho, \partial_t m$.

\subsubsection{Same-site interactions}

With same-site interactions, as studied in Ref.~\cite{martin_transition_2025, scandolo_active_2023}, we find
\begin{subequations}
    \begin{align}
        \partial_t  \rho  & =  D\partial^2_{xx} \rho \\
    \partial_t m & = D\partial^2_{xx} m - g(\delta\rho, m+h(x)) 
\end{align} \label{eq:cg_hydro_DIM}%
\end{subequations}
with the functions
\begin{subequations}
    \begin{align}
        g(\rho , m) & = 2\gamma e^{-\beta +(\cosh{\beta}-1)\rho} \left(\cosh{(\sinh(\beta)m)}m - \sinh{(\sinh(\beta)m)}\rho \right) \\ 
        \bar{g}(\rho , m) & = 2\gamma e^{-\beta +(\cosh{\beta}-1)\rho}
    \end{align}
\end{subequations}
Decomposing $\rho=\rho_0 + \delta\rho$, we recover the deterministic part of Eq.~\eqref{eq:fluct_hydro_full}.

\subsubsection{Nearest-neighbor interactions}

To investigate the effect of nearest-neighbor interactions, we consider a reaction term where the flip rate at site $i$ from $+$ to $-$ is $g_+^i = \gamma e^{\beta \sum_{\langle ij\rangle} (n_j^+ - n_j^-)}$ where $\sum_{\langle i,j\rangle}$ denotes the sum over neighboring sites (not including $i$ itself, which does not change the results qualitatively).

In this case, the reaction rate averaged over the factorized Poisson measure reads 
\begin{align}
    \langle n_i(t_j)^+ e^{-\beta\sum_{\langle i,k \rangle}[n^+_k(t_j) - n_k^-(t_j)]}\rangle & = \rho_i^+(t_j) \prod_{\langle i,k \rangle}\sum_{m=0}^\infty\sum_{l=0}^\infty  \frac{(\rho_k^+(t_j))^m}{m!}\frac{(\rho_k^-(t_j))^l}{l!} e^{-\rho_k^+(t_j) -\rho_k^-(t_j) - \beta(m-l)} \nonumber\\
    &= \rho_i^+(t_j) \prod_{\langle i,k\rangle} e^{-\rho_k^+(t_j) -\rho_k^-(t_j) + e^{-\beta}\rho_k^+(t_j) + e^{\beta}\rho_k^-(t_j)}
\end{align}
which lead to a continuum action
\begin{align}
    S =  &\hat{\rho} \partial_t\rho + \hat{m} \partial_t m  + D(\partial_x \hat{\rho})(\partial_x \rho)  + D(\partial_x \hat{m})(\partial_x m) \nonumber \\ & + \frac{D}{2}[(\partial_x \hat{\rho} + \partial_x\hat{m})^2 + (\partial_x \hat{\rho} - \partial_x\hat{m})^2 ]\rho + \frac{D}{2}[(\partial_x \hat{\rho} + \partial_x\hat{m})^2 - (\partial_x \hat{\rho} - \partial_x\hat{m})^2 ] m \nonumber \\
    & + \gamma \left(e^{2\hat{m}}-1 \right)\left(\frac{\rho+m}{2}\right) e^{(\cosh{\beta}-1)z\rho -z\sinh{\beta}m}f(\nabla^2 \rho, \nabla^2m) \nonumber\\
    &+ \gamma \left(e^{-2\hat{m}}-1 \right)\left(\frac{\rho-m}{2}\right) e^{(\cosh{\beta}-1)z\rho +z\sinh{\beta}m} f(\nabla^2 \rho, -\nabla^2m).
\end{align}
where $z$ is the coordination number of the square lattice ($z=2^d$ in $d$-dimensions) with the shorthand $f(\nabla^2 \rho, \nabla^2m) = 1 + a^2z(\cosh{\beta}-1)\nabla^2\rho + a^2z \sinh{\beta} \nabla^2m$. The corresponding dynamical equations are then given by
\begin{subequations}
    \begin{align}
        \partial_t \rho = &D \nabla^2 \rho \\
        \partial_t m = &D \nabla^2 m + 2\gamma e^{(\cosh\beta-1)z\rho}\left[ m\cosh(zm\sinh{\beta}) - \sinh(zm\sinh{\beta})\rho\right] \nonumber\\
        &+a^2 D_{\rho \rho}\rho \nabla^2\rho  +a^2 D_{\rho m}\rho \nabla^2 m  + a^2 D_{m\rho} m \nabla^2\rho  +a^2 D_{m m} m \nabla^2 m
    \end{align}
\end{subequations}
with the nonlinear diffusion coefficients
\begin{subequations}
    \begin{align}
        D_{\rho\rho} = & 2\gamma z e^{(\cosh\beta-1)z\rho}\sinh(zm\sinh{\beta})(1-\cosh{\beta}) \\
        D_{\rho m} = & 2\gamma z e^{(\cosh\beta-1)z\rho}\cosh(zm\sinh{\beta})\sinh{\beta}\\
        D_{m\rho} = & 2\gamma z e^{(\cosh\beta-1)z\rho}\cosh(zm\sinh{\beta})(1-\cosh{\beta})\\
        D_{mm} = & -2\gamma ze^{(\cosh\beta-1)z\rho}\sinh(zm\sinh{\beta})\sinh{\beta}
    \end{align}
\end{subequations}
\end{widetext}
At the deterministic level if $a\rightarrow0$, we recover the result of Eq.~\eqref{eq:cg_hydro_DIM} with a slightly different reaction term: In the limit $m\rightarrow 0$, $\delta\rho -\rho_0\rightarrow 0$ and introducing $\bar{\gamma} = 2\gamma e^{(\cosh\beta-1)z\rho_0}$ we now have at cubic order
\begin{subequations}
    \begin{align}
        \partial_t \delta\rho = &D \nabla^2 \delta\rho \\
        \partial_t m = &D\nabla^2 m + \bar{\gamma}(1-z\rho_0 \sinh{\beta}) m - \mu m \delta\rho - um^3
    \end{align}
\end{subequations}

If density fluctuations can be ignored (if $D\sim a^2 \gamma z$), the dynamics of the magnetization are given by
\begin{equation}
    \partial_t m = (D+a^2 \bar{\gamma}\rho_0z\sinh{\beta})\nabla^2 m + \bar{\gamma}(1-z\rho_0 \sinh{\beta}) m - um^3
\end{equation}
We recover a Landau-Ginzburg-type theory with a diffusion term that accounts for both hopping and color-switching (spin) dynamics.

\subsection{Coarse-graining the Wave-Pinning Ising}
\label{sec:wp_cg_derivations}

We here extend this coarse-graining approach to our lattice model of wave-pinning systems introduced in the Sec.~\ref{sec:nonlocal}. We now have a different set of moves, with a reservoir with $n^r_+, n^r_-$ particles of types $+$ and $-$. For our set of reactions, there are now six cases:
\begin{enumerate}
    \item[(i)] A particle hops from site $i$ to site $i+1$: $J_i^\pm(t_j) = -1$, $J_{i+1}^\pm(t_j) = 1$. 
    \item[(ii)] A particle hops from site $i$ to site $i-1$: $J_i^\pm(t_j) = -1$, $J_{i-1}^\pm(t_j) = 1$. 
    \item[(iii)] A $+$ particle unbinds at $i$ and goes into the reservoir: $J_r^+(t_j)  = +1 = - J_i^+(t_j)$.
    \item[(iv)] A $-$ particle unbinds at $i$ and goes into the reservoir: $J_r^-(t_j)  = +1 = - J_i^-(t_j)$.
    \item[(v)] A $+$ particle from the reservoir binds at site $i$: $J_i^+(t_j)  = +1 = - J_r^+(t_j)$.
    \item[(vi)] A $-$ particle from the reservoir binds at site $i$: $J_i^-(t_j)  = +1 = - J_r^-(t_j)$.
\end{enumerate}

The time-dependent and diffusion terms in the action are unchanged, but the reaction terms now read as 
\begin{align}
    S_{wp} = &\,\left(e^{\hat{n}_r^+(t_j)-\hat{n}_i^+(t_j)} -1 \right)r n_r^-(t_j)\mathrm{dt} \nonumber\\ &+ \left(e^{\hat{n}_r^-(t_j)-\hat{n}_i^-(t_j)} -1 \right)r n_r^-(t_j)\mathrm{dt} \nonumber \\
    & + \left(e^{\hat{n}_i^+(t_j)-\hat{n}_r^+(t_j)} -1 \right)R_+^{i,j} n_i^+(t_j)\mathrm{dt} \nonumber\\ &+ \left(e^{\hat{n}_i^-(t_j)-\hat{n}_r^-(t_j)} -1 \right)R_-^{i,j} n_i^-(t_j)\mathrm{dt}
\end{align}
where the reaction rates are $R_\pm^{i,j} = r + k n^\mp_i(t_j) (n^\mp_i-1)$. As before, we notice that as $D/a^2 \rightarrow+\infty$ diffusion dominates, leading to our Poisson ansatz such that 
\begin{subequations}
    \begin{align}
        \langle n_i(t_j)^\pm \rangle &= \rho_i^\pm(t_j) \\
        \langle R_\pm^{i,j} n^\pm_i(t_j)\rangle & = r\rho_i^\pm(t_j) - k [\rho_i^\mp(t_j)]^2 \rho_i^\pm(t_j)
    \end{align}
\end{subequations}
By number conservation, we also have
\begin{equation}
    \langle n_r^\pm(t_j)\rangle = \left\langle N_0^\pm - \sum_i n^\pm_i(t_j)\right\rangle = N_0^\pm - \sum_i \rho_i^\pm(t_j) \equiv \rho_r^\pm(t_j).
\end{equation}
After Taylor-expanding the particle and response fields in space and time, by variation of the action we find the dynamics
\begin{subequations}
    \begin{align}
        \partial_t \rho_+ & = D \nabla^2 \rho_+ - r\rho_+ - k\rho_-^2 \rho_+ + r\rho_r^+ \\
        \partial_t \rho_- & = D \nabla^2 \rho_- - r\rho_- - k\rho_+^2 \rho_- + r\rho_r^-
    \end{align}
\end{subequations}
with $\rho_r^\pm = N_0^\pm - (1/a)\int \mathrm{dx} \,\rho_\pm(x)$.

\subsection{Fluctuating hydrodynamics}

 Here we detail the derivation of the fluctuating hydrodynamics for the same-site interacting Diffusive Ising model~\cite{martin_transition_2025}; the fluctuating hydrodynamics for the other systems discussed above can be derived with a similar approach.  To proceed, we Taylor-expand to second order the action $\langle S\rangle$ in the response fields $\hat{\rho}, \hat{m}$ about $\hat{\rho}=\hat{m}=0$, keeping terms up to order $a$.
 \begin{widetext}
     
We find the action
\begin{equation}
    \langle S\rangle  =  \begin{pmatrix}
        S_0^\rho & S_0^m
    \end{pmatrix} \begin{pmatrix}
        \hat{\rho} \\ \hat{m}
    \end{pmatrix} + \frac{1}{2} \begin{pmatrix}
        \partial_x\hat{\rho} & \partial_x \hat{m} &\hat{m}
    \end{pmatrix} \bar{M} \begin{pmatrix}
        \partial_x\hat{\rho} \\ \partial_x \hat{m} \\\hat{m}
    \end{pmatrix},
\end{equation}
in which we introduced the deviations from the deterministic hydrodynamics
\begin{subequations}
    \begin{align}
        S_0^\rho & = \partial_t \rho - D\nabla^2\rho \\
        S_0^m &= \partial_t m - D\nabla^2 m + 2\gamma m e^{-\beta+\rho(\cosh{\beta}-1)}(\cosh[\sinh(\beta)m]m - \sinh[\sinh(\beta)m]\rho)
    \end{align}
\end{subequations}
with the shorthands $\hat{\gamma} = 2\gamma e^{-\beta+\rho(\cosh{\beta}-1)}$ and 
\begin{equation}
    S_f^m = \frac{1}{2}\left(\cosh[\sinh(\beta)m]\rho - \sinh[\sinh(\beta) m] m\right),
\end{equation}
and where we define the matrix
\begin{equation}
    \bar{M} = \begin{pmatrix}
        2D\rho & 2Dm & 0 \\
        2Dm & 2D\rho & 0\\
        0 & 0 & \hat{\gamma}S_f^m
    \end{pmatrix}.
\end{equation}
We can then rewrite the probability density as
\begin{equation}
    P[\rho,m] =\int \mathcal{D}\rho \mathcal{D}\hat{\rho}\mathcal{D}m\mathcal{D}\hat{m}\,\exp \left(\int \mathrm{d}t \mathrm{d}x \frac{1}{a}\begin{pmatrix}
        S_0^\rho & S_0^m
    \end{pmatrix} \begin{pmatrix}
        \hat{\rho} \\ \hat{m}
    \end{pmatrix} + \frac{1}{2a} \begin{pmatrix}
        \partial_x\hat{\rho} & \partial_x \hat{m} &\hat{m}
    \end{pmatrix} \bar{M} \begin{pmatrix}
        \partial_x\hat{\rho} \\ \partial_x \hat{m} \\\hat{m}
    \end{pmatrix}\right).
\end{equation}
Using a Hubbard-Stratonovich transform, we introduce three Gaussian fields $\eta_1, \eta_2, \eta_3$ such that
\begin{equation}
    P[\rho, m] =\int \mathcal{D}\rho \mathcal{D}\hat{\rho}\mathcal{D}m\mathcal{D}\hat{m}\eta_1\mathcal{D}\eta_2\mathcal{D}\eta_3 \exp \left(\int \mathrm{d}t \mathrm{d}x \frac{1}{a}\begin{pmatrix}
        S_0^\rho & S_0^m
    \end{pmatrix} \begin{pmatrix}
        \hat{\rho} \\ \hat{m}
    \end{pmatrix} + \frac{1}{\sqrt{a}}\left(
        \eta_1\;\eta_2 \; \eta_3
    \right)  \begin{pmatrix}
        \partial_x\hat{\rho} \\ \partial_x \hat{m} \\\hat{m}
    \end{pmatrix} - \frac{1}{2}\left(
        \eta_1\;\eta_2 \; \eta_3
    \right) M^{-1} \begin{pmatrix}
        \eta_1 \\ \eta_2 \\ \eta_3
    \end{pmatrix}\right).
\end{equation}
By integrating $\eta_1, \eta_2$ by parts, then integrating out the response fields $\hat{\rho}, \hat{m}$, the probability density reads
\begin{equation}
    P[\rho, m] =\int \mathcal{D}\rho\mathcal{D}m \mathcal{D}\eta_1\mathcal{D}\eta_2\mathcal{D}\eta_3\,\delta(S_0^\rho - \sqrt{a}\partial_x \eta_1)\delta(S_0^m - \sqrt{a}\partial_x \eta_2 -\sqrt{a}\eta_3 ) e^{-\int \mathrm{d}x \mathrm{d}t  \frac{1}{2}\left(
        \eta_1\;\eta_2\; \eta_3
    \right) \bar{M}^{-1} \begin{pmatrix}
        \eta_1 \\ \eta_2 \\ \eta_3
    \end{pmatrix}}.
\end{equation}
which leads to the final SPDEs
\begin{subequations}
\begin{align}
    \partial_t \rho & = D\nabla^2\rho + \sqrt{a}\partial_x \eta_1 \\
    \partial_t m & = D\nabla^2 m - 2\gamma m e^{-\beta+\rho(\cosh{\beta}-1)}(\cosh[\sinh(\beta)m]m - \sinh[\sinh(\beta)m]\rho) + \sqrt{a} \partial_x \eta_2 + \sqrt{a}\eta_3
\end{align}
\end{subequations}
with $\langle \eta_i(x,t)\eta_j(x',t')\rangle = \bar{M}_{ij}\delta(x-x')\delta(t-t')$, and the noise is to be interpreted in the It\^o sense due to the time-discretization employed in the model. For further RG work we will then rewrite this Langevin SPDE into a Martin-Siggia-Rose action [Eq.~\eqref{eq:MSR_action_RG}].
 \end{widetext}

\section{Self-consistency of fluctuating hydrodynamics by dynamical renormalization group analysis}
\label{sec:RG_calc}

 In this section, we derive the effective deterministic hydrodynamic equations for the diffusive Ising model in the presence of finite particle numbers by using the dynamical renormalization group (RG) approach. After rewriting our hydrodynamic equation in the Martin-Siggia-Rose formalism, we will derive the renormalization group equations in the absence of noise. We will then compute perturbatively the corrections to the RG flow equations due to finite noise and compute the effective correlation function used in Fig.~\ref{fig:from_lattice_to_continuum}e.
 
 We will work using the cubic order model of Eq.~\ref{eq:fluct_hydro_cube} applicable in the disordered phase and in the vicinity of the phase transition for small enough $a$. As $\rho_0 = 1$, this implies the condition $\sinh{\beta}\approx 1$, which implies $\cosh{\beta} \approx  2+\sqrt{2}$. 

\paragraph*{MSR action} The probability distribution for the fields $\rho, m$ can be written introducing response fields $\hat{\rho}, \hat{m} $ as 
\begin{equation}
    P[\rho, m] \propto \int \mathcal{D}[i\hat{\rho}]\mathcal{D}[i\hat{m}] \exp{\left(- S[\rho, m, \hat{\rho}, \hat{m}]\right)}
\end{equation}
\begin{widetext}

Breaking down Gaussian, anharmonic and noise contributions, we rewrite the Langevin PDEs Eq.~\eqref{eq:fluct_hydro_cube} as the MSR action
\begin{equation}
    S[\rho, m, \hat{\rho}, \hat{m}] = \int \mathrm{d}x \mathrm{d}t \left[ (S_0^\rho \;\; S_0^m ) \begin{pmatrix}
        \delta\hat{\rho} \\\hat{m}
    \end{pmatrix}  + \mathcal{A}[\delta \rho, m, \delta\hat{\rho}, \hat{m}] + \mathcal{A}_\text{ext}- \frac{a}{2} (\partial_x \delta\hat{\rho} \;\; \partial_x \hat{m} \;\; \hat{m} ) M \begin{pmatrix}
        \partial_x \delta\hat{\rho} \\ \partial_x \hat{m} \\ \hat{m} 
    \end{pmatrix}  \right] \label{eq:MSR_action_RG}
\end{equation}
where the Gaussian action density is given by
\begin{align}
    (S_0^\rho \;\; S_0^m ) \begin{pmatrix}
        \delta\hat{\rho} \\\hat{m}
    \end{pmatrix} = & \left[\partial_t \delta \rho - D \partial_{xx}^2 \delta \rho  \right]\hat{\rho} + \left[ \partial_t m - D \partial_{xx}^2 m + r m \right]\hat{m} 
\end{align}
The anharmonic contribution is given by
\begin{equation}
    \mathcal{A}[\delta \rho, m, \delta\hat{\rho}, \hat{m}] = \left[ \mu \delta \rho m + u m^3 \right]\hat{m} 
\end{equation}
while the noise terms up to first order noise are given by
\begin{align}
    \frac{a}{2} (\partial_x \delta\hat{\rho} \;\; \partial_x \hat{m} \;\; \hat{m} ) M \begin{pmatrix}
        \partial_x \delta\hat{\rho} \\ \partial_x \hat{m} \\ \hat{m} \end{pmatrix} = \frac{a}{2} \Big[ & 2D\rho_0 (\partial_x\delta\hat{\rho})^2  + 2D\rho_0 (\partial_x \hat{m})^2   + 2\hat{\gamma}\rho_0\hat{m}^2\Big].
\end{align}
\end{widetext}
The external field term contributes to the action as 
\begin{equation}
    \mathcal{A}_\text{ext} = - \hat{m} h
\end{equation}
The Gaussian propagators of the density and magnetization fields are given by
\begin{subequations}
    \begin{align}
        \langle \rho(x,t) \hat{\rho}(x',t') \rangle_0 & = G_0^\rho(x-x', t-t') \\
        \langle m(x,t) \hat{m}(x',t') \rangle_0 & = G_0^m(x-x', t-t')
    \end{align}
\end{subequations}
which are given in Fourier space by
\begin{subequations}
    \begin{align}
        G_0^\rho(q,\omega) & = \frac{1}{-i\omega + Dq^2} \\
        G_0^m(q,\omega) & = \frac{1}{-i\omega + Dq^2 + r}
    \end{align}
\end{subequations}
with $r = \hat{\gamma}(1-\sinh{\beta})$. We will choose the Fourier transform convention
\begin{subequations}
\begin{align}
    \phi(q, \omega) & = \int \mathrm{d}^d\mathbf{x}\mathrm{d}t \,e^{i(\mathbf{q}\cdot\mathbf{x}-\omega t)} \phi(\mathbf{x},t) \\ \phi(\mathbf{x}, t) &= \int \frac{\mathrm{d}^d\mathbf{q}}{(2\pi)^d}\frac{\mathrm{d}\omega }{2\pi}\, e^{-i(\mathbf{q}\cdot\mathbf{x}-\omega t)} \phi(\mathbf{q},\omega).
\end{align}
\end{subequations}
Since $m(x,t)$ and $\rho(x,t)$ are unitless, this implies that $m(q,\omega)$ and $\rho(q,\omega)$ have units of $L^d T$.

\subsection{0-loop: scaling in the absence of fluctuations}

We consider a renormalization group transformation in the absence of fluctuations ($a=0$) in $d$ dimensions. We introduce the extra parameters $\lambda_m, \lambda_\rho$ such that
\begin{align}
    (G_0^m)^{-1} = & -i\lambda_m \omega + D_m q^2 +r, \\
    (G_0^\rho)^{-1} = & -i\lambda_\rho \omega + D_\rho q^2.
\end{align}
while the noise vertices are given by
\begin{align}
    \Gamma_{\hat{m}\hat{m}} = & 2\Tilde{\gamma} + 2\Tilde{D}_m q^2, \\
    \Gamma_{\hat{\rho}\hat{\rho}} = & 2\Tilde{D}_\rho q^2.
\end{align}
In the bare theory $\lambda_m  = \lambda_\rho = 1$, $\Tilde{D}_m = \Tilde{D}_\rho= aD$, $D_m = D_\rho = D$ and $\tilde{\gamma} = a\hat{\gamma}$.
Momentum and time are rescaled by $q \rightarrow q/b$. As is the case for models C and D of the Hohnenberg-Halperin classification~\cite{hohenberg_theory_1977,tauber_critical_2014}, we allow for two distinct dynamical scaling exponents $z_\rho, z_m$ with $\omega \rightarrow b^{z_m} \omega$ in each integrand.  The fields are renormalized in Fourier space (momentum and frequency) as
\begin{equation}
    m\rightarrow \zeta_m m, \quad \hat{m}\rightarrow \hat{\zeta}_m \hat{m}, \quad \rho \rightarrow \zeta_\rho \rho, \quad \hat{\rho}\rightarrow \hat{\zeta}_\rho \hat{\rho}.
\end{equation}
We thus find, denoting by $u$ the coefficient of the cubic term
\begin{subequations}
\begin{align}
    \lambda_m' & = b^{-d} \zeta_m \hat{\zeta}_m \lambda_m \\
    \lambda_\rho' & = b^{-d} \zeta_\rho \hat{\zeta}_\rho \lambda_m \\
    D'_m & =   b^{-d+z_m -2} \zeta_m \hat{\zeta}_m D_m\\
    D'_\rho & = b^{-d+z_m -2} \zeta_\rho \hat{\zeta}_\rho D_\rho \\
    r' & = b^{-d+z_m} \zeta_m \hat{\zeta}_m r \\
    \mu' & = b^{-2d+z_m} \zeta_m \zeta_\rho \hat{\zeta}_m \mu \\
    u' & = b^{-3d+ z_m} \zeta_m^3 \hat{\zeta}_m u \\
\Tilde{D}_\rho' & = b^{-d+z_m-2} \hat{\zeta}_\rho^2  \Tilde{D}_\rho \\
\Tilde{\gamma}'& = b^{-d+z_m} \hat{\zeta}_m^2 \Tilde{\gamma}\\
\Tilde{D}_m' & =  b^{-d+z_m -2} \hat{\zeta}_m^2 \Tilde{D}_m
\end{align} \label{eq:gaussian_RG_scaling}%
\end{subequations}
We choose for the $\lambda$'s to stay constant such that 
\begin{equation}
    \zeta_m \hat{\zeta}_m = \zeta_\rho \hat{\zeta}_\rho  = b^d
\end{equation}
Choosing $z_m$ such that the diffusion constants stay finite, we have $z_m = 2$. We now have
\begin{subequations}
\begin{align}
    r' &=  b^{z_m} r \\
    \mu' & =  b^{-d+z_m} \zeta_\rho \mu \\
    u' & =  b^{-2d + z_m} \zeta_m^2  u \\
\Tilde{D}_\rho' & = b^{-d+z_m-2} \hat{\zeta}_\rho^2  \Tilde{D}_\rho \\
\Tilde{\gamma}'& = b^{-d+z_m} \hat{\zeta}_m^2 \Tilde{\gamma}\\
\Tilde{D}_m' & =  b^{-d+z_m -2} \hat{\zeta}_m^2 \Tilde{D}_m
\end{align}
\end{subequations}
To proceed, we need to find the scaling dimensions of the Gaussian fields. 
For the linear correlator to stay finite, we choose $\zeta_m, \zeta_\rho$ so that the conservative noises stay finite, which implies $\hat{\zeta}_m = b^{(d-z_m)/2+1}$ and $\hat{\zeta}_\rho = b^{(d-z_m)/2 +1}$. This leads to 
\begin{subequations}
    \begin{align}
        \zeta_m & = b^{(d+z_m)/2-1} \\
        \zeta_\rho & = b^{(d+z_m)/2 - 1} \\
    \end{align}
\end{subequations}
Without the noise, we thus find the recursion relations
\begin{subequations}
\begin{align}
    r' &= b^{z} r = b^2 r \\
    \mu' & = b^{3z/2-1-d/2} \mu = b^{2-d/2} \mu\\
    u'&  =  b^{ 2z-2 -d}  u = b^{2-d} u \\
    \Tilde{D}_m' & = b^{2} \Tilde{D}_m
\end{align}%
\end{subequations}

\subsection{Perturbative calculations} 
\begin{figure*}[t]
    \centering
\includegraphics[scale=1]{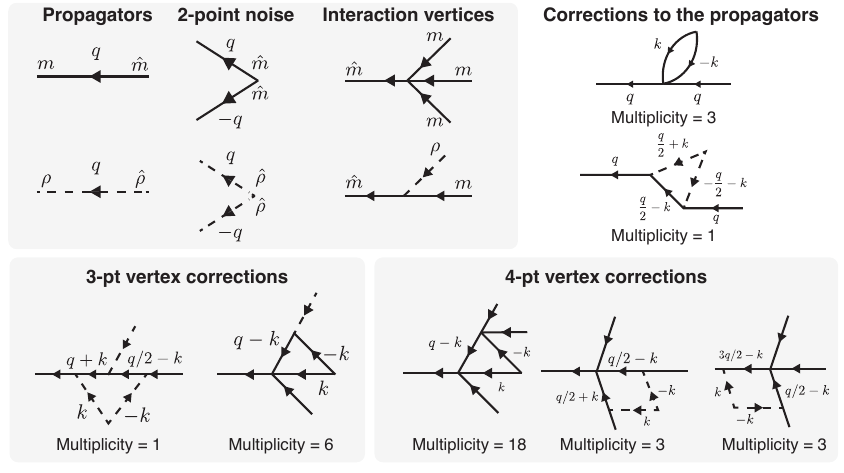}
    \caption{Elements of diagrammatic perturbation theory and first order diagrams for finite-$a$ effects.}
    \label{fig:diag_additive}
\end{figure*}

In what follows, we will set $\rho_0 = 1$ and use the notation $\rho, \hat{\rho}$ for $\delta\rho, \delta\hat{\rho}$. We use the MSR action to compute perturbed propagator and two-point correlation functions \cite{tauber_critical_2014}. 
The Gaussian propagators of the density and magnetization fields are given by
\begin{subequations}
    \begin{align}
        \langle \rho(x,t) \hat{\rho}(x',t') \rangle_0 & = G_0^\rho(x-x', t-t') \\
        \langle m(x,t) \hat{m}(x',t') \rangle_0 & = G_0^m(x-x', t-t')
    \end{align}
\end{subequations}
which are given in Fourier space by
\begin{subequations}
    \begin{align}
        G_0^\rho(q,\omega) & = \frac{1}{-i\omega + D_\rho q^2} \\
        G_0^m(q,\omega) & = \frac{1}{-i\omega + D_m q^2 + r}
    \end{align}
\end{subequations}
with $r = \hat{\gamma}(1-\sinh{\beta})$. 
The 2-pt (additive) noise vertices lead to Gaussian correlators
\begin{subequations}
    \begin{align}
        \langle m(q,\omega)m(q',\omega')\rangle_0 = & \delta(q+q')\delta(\omega+\omega')\nonumber\\ &\;\times |G_0^m(q,\omega)|^2  2\Tilde{\gamma}\\
        \langle \rho(q,\omega)\rho(q',\omega')\rangle_0 = & \delta(q+q')\delta(\omega+\omega') \nonumber\\ &\;\times|G_0^\rho(q,\omega)|^2 2\Tilde{D}_\rho q^2
    \end{align}
\end{subequations}
We will now proceed to do a perturbative treatment of the corrections induced by a small but finite value of $a = \Tilde{D}_\rho/D_\rho$ (or in terms of dimensionless number $a^2 \gamma/ D \ll 1$). At the first order in $a$, multiplicative noise corrections are negligible and the first order corrections are given by the two diagrams in Fig.~\ref{fig:diag_additive}.

\subsubsection{Self-energy}
\begin{widetext}
To 1-loop order, the effective magnetization propagator (vertex function $\Gamma_{\hat{m}m}(q,\omega)$) is given by 
\begin{align}
    \Gamma_{\hat{m}m}(q,\omega)= G^m(q,\omega)^{-1} & = G_0^m(q,\omega)^{-1} - \Sigma(q,\omega) \\
        &= -i\omega +r + D_mq^2 - \Sigma(q,\omega)
\end{align}
with $\Sigma(q,\omega)$ the self energy given by

\begin{align}
     \Sigma(q,\omega)  = & \,  -3 u  \int \frac{\mathrm{d}^d k}{(2\pi)^d}\frac{\mathrm{d}\nu}{2\pi} G_0^m(k,\nu)G_0^m(-k,-\nu) (2\Tilde{D}_mk^2+2\Tilde{\gamma})  \nonumber \\ & \left. + \mu^2 \int \frac{\mathrm{d}^d k}{2\pi}\frac{\mathrm{d}\nu}{2\pi} G_0^\rho\left(\frac{q}{2}+k,\frac{\omega}{2}+\nu\right)G_0^\rho\left(-\frac{q}{2}-k,-\frac{\omega}{2}-\nu\right)G_0^m\left(\frac{q}{2}-k,\frac{\omega}{2}-\nu\right)  2\Tilde{D}_\rho\left(\frac{q}{2}+k\right)^2\right]
\end{align}

We will now turn to each diagram in order.
\paragraph{Phi-4 process} We denote this contribution
\begin{equation}
    I_0 = -  3 u  \int \frac{\mathrm{d}^d k}{(2\pi)^d}\frac{\mathrm{d}\nu}{2\pi} \frac{2\Tilde{D}_mk^2+2\Tilde{\gamma}}{\left[D_mk^2+r\right]^2 +\nu^2} 
\end{equation}
We first integrate the frequency integral through elementary means in this one-dimensional system
\begin{equation}
    \int_{-\infty}^\infty\frac{\mathrm{d}\nu}{2\pi} \frac{2\tilde{\gamma}}{\left[D_mk^2+r\right]^2 +\nu^2}  = \frac{\tilde{D}_mk^2+\tilde{\gamma} }{D_mk^2+r}.
\end{equation}
Notice that this integral is singular at the critical point -- we will thus need to renormalize the correction in the vicinity of the critical point.

\paragraph{Coupling to density fluctuation }
We have, accounting for the factor $2\times 1/2!$ from the two interaction vertices,

\begin{equation}
    I_1(q,\omega) = \mu^2 \int \frac{\mathrm{d}^dk}{(2\pi)^d}\frac{\mathrm{d}\nu}{2\pi} \frac{2\Tilde{D}_\rho \left(\frac{q}{2}+k\right)^2}{D_\rho^2\left(\frac{q}{2}+k\right)^4 + \left(\frac{\omega}{2}+\nu \right)^2} \frac{1}{D_m\left(\frac{q}{2}-k \right)^2 -i\left(\frac{\omega}{2}-\nu \right)+r}
\end{equation}

We evaluate the integral over frequency $\int \mathrm{d}\nu$ using the residue theorem. The integrand has a pole at $\nu = \omega/2 +i D_m(q/2-k)^2 + ir$ which in the disordered phase ($r>0$) lies in the upper complex plane, along with 2 other poles at $\nu = -\omega/2 \pm iD_m(q/2+k)^2 $. Integrating over the lower complex plane, we find 
\begin{equation}
    I_1(q,\omega) =  \frac{\mu^2\Tilde{D}_\rho}{D_\rho} \int \frac{\mathrm{d}^dk}{(2\pi)^d} \frac{2}{4D_mk^2 + D_mq^2+2r - 2i\omega}
\end{equation}

We thus have a self energy $\Sigma(q,\omega) = I_0 + I_1(q, \omega)$, which leads to contributions 
\begin{align}
    \Gamma_{\hat{m}m}(0,0) & = r + 3u \int\frac{\mathrm{d}^dk}{(2\pi)^d} \frac{\Tilde{D}_mk^2+\Tilde{\gamma}}{D_mk^2+r} - \frac{\mu^2\Tilde{D}_\rho}{D_\rho}\int \frac{\mathrm{d}^dk}{(2\pi)^d} \frac{1}{D_m[1+w]k^2+r} \\
    \frac{\partial\Gamma_{\hat{m}m}}{\partial q^2}(0,0) & = D_m\left[1  + \frac{\mu^2\Tilde{D}_\rho}{4D_\rho}\int\frac{\mathrm{d}^dk}{(2\pi)^d} \left( \frac{D_m(1+w) }{(D_m(1+w)k^2+r)^2} - \frac{4}{d}\frac{D^2_m[1-w]^2k^2  }{(D_m[1+w]k^2+r)^3} \right)\right] \\
    \frac{\partial\Gamma_{\hat{m}m}}{\partial (-i\omega)}(0,0) & = 1 +\frac{\mu^2\Tilde{D}_\rho}{D_\rho} \int\frac{\mathrm{d}^dk}{(2\pi)^d} \frac{1}{(D_m[1+w]k^2+r)^2}
\end{align}
with $w=D_\rho/D_m$.

\subsubsection{Noise vertex correction}

We are interested in first-order perturbations at order $O(a)$ - since the noise vertex is already $\propto a$, we do not need to consider its $1$-loop order correction which contributes at order $O(a^2)$.

\subsubsection{3-pt vertex correction}
\emph{Density coupling:} From the diagram pertaining to the coupling to density fluctuation we find
\begin{align}
    I_3^\rho(q, p)  & = \mu^3  \int \frac{\mathrm{d}^d k}{(2\pi)^d} \frac{\mathrm{d}\nu}{2\pi} 2\Tilde{D}_\rho k^2 G_0^\rho(k,\nu) G_0^\rho(-k,-\nu) G_0^m(p-k,\omega-\nu)G_0^m(p+q-k,\omega+\omega'-\nu) \\
    & = \mu^3 \frac{\Tilde{D}_\rho}{D_\rho} \int \frac{\mathrm{d}^d k}{(2\pi)^d} \frac{\mathrm{d}\nu}{2\pi} \frac{2D_\rho k^2}{\nu^2 +D^2_\rho k^4} \frac{1}{D_m(p-k)^2 +r  -i(\omega-\nu)} \frac{1}{D_m(p+q-k)^2 + r  -i(\omega'-\nu)}
\end{align}
which after integration over the internal frequency $\nu$ and Taylor-expanding for $\omega, \omega', p, q \rightarrow 0$ gives
\begin{equation}
    I_3^\rho(q, p)  = \mu^3 \frac{\Tilde{D}_\rho}{D_\rho}  \int \frac{\mathrm{d}^d k}{(2\pi)^d} \frac{1}{(D_m[1+w] k^2+r)^2}.
\end{equation}
\emph{Magnetization coupling:} This diagram has multiplicity 6: 3 ways to choose which leg of the 4-pt vertex to attach to the external momentum, and then 2 ways each to attach the density external leg. Thus, the second diagram gives for vanishing external legs
\begin{align}
    I_3^m(q, p)  & = -6\mu u \int \frac{\mathrm{d}^d k}{(2\pi)^d} \frac{\mathrm{d}\nu}{2\pi} \left(2\Tilde{\gamma}+2\Tilde{D}_m k^2 \right)G_0^\rho(k,\nu) G_0^\rho(-k,-\nu) G_0^m(-k,-\nu) \\
    & = -6\mu u \int \frac{\mathrm{d}^d k}{(2\pi)^d} \frac{\mathrm{d}\nu}{2\pi} \frac{2\Tilde{\gamma}+2\Tilde{D}_m k^2}{\nu^2 +(D_m k^2+r)^2} \frac{1}{D_m(-k)^2 +r  -i(-\nu)}
\end{align}
We thus find the 3-pt vertex function
\begin{equation}
    \Gamma_{\hat{m}\rho m}(0,0) = \mu \left[ 1 + \frac{\mu^2 \Tilde{D}_\rho}{D_\rho} \int \frac{\mathrm{d}^d k}{(2\pi)^d} \frac{1}{(D_m[1+w] k^2+r)^2} - 3 u \int \frac{\mathrm{d}^d k}{(2\pi)^d} \frac{\Tilde{\gamma}+\Tilde{D}_m k^2}{(D_m k^2+r)^2} \right ]
\end{equation}

\subsubsection{4-pt vertex correction}
 We have 3 diagrams here, that we will evaluate at symmetrized external momenta. The first one comes from $\phi^4$
\begin{align}
    I_4^u = -18 u^2  \int \frac{\mathrm{d}^d k}{(2\pi)^d} \frac{\mathrm{d}\nu}{2\pi} \frac{2\Tilde{\gamma}+2\Tilde{D}_m k^2}{(D_mk^2+r)^2+\nu^2}\frac{1}{D_m(q-k)^2 + r - i (\omega-\nu) }
\end{align}
Integrating the internal frequency, we find
\begin{equation}
    I_4^u = -\frac{18}{2} u^2 \int \frac{\mathrm{d}^d k}{(2\pi)^d} \frac{\Tilde{D}_m k^2+\Tilde{\gamma}}{(D_m k^2+r)^2}
\end{equation}
Then we have 2 different sets of integrals which differ by the factors of external momenta, which give, at $q, \omega =0$ 
\begin{align}
    I_4^{\mu,1} & = 3 u \frac{\mu^2 \Tilde{D}_\rho}{D_\rho} \int \frac{\mathrm{d}^d k}{(2\pi)^d} \frac{\mathrm{d}\nu}{2\pi} 2D_\rho k^2 G_0^\rho(k,\nu) G_0^\rho(-k,-\nu) G_0^m(+k,\nu)G_0^m(-k,-\nu) \\
    & = 3u \frac{\mu^2 \Tilde{D}_\rho}{D_\rho} \int \frac{\mathrm{d}^d k}{(2\pi)^d} \frac{\mathrm{d}\nu}{2\pi}\frac{2D_\rho k^2}{D^2_\rho k^4+\nu^2} \frac{1}{(D_m k^2+r)^2+\nu^2}
\end{align}
and 
\begin{equation}
    I_4^{\mu,2}= 3u \frac{\mu^2 \Tilde{D}_\rho}{D_\rho} \int \frac{\mathrm{d}^d k}{(2\pi)^d} \frac{\mathrm{d}\nu}{2\pi}\frac{2D_\rho k^2}{D^2_\rho k^4\nu^2} \frac{1}{(D_m k^2+r-i\nu)^2}
\end{equation}
which after integration over the internal frequency give

\begin{equation}
    I_4(q, p) = I_4^{\mu,1}+I_4^{\mu,2} =  3 u\frac{\mu^2 \Tilde{D}_\rho}{D_\rho} \int \frac{\mathrm{d}^d k}{(2\pi)^d} \left( \frac{1}{(D_m[1+w]k^2+r)^2}+ \frac{1}{(D_mk^2+r)(D_m[1+w]k^2+r)} \right) 
\end{equation}

We thus find the vertex function

\begin{align}
    \Gamma_{\hat{m}mmm}(0,0)=  u\bigg[1 & - \frac{18}{2}u \int \frac{\mathrm{d}^d k}{(2\pi)^d} \frac{\Tilde{D}_m k^2+\Tilde{\gamma}}{(D_mk^2+r)^2}   \nonumber \\ & \left. + 3\frac{\mu^2 \Tilde{D}_\rho}{D_\rho} \int \frac{\mathrm{d}^d k}{(2\pi)^d} \left( \frac{1}{(D_m[1+w]k^2+r)^2}+ \frac{1}{(D_mk^2+r)(D_m[1+w]k^2+r)} \right)    \right]
\end{align}
\end{widetext}

\subsection{Momentum-shell renormalization}

Our lattice model provides us with a cutoff momentum $\Lambda = 2\pi/a$. To obtain the effective response at intermediate scales $0<q<\Lambda$, we compute the corrections to the propagators by using the momentum-shell renormalization group to regularize our singular perturbation theory. We integrate the short-wavelength modes contained in a shell with momentum $\Lambda/b < k < \Lambda$, such that $b = e^s$ with $s \ll 1$. In the regime we consider, the density has linear (Gaussian) dynamics and its propagator thus does not renormalize, which imposes that $z_\rho =2$.

The self-energy corrections are given to 1-loop order by
\begin{align}
    G^m(q,\omega)^{-1} = & -i\omega + D_m q^2 +r - I_0 -I_1 \\
    G^\rho(q,\omega)^{-1} = & -i\omega + D_\rho q^2 
\end{align}
consistent with the fact that the density propagator does not renormalize.
The contributions $I_0$ and $I_1$ to the self energy are then 
\begin{align}
    I_0 & = -  3u  \int^\Lambda_{\Lambda/b} S_d \Lambda^{d-1}\frac{\mathrm{d}k}{(2\pi)^d}\frac{\Tilde{\gamma}}{D_m k^2+r} \nonumber\\ & = - 3 u K_d \Lambda^{d}   \frac{\Tilde{\gamma}}{D_m\Lambda^2+r} \log b  
\end{align}
The regularization of the momentum integrals will lead generically to approximations of the form
\begin{equation}
    \int_{\Lambda/b}^\Lambda\frac{\mathrm{d}^dk}{(2\pi)^d} f(k) \approx K_d \Lambda^d f(\Lambda) \ln b
\end{equation}
as $
\log b\ll 1$ with $K_d = (2\pi)^{-d}  S_d$, where $S_d = 2\pi^{d/2}/\Gamma(d/2)$ is the $d$-dimensional solid angle (area of the unit sphere). 
These contributions lead to renormalized parameters at $r=0$ (via the vertex functions)
\begin{align}
    -i \omega_R &=  -i\omega \left[1  + \delta \Omega \ln b \right] \nonumber\\
    r_R & =  r \left[1 +  \delta r \ln b \right] \nonumber\\
    D_R & =  D \left[ 1 + \delta D_m \ln b \right] \\
    \mu_R & = \mu \left[ 1 + \delta \mu \ln b\right] \nonumber\\
    u_R & = u \left[ 1 + \delta u \ln b\right]\nonumber\\
    \Tilde{\gamma}_R & = \Tilde{\gamma} \left[1 + \delta \Tilde{\gamma}\ln b  \right]\nonumber
\end{align}
with contributions $\delta X$ summarized in Table~\ref{tab:corrections_1loop}.

\begin{table*}[t]
    \centering
    \begin{tabular}{|c|c|}
    \hline
    Correction & Expression \\
    \hline
         $\delta\Omega $ &  $\frac{\mu^2 a^d \rho_0}{(D_m[1+w] \Lambda^2 +r)^2} K_d \Lambda^{d-4}$ \\[10pt]
         $\delta r $ &  $ \frac{3ua^d \rho_0}{r} \frac{\Tilde{\gamma} K_d \Lambda^d}{D_m \Lambda^2+r} - \frac{\mu^2 a}{r}\frac{K_d \Lambda^d}{D_m[1+w]\Lambda^2+r}$ \\[10pt]
     $\delta D_m $ &  $\frac{\mu^2 a^d \rho_0 }{4}K_d \Lambda^d \left( \frac{D_m(1+w) }{(D_m[1+w]\Lambda^2+r)^2}-\frac{4}{d}\frac{D_m[1-w]^2\Lambda^2 }{(D_m[1+w]\Lambda^2+r)^3}\right) $\\[10pt]
     $\delta \mu $ & $ \mu^2 a^d \rho_0 K_d \Lambda^d\frac{1}{(D_m[1+w] \Lambda^2+r)^2} - 3u K_d \Lambda^d \frac{\Tilde{\gamma}}{(D_m \Lambda^2+r)^2} $ \\[10pt]
     $\delta u $ & $- 9u a^d \rho_0K_d\Lambda^d \frac{\Tilde{\gamma}}{(D_m\Lambda^2+r)^2} + 3\mu^2 a^d \rho_0  K_d\Lambda^d \left( \frac{1}{(D_m[1+w]\Lambda^2+r)^2}+ \frac{1}{(D_m\Lambda^2+r)(D_m[1+w]\Lambda^2+r)}\right)$ \\[10pt]
     $\delta \Tilde{\gamma}$ &  0 \\[10pt]\hline
    \end{tabular}
    \caption{Corrections to scaling introduced by fluctuations at 1-loop. Here $w=D_\rho/D_m$}
    \label{tab:corrections_1loop}
\end{table*}

\subsubsection{Rescaling of space and time}

We rescale space and time by $k \rightarrow bk$ and $\omega \rightarrow b^{z_m} \omega$, where $z_x$ are the (yet unknown) dynamical exponents for each fields. Under this rescaling with the same field renormalization as in the Gaussian case, the bare coefficients rescale as given by the equations in Eq.~\eqref{eq:gaussian_RG_scaling}
and the renormalization procedure leads to 
\begin{subequations}
    \begin{align}
    \omega' & =  b^{-d}\omega \hat{\zeta}_m \zeta_m \left[1  + \delta \Omega \ln b \right] =  b^{-d}\omega \hat{\zeta}_\rho \zeta_\rho \\
    r' & =  b^{-d+z_m} \hat{\zeta}_m \zeta_m r\left[ 1 + \delta r \ln b  \right ], \\
    D'_m & =  b^{-d+z_m-2}D_m \hat{\zeta}_m \zeta_m\left[ 1 + \delta D_m \ln b \right] \\
    D'_\rho &=  b^{-d+z_m-2}D_\rho \hat{\zeta}_\rho \zeta_\rho. \\
    \mu' & =  b^{-2d+z_m} \zeta_\rho \zeta_m \hat{\zeta}_m \mu \left[1 + \delta \mu \ln b \right]\\
    u' & = b^{-3d+z_m} \zeta_m^3 \hat{\zeta}_m u \left[1 + \delta u \ln b \right] \\
    \Tilde{D}'_\rho & = b^{-d+z_m-2}\hat{\zeta}_\rho^2 \Tilde{D}_\rho \\
    \Tilde{\gamma} & = b^{-d+z_m} \hat{\zeta}_m^2\Tilde{\gamma} \left[ 1  + \delta \Tilde{\gamma} \ln  b \right] \\
    \Tilde{D}_m' & = b^{-d+z_m-2} \hat{\zeta}_m^2\Tilde{D}_m \left[ 1  + \delta \Tilde{D}_m \ln  b \right]
    \end{align}
\end{subequations}
We choose the $\zeta$'s such that the frequency does not renormalize and for both conservative noise amplitudes to remain constant. Since the dynamical action of the density does not renormalize, we find again 
\begin{subequations}
    \begin{align}
    \zeta_\rho & =  b^{d/2+z_m/2 -1} \\
    \hat{\zeta}_\rho & = b^{d/2-z_m/2+1} 
    \end{align}
\end{subequations}
Does the magnetization field picks up an anomalous scaling exponent? The requirement that the conservative noise stays finite gives
\begin{equation}
    \hat{\zeta}_m = b^{d/2-z_m/2+1}\left[1  - \frac{1}{2}\delta \Tilde{D}_m \ln  b\right]
\end{equation}
while the time-normalization imposes
\begin{align}
    b^{-d} \hat{\zeta}_m \zeta_m \left[1  + \delta \Omega \ln b\right] =1
\end{align}%
which leads to
\begin{equation}
   \zeta_m = b^{d/2+z_m/2-1}\left[1  - \delta \Omega  \ln  b  + \frac{1}{2}\delta\Tilde{D}_m \ln  b   \right]
\end{equation}
where the dynamical exponent $z_m$ is still to be determined by the requirement that $D_m$ stays finite under the renormalization group flow \cite{cavagna_renormalization_2019}. Inserting the field scaling factors, we find
\begin{equation}
    D'_m  = b^{z_m-2} D_m \left[ 1 +\delta D_m \ln b - \delta \Omega \ln b \right]
\end{equation}
Which leads to a dynamical exponent
\begin{equation}
    z_m = 2  - \delta D_m^* + \delta \Omega^*
\end{equation}
where the starred quantities are taken at the RG fixed point values.
We can thus rewrite the recursion relations replacing the field scaling functions by their values
\begin{subequations}
    \begin{align}
    r' & =  b^{2}  r\left[ 1 + \delta D_m \ln b + \delta r \ln b  \right ], \\
    \mu' & =  b^{2 -d/2} \mu \left[1 + \frac{1}{2}\delta \Omega \ln b - \frac{3}{2}\delta D_m + \delta \mu \ln b \right]\\
    u' & = b^{2-d} u \left[1 - \delta \Omega \ln b - 2 \delta D_m\ln b +\delta \Tilde{D}_m\ln b + \delta u \ln b \right] \\
    \Tilde{\gamma}' & = b^{2} \Tilde{\gamma} \left[1+ \delta \Tilde{\gamma}\ln b  - \delta \Tilde{D}_m \ln b\right]
    \end{align}
\end{subequations}
The anomalous exponent $\eta$ such that $\zeta_m = b^{d/2 +1 - \eta/2}$ is thus given by
\begin{equation}
    \eta = \delta D_m^* +  \delta \Omega^*  - \delta\Tilde{\gamma}^*
\end{equation}

\subsubsection{Additive mass renormalization}
The critical point is now reached for a value $r=r_c$ where the vertex function $\Gamma_{\hat{m}m}(0,0) = 0$, which is at our 1-loop order given by
\begin{align}
    r_c & = - 3 u K_d \Lambda^d \frac{\Tilde{\gamma}}{D\Lambda^2 +r_c }\ln b + \mu^2 a \frac{K_d \Lambda^d}{2D\Lambda^2 +r_c }\ln b \nonumber \\ & = -\frac{3u \Tilde{\gamma}}{D} K_d \Lambda^{d-2}\ln b + \frac{\mu^2 a}{2} \frac{K_d \Lambda^{d-2}}{D} \ln b 
\end{align}

\paragraph{Dimensionless variables} We introduce an effective coupling constant and other dimensionless variables in which we will recast our theory
\begin{align}
    g &= \frac{\mu^2 a^d \rho_0}{4D^2_m} K_d\Lambda^{d-4},  \nonumber\\ f &= \frac{3u \Tilde{\gamma}}{D^2_m} K_d\Lambda^{d-4}, \nonumber\\ \alpha &= \frac{r}{D_m\Lambda^2}, \nonumber\\ w &= \frac{D_\rho}{D_m}
\end{align}

\subsection{Flow equations}

We thus write down our flow equations into differential $\beta$-functions with $s=\log b$
\begin{subequations}
    \begin{align}
        \frac{\mathrm{d}r}{\mathrm{d}s} & =  r \left[ 2  +\delta D_m + \delta r \right] \\
        \frac{\mathrm{d}\mu}{\mathrm{d}s} & = \mu \left[ 2-\frac{d}{2} +\frac{1}{2}\delta\Omega - \frac{3}{2}\delta D_m + \delta \mu \right]\\
        \frac{\mathrm{d}u}{\mathrm{d}s} & = u \left[ 2-d - \delta \Omega - 2\delta D_m + \delta\tilde{\gamma} + \delta u \right] \\
        \frac{\mathrm{d}\tilde{D}_m}{\mathrm{d}s} & = 0 \\
        \frac{\mathrm{d}\tilde{D}_\rho}{\mathrm{d}s} & =  (\delta\Omega-\delta D_m)D_m \\
        \frac{\mathrm{d}\tilde{\gamma}}{\mathrm{d}s} & =  \tilde{\gamma} \left[2 + \delta \tilde{\gamma} \right]\\
        \frac{\mathrm{d}\tilde{D}_\rho}{\mathrm{d}s} & = 0 \\
        \frac{\mathrm{d}\tilde{D}_m}{\mathrm{d}s} & = 0
    \end{align}
\end{subequations}
which we integrate up to a scale $s =\ln(\Lambda/q)$ to obtain scale-dependent parameters $r(s), \mu(s),D_m(s),\tilde{D}_m(s),D_\rho(s), \tilde{D}_\rho(s), \tilde{\gamma}$. 
With the rescaled momenta $q' = qe^s$, the power spectrum is then given by
\begin{equation}
    \langle |m_q|^2\rangle  = a^d\frac{\tilde{D}_m(s) (q'(s))^2 +\tilde{\gamma}(s)}{D_m(s) (q'(s))^2 +r(s)} = a^d\frac{\tilde{D}_m^R(q) q^2 +\tilde{\gamma}_R(q)}{D_m^R(q) q^2 +r_R(q)}
\end{equation}
Replacing $q'$ by its expression in terms of $q$ and $s$, we find the renormalized coefficients $\tilde{D}_m^R = \tilde{D}_m(0), D_m^R = D_m(0), \tilde{\gamma}_R(q) = \tilde{\gamma}(s)e^{-2s}, r_R(q) = r(s)e^{-2s}$ from which we can understand the effective scale-dependent response of the fluctuating hydrodynamic theory.

\subsection{Scaling analysis and interpretation}
\label{sec:rg_scaling_analysis}

We can recover the RG result that $g= \rho_0 \left(\hat{\gamma} a^2/D\right) \ll 1$ controls the convergence to the mean-field reaction-diffusion equations by a simplified scaling analysis. Consider the cubic order version of the model ignoring density fluctuations, which reduces to the Landau-Ginzburg dynamics
\begin{equation}
    \partial_t m = D \nabla^2 m + rm - um^3 + \sqrt{\hat{\gamma}a^d}\zeta(x,t)
\end{equation}
The goal is to understand the relevance of noise at the smallest available spatial scale, which is $L = a$. Introducing rescaled space, time, and fields by $x = a\Bar{x}$, $t = T \Bar{t}$ and $m = M \Bar{m}$, we have
\begin{equation}
    \partial_{\Bar{t}} \Bar{m} = \frac{DT}{L^2} \Bar{\nabla}^2 \Bar{m} + rT \Bar{m} - u M^2 T \Bar{m}^3 + \sqrt{T\frac{\hat{\gamma}a^d}{L^d M^2}} \Bar{\zeta}
\end{equation}
by choosing $T = L^2/D = a^2 /D $ the corresponding diffusion timescale and $M^2 = D/(a^2 u)$, we have 
\begin{equation}
    \partial_{\Bar{t}} \Bar{m} = \Bar{\nabla}^2 \Bar{m} + \frac{r a^2}{D} \Bar{m} - \Bar{m}^3 + \sqrt{\left(\frac{\hat{\gamma}u a^4}{D^2}\right)}  \Bar{\zeta}
\end{equation}
The mass is thus increased at large scales as in the Gaussian RG and, since $u\sim \hat{\gamma}/\rho_0 \sim \hat{\gamma}$, the dimensionless noise amplitude is as expected now $g = \left(\frac{\hat{\gamma} a^2}{D}\right)^2 \rho_0$, where $\rho_0$ is the average occupation number per site.

In the presence of conserved noise due to transport, the corresponding term has dimensionless amplitude
\begin{equation}
    \sqrt{\rho_0T\frac{Da^d}{M^2L^d}}\frac{1}{L} = \sqrt{\rho_0\left(\frac{\hat{\gamma} a^2}{D}\right)}
\end{equation}
which indicates a slightly different scaling with $f = \rho_0\left(\frac{\hat{\gamma} a^2}{D}\right)$. Whenever $\left(\frac{\hat{\gamma} a^2}{D}\right) < 1$, which should be common in chemical systems, $f$ is in fact the dominant noise term. In very dilute systems, non-conserved noise dominates.

The situation can be understood as a crossover away from Gaussian universality to a strongly-interacting fixed point \cite{cardy_scaling_1996}: in the same manner as the Ginzburg criterion quantifies the departure away from mean-field behavior as a nonlinearity appears and interacts with fluctuations, here number fluctuations interact with nonlinearities to bring the system away from the Gaussian RG fixed point. In this system, the noise source is due to the finite microscopic scale inducing dimension-dependent fluctuations. 

The dimensionless number $\mathrm{Da} = \left( \hat{\gamma} a^2 / D_m\right)$ can be understood as a Damk\"ohler number quantifying reaction timescale to residency time: in that picture, it indicates the convergence to local equilibrium within the microscopic-ranged `box' $a$. Alternatively, it is the ratio of the microscopic lengthscale $a$ to the typical distance traveled between reaction events.
In the limit where $\mathrm{Da} \ll 1$, particles travel far between interactions and the interactions are thus effectively long-ranged: as a long-ranged Ising model, we expect the mean-field behavior to hold. As a final remark, $\hat{\gamma} = \gamma e^{-\beta + \rho_0(\cosh{\beta}-1)} \geq \gamma$, with equality when $\sinh{\beta} = 1/\rho_0$: The effective reaction rate setting the validity range of the perturbative expansion is smallest at the critical point of the mean-field theory.

\bibliography{biblio}

\end{document}